\PassOptionsToPackage{pdfpagelabels=false}{hyperref}
\documentclass[fleqn,usenatbib]{mnras}
\usepackage[T1]{fontenc}
\usepackage{ae,aecompl}
\usepackage{graphicx}% Including figure files   
\usepackage{amsmath}% Advanced maths commands  
\usepackage{amssymb}% Extra maths symbols  
\usepackage{txfonts}
\usepackage[symbol]{footmisc}

\title[Bar and spiral strength]{Clash of the Trident and Tuning Fork: insights from bar and spiral strength in the (massive black hole)-stellar mass diagrams, and the `Triangal' galaxy evolution schema}
\author[Graham]
{
Alister W.\ Graham$^1$\thanks{E-mail: AGraham@swin.edu.au}
\\
$^1$ Centre for Astrophysics and Supercomputing, Swinburne University of Technology, Hawthorn, VIC 3122, Australia
}

\date{Accepted XXX. Received YYY; in original form ZZZ}
\pubyear{2026}

\begin{document}
\label{firstpage}
\pagerange{\pageref{firstpage}--\pageref{lastpage}}
\maketitle

\begin{abstract}
The `Triangal' galaxy evolution schema is used to assess whether the Tuning Fork (bar strength) or the van~den~Bergh Trident and ATLAS$^{3D}$ Comb (spiral strength) offer greater evolutionary insight. A new catalogue of quantitative bar strengths (measured by the bar-to-total luminosity ratio, $P$), refined galaxy morphologies, and dust bin classifications is presented.  It contains 137 galaxies with spheroid stellar masses, obtained from multi-component decompositions, and directly measured black hole masses, $M_{\rm bh}$. By placing these galaxies within the $M_{\rm bh}$--($M_{\rm\star,sph}$, $M_{\rm\star,gal}$) parameter space, an evolutionary reference frame reflecting integrated growth is established. 
Galaxies with varying bar strengths, and double bars, are observed to not occupy preferred locations, highlighting that bars are products of secular evolution---and can be transient or recurrent phenomena---and that they track neither hierarchical mass assembly nor galaxy speciation. In contrast, three physically distinct formation channels for S0/a galaxies are identified: (primeval S0)-to-S transitions; faded spiral galaxies; and, most commonly, wet-major-merger-built dust-rich S0 galaxies (on the `green mountain').   Galaxies with particularly strong spirals appear on the right-hand side of the spiral galaxy distribution.
Furthermore, a `Dust Attrition/Retention' sequence places S0 (and compact massive ES,b) galaxies with dusty nuclear discs between the dust-poor and dust-rich S0 galaxies, and a `Disc Down-sizing' sequence is revealed, in which E galaxies with dusty nuclear discs---potentially formed through `damp' mergers---bridge the ES,e (ellicular) galaxies with intermediate-scale stellar discs and the dust-poor pure E galaxies.
Extensive historical context is provided, and, finally, suspected biases in precision cosmology stemming from neglected precision galaxy morphology are discussed. 
\end{abstract}

% https://academic.oup.com/DocumentLibrary/mnras/keywords.pdf


\section{Introduction}

Galaxy morphology has long been used to interpret galaxy evolution, but the extent to which different classification schemas reflect underlying physical processes remains debated. Although early evolutionary interpretations based on the nebular hypothesis of the 1700s and 1800s were eventually abandoned, they firmly established the classical sequence from elliptical (E) to lenticular (S0--S0/a) to spiral (S) to irregular \citep[see][]{1919pcsd.book.....J, 1920MNRAS..80..746R} that has remained the backbone of schemas today. The recognition of barred structures \citep{1917ApJ....46...24P, 1918PLicO..13....9C} subsequently led \citet{1926ApJ....64..321H} to bifurcate this sequence, leading to the two-pronged Tuning Fork diagram \citep{1928asco.book.....J, 1936rene.book.....H}, in which the presence of a bar serves as a secondary classifier.
Through dynamical instabilities, bars are known to emerge from discs and subsequently evolve by redistributing angular momentum and mass within a galaxy, and   
\citet{1959HDP....53..275D} effectively introduced a third prong to the Fork, 
with the recognition of weak bars, giving a ternary bar-strength classification schema: absent (A), weak/intermediate (AB), and barred
(B) for the S {\it and} S0 galaxies.  
Among other intricate additions, \citet{1959HDP....53..275D} also recognised the S0/a galaxies with weak spiral patterns.

Addressing the unknown evolutionary connection between the two main broad types of galaxy noted by \citet{1920MNRAS..80..746R} --- namely, the amorphous cloudy type sometimes referred to as the armless spiral nebulae, i.e. the ellipticals, and the actual spiral galaxies containing granular knots thought
to have condensed out of the amorphous nebulae ---, \citet{1951ApJ...113..413S} proposed/reasoned that if two S galaxies collided, the gas particles (with their much greater number density) would collide while the stars (with their much lower number density) would pass by each other like the proverbial ships in the night. This encounter would effectively stop the gas in its tracks, while the two emergent (now gas-free) galaxies were expected to fade and lose their spiral patterns, and, in so doing, birth S0 galaxies. 
This dramatically reversed the evolutionary path heralded by Jeans and Reynolds. 
Although it is now known that colliding spiral galaxies, with their high dark matter content, tend instead to fall back onto each other and merge, the \citet{1976ApJ...206..883V} `Trident' was a new classification schema motivated, in part, by this idea from \citet{1951ApJ...113..413S}.  
While the \citet{1976ApJ...206..883V} Trident maintained the E-S0-Sa-Sb-Sc (portion of the) backbone from \citet{1920MNRAS..80..746R}, it contained a significant revision that clashed with the Tuning Fork because the strength of the spiral pattern, rather than the strength of the bar, was now proposed as the key discriminant among disc galaxies. This gave rise to an alternate set of parallel prongs.
\citet{1976ApJ...206..883V}
interpreted the relationships between his parallel sequences (Normal Spirals, Anemic Spirals, and Lenticulars) as evolutionary stages driven by gas depletion.
Favouring spiral strength over bar strength, this classification was echoed in the ATLAS$^{3D}$ Comb \citep{2011MNRAS.416.1680C}, 
which uses the angular momentum, quantified by the spin parameter, $\lambda$, as a measure of the stellar rotation-to-pressure support at or within some fixed galaxy radius, commonly $\lesssim$1~$R_{\rm e}$ due to observing constraints.  This spin parameter roughly correlates with the disc-to-bulge mass ratio but can be problematic 
because its strength may vary with radius in some galaxies \citep[e.g.,][]{2017MNRAS.470.1321B, 2017ApJ...840...68G}. 

Different structural features (bars, spiral arms, bulge-to-disc ratios) and kinematics have, thus, at different times been elevated to primary evolutionary signposts. The persistence of multiple, partially conflicting schemata highlights an unresolved question: which morphological features most faithfully trace a galaxy’s evolutionary state, rather than merely its instantaneous appearance?
To address this, a promising approach is to follow the lead of \citet{2011MNRAS.416.1680C} and move beyond purely 
taxonomic diagrams. This strategy is adopted here by examining galaxy morphology within parameter
spaces whose axes encode cumulative growth. For this purpose, the black hole
mass–spheroid stellar mass ($M_{\rm bh}$--$M_{\rm \star,sph}$) diagram is employed. Both
$M_{\rm bh}$ and $M_{\rm \star,sph}$ are expected to grow monotonically, or remain approximately
constant\footnote{The exceptions here are when gravitational tidal forces/stripping directs/pulls stars away from a galaxy, and also stellar winds and the production of compact stellar remnants, such as black holes and neutron stars.}, over cosmic time for most galaxies, thereby offering a natural evolutionary ordering. 
While both $M_{\rm bh}$ and $M_{\rm \star,sph}$ 
are expected to grow monotonically (to first order) over cosmic time, it is important to clarify what is meant here by an `evolutionary' diagram. The term is used in the sense that these quantities encode the integrated outcome of a galaxy’s growth history, rather than its instantaneous appearance. Consequently, galaxies occupying different regions of these diagrams are not merely morphologically distinct but have experienced systematically different assembly pathways.

While the sample size of galaxies with direct black hole mass measurements has steadily increased over the last decade \citep[e.g.,][]{2016ApJ...818...47S, 2016ApJ...817...21S, 2016ApJ...831..134V, 2019ApJ...887..195M, 2019ApJ...887...10S}, most studies have typically employed broad morphological classifications. This lack of granular detail---which serves as a fossil record of formation history---has effectively obscured the distinct evolutionary pathways evident in the Universe. By treating, for example, all S0 galaxies as a monolithic class, or overlooking the structural distinction between large- and intermediate-scale discs, or intermediate-scale discs (these do not dominate the light at large radii) and smaller nuclear discs (10s-100s of parsec), such works inadvertently smear out the morphology-dependent scaling relations required to track galaxy speciation.

Recent work has shown that galaxies do
not populate the $M_{\rm bh}$--$M_{\rm \star,sph}$ diagram along a single sequence. Instead, differing galaxy types define a set of somewhat parallel relations.  This, in turn, led to the ‘Triangal’ galaxy evolution
schema \citep{Graham-triangal}, which unifies the accretion-dominated and merger-dominated regimes. 
The ‘Triangal’ is a schematic representation of galaxy evolution within the $M_{\rm bh}$--$M_{\rm \star,sph}$ parameter space and others, in which distinct regions correspond to different dominant growth channels (e.g., major mergers versus disc growth). It is not a purely morphological classification, but a physically motivated framework grounded in how galaxies populate these mass-scaling relations.
The `Triangal' is already known to capture multiple evolutionary pathways, 
including disc growth, fading, and major mergers.  Furthermore, it is consistent with ultra-massive black holes, has already revealed three physically distinct classes of S0 galaxies linked to their formation histories \citep[e.g.][]{2024MNRAS.531..230G}, and even extends its utility to precision cosmology: exposing potential merger-driven biases in Baryon Acoustic Oscillation (BAO) measurements and offering a physically motivated means to refine Type~Ia supernova (SN~Ia) as standard candles (see Section~\ref{Sec_Cosmology}).
In contrast, aside from measures of different bulge-to-disc ratios, early classification schemas essentially had one class of S0 galaxy, with more recent work \citep[e.g.,][]{2014MNRAS.440..889S, 2021MNRAS.508..895D} recognising the two traditional classes based on ram-pressure-stripped and faded spirals \citep{1972ApJ...176....1G} and S0 galaxies built from collisions involving spirals \citep[e.g.,][]{1977egsp.conf..401T, 1979A&A....76...75R}.
The identification \citep{Graham-triangal} of a third class of `primeval' S0 galaxy\footnote{Here, `primeval' S0 galaxies refers to systems with relatively low stellar mass that are associated with a simple, rather than layered, formation.  They are not expected to have once had a spiral pattern.} offset to lower masses in the $M_{\rm bh}$--$M_{\rm \star,sph}$ and $M_{\rm bh}$--$M_{\rm \star,gal}$ diagrams suggests substantial unutilised potential in these diagrams that is only just starting to be realised elsewhere, for example, in refined interpretations of colour-magnitude/mass diagrams \citep{2024MNRAS.531..230G} and mass-(star formation rate) diagrams \citep{2024MNRAS.52710059G}. 

To further these investigations, 
this paper uses the $M_{\rm bh}$--$M_{\rm \star,sph}$ (and $M_{\rm bh}$--$M_{\rm \star,gal}$) diagram as a diagnostic tool to re-examine the long-standing debate over the evolutionary significance of bars and spiral structure. Specifically, it is investigated whether galaxies with strong versus weak bars, and with strong versus weak spiral patterns, occupy distinct regions of these diagrams. If bars or spiral arms represent key evolutionary stages, then their strength should correlate with position in a space that traces cumulative galaxy growth. Conversely, if bars and spirals are largely transient or recurrent phenomena, their presence may cut across evolutionary tracks rather than define evolutionary processes.

\begin{figure}
\begin{center}
\includegraphics[width=1.0\columnwidth]{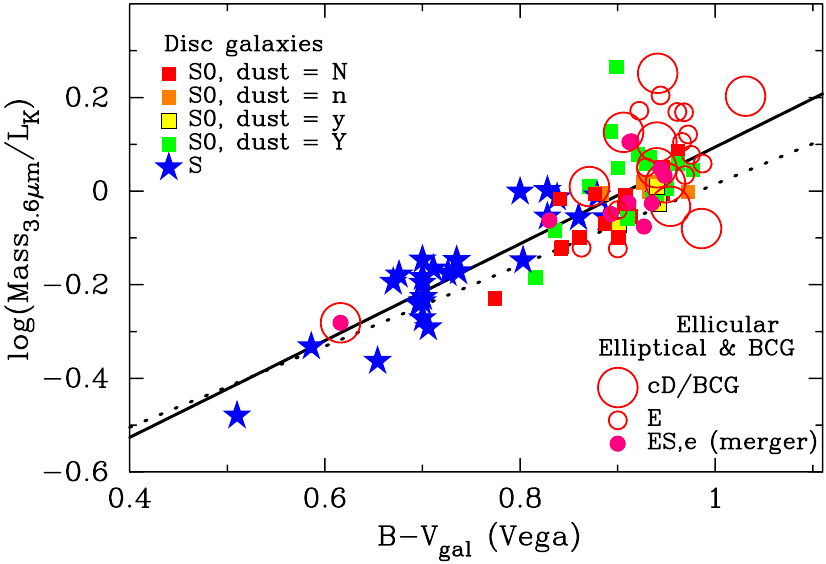}
\caption{Visual of the \textit{2MASS} $K_{\rm s}$-band stellar mass-to-light ratios, versus $B-V$ colour, that make the \textit{2MASS}-based stellar masses consistent with the \textit{SST}-based stellar masses from \citet{Graham:Sahu:22a}. The lines are explained in Section~\ref{Sec_mass}.
}
\label{FigML}
\end{center}
\end{figure}

The paper is organised as follows. Section~\ref{Sec_Data} describes the data and measurements used to populate the black hole mass scaling diagrams, including the derivation of stellar masses and the compilation of black hole masses (Section~\ref{Sec_mass}), the galaxy morphologies (Section~\ref{Sec_morph}), and the `dust bin' classifications that trace the origin of S0 galaxies (Section~\ref{Sec_dust}). Section~\ref{Sec_Data} also introduces the quantitative measure of bar strength via the bar-to-total stellar mass ratio ($P$, Section~\ref{Sec_weak_n_strong}), and defines the classification of weak, normal, and strong spiral structure (S0/a, S, and S$+$, Section~\ref{Sec_S0a}). 
Section~\ref{Sec_Anal} presents the results, first examining the distribution of weak and strong spiral patterns within the $M_{\rm bh}$--$M_{\star}$ diagrams (Section~\ref{Sec_spiral}), and then assessing whether bar strength correlates with position in these diagrams, including comparisons between weak, strong, and absent bars across both spiral and lenticular systems (Section~\ref{Sec_bars}). 

Section~\ref{Sec_d_and_c} discusses the implications of these findings, beginning with the development, interpretation, and predictive power of the black hole--stellar mass relations (Section~\ref{Sec_prog}), followed by an assessment of the role of stellar bars in galaxy evolution (Section~\ref{Sec_stellar_bar}). Subsequent subsections explore the tripartite nature of S0/a galaxies (Section~\ref{Sec_tripartite}), introduce the `Disc Down-sizing' and `Dust Attrition/Retention' sequences (Section~\ref{Sec_down_sizing}), and culminate in a resolution of the competing classification frameworks, namely the Tuning Fork, Trident, and Triangal (Section~\ref{Sec_Resolution}), before outlining avenues for future work in Section~\ref{Sec_future}. These include 
the presence of rings and nuclear bars (Section~\ref{Sec_Rings}), the potential role of bar length and spiral arm class (Section~\ref{Sec_barlength}), and the placement of barred systems within the $M_{\rm bh}$--$\sigma$ diagram and fundamental planes (Section~\ref{Sec_fun}). These subsections collectively assess whether secondary structural components introduce systematic offsets or increased scatter in the scaling relations. The discussion then broadens to what insight and implications can be transferred to cosmology, in particular the use of Type~Ia supernova as a probe of dark energy (Section~\ref{Sec_Cosmology}).

Readers interested in pioneering works---often ignored or long-forgotten---that discovered key components of galaxies and introduced, or were early adopters of, much of the nomenclature used herein may appreciate Appendix~\ref{AppdxA}. The dependence on galaxy morphology for interpreting speciation makes it useful to recognise how specific structures---bars, spirals, lenticulars, and ellipticals---entered the astronomical lexicon. Many assumptions embedded in modern classification schemas trace back to early twentieth-century interpretations explicitly intended to be evolutionary in nature, even though most were later softened or abandoned. After reviewing these origins, Appendix~\ref{AppdxA} presents a modern framework for the development of galaxy growth, into which the classical nomenclature is integrated.

Finally, Appendix~\ref{App_fits} provides updated multi-component decompositions for eight galaxies used in this study.

\section{Data}
\label{Sec_Data}

\subsection{Stellar and black hole masses}
\label{Sec_mass}

\subsubsection{Spitzer Space Telescope (\textit{SST}) data}

The supermassive black hole masses, spheroid stellar masses, and galaxy stellar masses for 102 galaxies in Table~\ref{Table-data} are taken from \citet[][and references therein]{Graham:Sahu:22a}. 
The gravitationally stripped galaxy NGC~4342 and NGC~5055 (no black hole mass free from substantial contamination from the surrounding stars) are excluded from the sample of 104 in \citet{{Graham:Sahu:22a}}.\footnote{Gravitationally stripped galaxies, including NGC~4486B and M32, are excluded from the analysis; they can be seen in \citet[][their figure~6]{Graham:Sahu:22b} to reside to the left of the $M_{\rm bh}$--$M_\star$ relations.}
Updates for the stellar masses of nine early-type galaxies (ETGs) have come from \citet{2024MNRAS.535..299G}, and updates for the black hole masses in NGC~5102 and NGC~5206 have come from \citet{2019ApJ...872..104N}. 
As noted by \citet{Graham:Sahu:22a}, the 3.6~$\mu$m galaxy light captured by the \textit{Spitzer Space Telescope's} (\textit{SST's}) Infrared Array Camera - channel 1 \citep[IRAC-1:][]{2004ApJS..154...10F} for these 102 galaxies 
was modelled using multi-component decompositions to separate distinct galaxy features 
\citep{2016ApJS..222...10S, 2019ApJ...876..155S, 2019ApJ...873...85D, Graham:Sahu:22b} 
rather than simply using the sum of $N$ S\'ersic functions.  
Stellar mass-to-light ($M/L$) ratios from \citet{2013MNRAS.430.2715I}, modified to the 3.6\,$\mu$m band by \citet{Graham:Sahu:22a}, were used.\footnote{\citet{2024MNRAS.530.3429G} clarifies that the $M/L$ ratios were calibrated to a diet-Salpeter initial mass function \citep[IMF:][]{2001ApJ...550..212B} and provides the conversion factors to other IMFs.}  
The \textit{SST}-based spheroid and galaxy 
stellar masses are supplemented here with an array of new information detailed later in Section~\ref{Sec_Data} and included in Table~\ref{Table-data}.

\begin{table*}
\centering
\caption{Galaxy parameters for 102 galaxies}\label{Table-data}
\begin{tabular}{lllcrrrrll}
\hline
Galaxy & \multicolumn{2}{l}{Morph.\ Type} & Dust & Dist & $\log(M_{\rm bh}/M_\odot)$ & $\log(M_{\rm \star,sph}/M_\odot)$  & $\log(M_{\rm \star,gal}/M_\odot)$  & $B/T$ & $P$  \\
Name     &  (old)  & (new)       & bin  & Mpc  &                 dex         &                       dex         &                       dex          &        &   Bar/Tot     \\
(1)      &   (2)   & (3)         & (4)  & (5)  &       (6)                   &           (7)                     &             (8)                    &  (9)   & (10)   \\
\hline
Circinus  &   SAb   &   same     & Y   & 4.21  &  6.25$\pm$0.11  &   9.46$\pm$0.29  &   9.97$\pm$0.22  &  0.31   & 0  \\ 
IC 1459   &  E3     &   same     &  y  & 28.1  &  9.38$\pm$0.20  &  11.69$\pm$0.17  &  11.69$\pm$0.17  &  1.00   & 0  \\
IC 2560  &   SBb   &     SABb   &  Y  & 32.9  &  6.52$\pm$0.11  &   9.66$\pm$0.23  &  10.69$\pm$0.18  &  0.09   & 0.057  \\
IC 4296   &  BCG E  &  same     &  n  & 46.9  &  9.10$\pm$0.09  &  11.72$\pm$0.14  &  11.74$\pm$0.14  &  0.96   & 0  \\ 
NGC 0224/M31 & SAb  &   same     &  Y  & 0.75  &  8.15$\pm$0.16  &  10.23$\pm$0.15  &  11.01$\pm$0.13  &  0.17   & 0  \\  
NGC 0253  &   SABc  &     SBc    &  Y  & 3.47  &  7.00$\pm$0.30  &   9.76$\pm$0.25  &  10.71$\pm$0.13  &  0.11   &  0.375  \\
NGC 0404  &   SA0   &   same     &  Y  & 3.06  &  5.74$\pm$0.10  &   8.03$\pm$0.50  &   9.19$\pm$0.16  &  0.07   & 0  \\
NGC 0524  &   SA0   &   same     &  Y  & 27.7  &  9.00$\pm$0.10  &  10.88$\pm$0.15  &  11.38$\pm$0.13  &  0.32   & 0  \\ 
NGC 0821  &   E6    &     ES,e   &  N  & 23.2  &  7.59$\pm$0.17  &  10.84$\pm$0.15  &  10.90$\pm$0.14  &  0.87   & 0  \\ 
NGC 1023  &   SB0   &      SAB0  &  N  & 11.0  &  7.62$\pm$0.05  &  10.33$\pm$0.16  &  10.89$\pm$0.14  &  0.28   & 0.083  \\ 
NGC 1097  &  SBb    & $^{\star}$same   &  Y  & 24.8  &  8.38$\pm$0.04  &  10.84$\pm$0.25  &  11.42$\pm$0.13  &  0.26   & 0.190 \\
NGC 1194  &  m/SA0  &    same    &  Y  & 54.1  &  7.82$\pm$0.04  &  10.78$\pm$0.16  &  11.01$\pm$0.14  &  0.59   & 0  \\ 
NGC 1275  &  BCG ES,e  &    same    &  Y  & 69.0  &  8.88$\pm$0.21  &  11.56$\pm$0.18  &  11.60$\pm$0.17  &  0.91   & 0  \\ 
NGC 1300  &  SBbc   &    same    &  Y  & 20.7  &  7.86$\pm$0.14  &   9.68$\pm$0.20  &  10.55$\pm$0.14  &  0.14   & 0.138   \\
NGC 1316  &  SAB0   &     m/SB0  &  Y  & 17.8  &  8.16$\pm$0.29  &  11.05$\pm$0.35  &  11.69$\pm$0.31  &  0.23   & 0.163  \\ 
NGC 1320  &  SAa    &    same    &  Y  & 36.8  &  6.77$\pm$0.22  &  10.13$\pm$0.16  &  10.70$\pm$0.14  &  0.27   & 0  \\ 
NGC 1332  & SA0/ES,b &   same    &  n  & 22.0  &  9.15$\pm$0.07  &  11.15$\pm$0.15  &  11.17$\pm$0.14  &  0.95   & 0  \\ 
NGC 1374  &   E      &  SA0      &  N  & 19.0  &  8.76$\pm$0.05  &  10.30$\pm$0.16  &  10.59$\pm$0.13  &  0.51   & 0  \\
NGC 1398  &  SBab    &   same    &  Y  & 24.8  &  8.03$\pm$0.11  &  10.76$\pm$0.29  &  11.44$\pm$0.20  &  0.21   & 0.104 \\
NGC 1399  &  BCG E1  &   same    &  N  & 19.2  &  8.67$\pm$0.06  &  11.46$\pm$0.16  &  11.46$\pm$0.16  &  1.00   & 0  \\
NGC 1407  &  E0     &    same    &  N  & 27.8  &  9.65$\pm$0.08  &  11.60$\pm$0.17  &  11.66$\pm$0.16  &  0.87   & 0  \\ 
NGC 1600  &  E3     &    same    &  N  & 71.7  & 10.25$\pm$0.04  &  12.06$\pm$0.13  &  12.06$\pm$0.13  &  1.00   & 0  \\ 
NGC 2273  &  SBa    &    same    &  Y  & 30.3  &  6.95$\pm$0.06  &  10.35$\pm$0.22  &  10.83$\pm$0.17  &  0.33   & 0.100  \\ 
NGC 2549  &  SA0    &       SB0  &  N  & 12.2  &  7.15$\pm$0.60  &   9.67$\pm$0.19  &  10.21$\pm$0.17  &  0.29   & 0.265  \\
NGC 2778  &   E   &       SAB0   &  N  & 22.1  &  7.18$\pm$0.35  &   9.49$\pm$0.23  &  10.15$\pm$0.17  &  0.22   & 0.056   \\
NGC 2787  & SB0 &         SAB0/a & y & 7.2  &  7.59$\pm$0.09  &   9.37$\pm$0.24  &  10.10$\pm$0.19  &  0.19   & 0.037  \\
NGC 2960  &  SAa   &     same    &  Y  & 73.0  &  7.07$\pm$0.05  &  10.44$\pm$0.16  &  10.86$\pm$0.14  &  0.38   & 0  \\ 
NGC 2974  &   E4   &     SA0/a   &  Y  & 21.5  &  8.23$\pm$0.07  &  10.48$\pm$0.22  &  10.98$\pm$0.16  &  0.32   & 0  \\ 
NGC 3031/M81 & SAab &    same    &  Y  & 3.48  &  7.83$\pm$0.09  &  10.34$\pm$0.25  &  10.83$\pm$0.13  &  0.32   & 0  \\ 
NGC 3079  &  SBc    &    same    &  Y  & 16.5  &  6.38$\pm$0.12  &   9.88$\pm$0.29  &  10.64$\pm$0.20  &  0.17   & 0.209  \\
NGC 3091  &   E3    &      SA0   & N$^{\#}$ & 58.6 &  9.62$\pm$0.08  &  11.25$\pm$0.22  &  11.70$\pm$0.20  &  0.35   & 0  \\ 
NGC 3115  & SA0/ES,b &   same    &  Y  & 9.3  &  8.94$\pm$0.25  &  10.87$\pm$0.14  &  10.95$\pm$0.13  &  0.83   & 0  \\ 
NGC 3227  & SABa   &     same    &  Y  & 25.7  &  7.97$\pm$0.14  &  10.31$\pm$0.22  &  11.07$\pm$0.16  &  0.17   & 0.023  \\
NGC 3245  &  SA0   &     same    &  n  & 20.1  &  8.30$\pm$0.12  &  10.12$\pm$0.17  &  10.70$\pm$0.15  &  0.26   & 0  \\
NGC 3368/M96 & SABab &     SBab  & Y  & 10.8  &  6.89$\pm$0.11  &   9.95$\pm$0.17  &  10.83$\pm$0.17  &  0.13   & 0.131, 0.060$^{**}$ \\
NGC 3377  &   E5-6  &   ES,e     &  y  & 10.8  &  7.89$\pm$0.03  &  10.30$\pm$0.14  &  10.36$\pm$0.13  &  0.87   & 0  \\
NGC 3379/M105 &  E1  &      SA0  & n$^{\#}$ & 10.9 &  8.62$\pm$0.13  &  10.27$\pm$0.20  &  10.89$\pm$0.18  &  0.24   & 0  \\ 
NGC 3384  &   SB0  &     same    &  N  & 11.2  &  7.23$\pm$0.05  &  10.14$\pm$0.16  &  10.67$\pm$0.14  &  0.30   & 0.100 \\
NGC 3414  &   SA0   &      ES,e  &  y  & 24.3  &  8.38$\pm$0.09  &  10.95$\pm$0.19  &  10.98$\pm$0.18  &  0.93   & 0  \\ 
NGC 3489  &  SAB0 &      SB0/a   &  Y  & 11.6 &  6.76$\pm$0.07  &   9.53$\pm$0.20  &  10.30$\pm$0.14  &  0.17   & 0.461  \\ 
NGC 3585  &    E6    &     ES,e  &  N  & 19.3  &  8.49$\pm$0.13  &  11.38$\pm$0.15  &  11.39$\pm$0.14  &  0.98   & 0  \\ 
NGC 3607  &    SA0    &      ES,e &  Y  & 25.0  &  8.16$\pm$0.18  &  11.29$\pm$0.18  &  11.37$\pm$0.17  &  0.83   & 0  \\ 
NGC 3608  &  E2     &    same    &  N  & 22.1  &  8.30$\pm$0.17  &  10.98$\pm$0.14  &  10.98$\pm$0.14  &  1.00   & 0  \\
NGC 3627/M66 & SABb &      SBb  &  Y  & 10.4  &  6.94$\pm$0.09  &   9.72$\pm$0.21  &  10.76$\pm$0.20  &  0.09   & 0.105 \\
NGC 3665  &   SA0   &    same    &  Y  & 34.7  &  8.76$\pm$0.10  &  11.14$\pm$0.16  &  11.39$\pm$0.14  &  0.56   & 0  \\ 
NGC 3842  &  BCG E  &    same    &  N  & 87.5  &  9.94$\pm$0.13  &  11.91$\pm$0.14  &  11.93$\pm$0.13  &  0.95   & 0  \\ 
NGC 3923  &  E4-5   &   same     &  N  & 22.1  &  9.47$\pm$0.13  &  11.55$\pm$0.17  &  11.55$\pm$0.17  &  1.00   & 0  \\ 
NGC 3998  &   SA0   &   same     &  y  & 13.6  &  8.33$\pm$0.43  &  10.12$\pm$0.26  &  10.61$\pm$0.15  &  0.32   & 0  \\ 
NGC 4026  &   SB0  &    same     &  y  & 13.2 &  8.26$\pm$0.12  &  10.19$\pm$0.22  &  10.44$\pm$0.17  &  0.56   & 0.126 \\
NGC 4151  &  SABab &     SBab  &  Y  & 19.0  &  7.69$\pm$0.37  &  10.28$\pm$0.22  &  10.62$\pm$0.17  &  0.46   & 0.122  \\
NGC 4258/M106 & SABbc &     SBbc & Y & 7.6  &  7.60$\pm$0.01  &  10.02$\pm$0.19  &  10.70$\pm$0.13  &  0.21   & 0.286  \\
NGC 4261  &  E2-3  &    same     &  n  & 30.4 &  9.20$\pm$0.09  &  11.52$\pm$0.16  &  11.54$\pm$0.15  &  0.96   & 0  \\
NGC 4291  &   E    &      ES,e   &  N  & 25.2  &  8.51$\pm$0.37  &  10.68$\pm$0.19  &  10.72$\pm$0.18  &  0.91   & 0  \\
NGC 4303/M61 & SABbc & $^{\star}$SBbc & Y & 19.3 &  6.78$\pm$0.17  &   9.60$\pm$0.25  &  10.66$\pm$0.13  &  0.09   & 0.121  \\
NGC 4339  &   E0    &       SA0  &  N  & 15.8  &  7.62$\pm$0.33  &   9.73$\pm$0.21  &  10.22$\pm$0.14  &  0.32   & 0  \\
\hline
\end{tabular}
\end{table*}

\setcounter{table}{0}

\begin{table*}
\centering
\caption{Continued}
\begin{tabular}{lllcrrrrll}
\hline
Galaxy & \multicolumn{2}{l}{Morph.\ Type} & Dust & Dist & $\log(M_{\rm bh}/M_\odot)$ & $\log(M_{\rm \star,sph}/M_\odot)$  & $\log(M_{\rm \star,gal}/M_\odot)$ & $B/T$ & $P$ \\
Name     &  (old)  &  (new)       & bin  & Mpc  &                 dex         &                       dex         &                       dex          &       &    Bar/Tot      \\
(1)      &   (2)   &  (3)         & (4)  & (5)  &       (6)                   &           (7)                     &             (8)                    &  (9)  &  (10)    \\
\hline
NGC 4350  &  SA0  &       SAB0    &  n  & 17.0 &  8.87$\pm$0.41  &  10.39$\pm$0.25  &  10.66$\pm$0.13  & 0.54 & 0.050  \\ 
NGC 4371  &  SB0  &        SAB0/a & n  & 16.4 &  6.84$\pm$0.12  &   9.99$\pm$0.30  &  10.70$\pm$0.22  & 0.19 & 0.051  \\
NGC 4374/M84 & E1     &    same   &  Y  & 17.7 &  8.95$\pm$0.05  &  11.61$\pm$0.15  &  11.61$\pm$0.15  & 1.00 & 0  \\
NGC 4388  &   SAb  &     SABab    &  Y  & 18.0 &  6.90$\pm$0.10  &  10.07$\pm$0.30  &  10.45$\pm$0.21  & 0.42 & 0.023 \\
NGC 4395  &   SAm  &     SABm     &  Y  & 4.56 &  5.62$\pm$0.17  &   ...            &   9.15$\pm$0.13  & 0.00 & 0.039 \\
NGC 4429  & SA0  &         SAB0/a &  Y & 16.6 &  8.18$\pm$0.08  &  10.60$\pm$0.20  &  11.04$\pm$0.13  &  0.36  & 0.008 \\ 
NGC 4434  &   SA0     &    same   &  N  & 22.3 &  7.85$\pm$0.17  &   9.95$\pm$0.20  &  10.22$\pm$0.13  & 0.54 & 0  \\
NGC 4459  &   SA0     &    same   &  Y  & 15.5 &  7.82$\pm$0.10  &  10.56$\pm$0.21  &  10.77$\pm$0.15  & 0.62 & 0  \\ 
NGC 4472/M49 & BCG E2   &   same  &  N  & 15.7 &  9.36$\pm$0.04  &  11.75$\pm$0.13  &  11.75$\pm$0.13  & 1.00 & 0  \\
NGC 4473  &   E5      &    ES,e   &  N  & 15.1 &  8.07$\pm$0.36  &  10.75$\pm$0.13  &  10.83$\pm$0.13  & 0.83 & 0  \\ 
NGC 4486/M87 & BCG E0-1 &   same  &  y  & 16.8 &  9.81$\pm$0.06  &  11.58$\pm$0.15  &  11.58$\pm$0.15  & 1.00 & 0  \\
NGC 4501/M88 & SAb  & $^{\star}$same &  Y & 17.0 & 7.31$\pm$0.08 &  10.46$\pm$0.25  &  11.02$\pm$0.13  & 0.28 & 0  \\
NGC 4526  &  SAB0   &      SAB0/a &  Y  & 16.3 &  8.65$\pm$0.04  &  10.79$\pm$0.26  &  11.13$\pm$0.15  & 0.46 & ... \\ 
NGC 4552/M89 &  E0-1   &    ES,e  &  y  & 14.8 &  8.67$\pm$0.05  &  11.01$\pm$0.16  &  11.07$\pm$0.16  & 0.87 & 0  \\ 
NGC 4564  &    E6  &         SA0  &  N  & 14.4 &  7.77$\pm$0.06  &  10.08$\pm$0.16  &  10.35$\pm$0.14  & 0.54 & 0 \\  
NGC 4578  &   SA0  &      same    &  N  & 16.2 &  7.28$\pm$0.35  &   9.79$\pm$0.15  &  10.24$\pm$0.13  & 0.35 & 0  \\ 
NGC 4594/M104 &  SAa  &     SA0/a &  Y  & 9.55 &  8.81$\pm$0.03  &  11.04$\pm$0.25  &  11.26$\pm$0.13  & 0.60 & 0  \\ 
NGC 4596  & SB0  &        SAB0/a & Y & 17.0 &  7.90$\pm$0.20  &  10.28$\pm$0.20  &  10.86$\pm$0.13  & 0.26 & 0.089 \\
NGC 4621/M59 &  E5    &      ES,e &  N  & 17.6 &  8.59$\pm$0.06  &  11.24$\pm$0.16  &  11.28$\pm$0.15  & 0.91 & 0  \\ 
NGC 4649/M60  &  E2   &    SA0    & N$^{\#}$ & 16.2 &  9.66$\pm$0.10 &  10.81$\pm$0.16 &  11.48$\pm$0.14  & 0.21 & 0  \\ 
NGC 4697  &   E6    &     SA0     & n$^{\#}$ & 11.3 &  8.26$\pm$0.04 &  10.20$\pm$0.16 &  10.86$\pm$0.14  & 0.22 & 0  \\ 
NGC 4699  &   SABb  &    same     &  Y  & 20.4 &  8.27$\pm$0.09  &  10.30$\pm$0.25  &  11.27$\pm$0.20  & 0.11 & 0.023 \\
NGC 4736/M94 & SAab &   same      &  Y  & 5.0 &  6.83$\pm$0.11  &  10.03$\pm$0.21  &  10.51$\pm$0.14  & 0.33 & ...  \\ 
NGC 4742  &    E4   &      SA0    &  N  & 14.9 &  7.13$\pm$0.18  &   9.78$\pm$0.16  &  10.07$\pm$0.14  & 0.51 & 0  \\  
NGC 4762  &   SB0   &   same      &  N  & 17.0 &  7.24$\pm$0.14  &   9.74$\pm$0.15  &  10.83$\pm$0.13  & 0.08 & 0.184  \\
NGC 4826/M64 & SAab &   same      &  Y  & 7.2 &  6.18$\pm$0.12  &   9.88$\pm$0.21  &  10.74$\pm$0.15  & 0.14 & 0  \\ 
NGC 4889  &  BCG E4 &   same      &  N  & 96.3 & 10.30$\pm$0.44  &  12.26$\pm$0.14  &  12.26$\pm$0.14  & 1.00 & 0  \\ 
NGC 4945  &  SBcd   &    same     &  Y  & 3.56 &  6.13$\pm$0.30  &   9.29$\pm$0.20  &  10.42$\pm$0.13  & 0.07 & ...  \\
NGC 5018  &   E3    &   m/SA0     &  Y  & 38.4 &  8.00$\pm$0.08  &  10.93$\pm$0.16  &  11.31$\pm$0.14  & 0.42 & 0  \\ 
NGC 5077  &  E3-4   &    same     &  y  & 39.8 &  8.85$\pm$0.23  &  11.27$\pm$0.21  &  11.37$\pm$0.20  & 0.79 & 0  \\ 
NGC 5128  &   m/SA0 &     same    &  Y  & 3.76 &  7.65$\pm$0.12  &  10.71$\pm$0.25  &  11.14$\pm$0.12  & 0.37 & 0  \\ 
NGC 5252  &   SA0   &      same   &  Y  & 104.0 &  9.03$\pm$0.40  &  10.97$\pm$0.27  &  11.50$\pm$0.15  & 0.30 & 0  \\ 
NGC 5419  &  BCG E  &      same   &  N  & 57.0 &  9.87$\pm$0.14  &  11.87$\pm$0.16  &  11.87$\pm$0.14  & 1.00 & 0  \\ 
NGC 5576  &  E3     &      same   &  N  & 24.5 &  8.19$\pm$0.10  &  10.90$\pm$0.16  &  10.90$\pm$0.16  & 1.00 & 0  \\ 
NGC 5813  &   E1-2   &   SA0      &  y  & 31.0 &  8.83$\pm$0.06  &  10.96$\pm$0.16  &  11.34$\pm$0.14  & 0.42 & 0  \\ 
NGC 5845  &  E    &     SA0/ES,b  &  n  & 25.0 &  8.41$\pm$0.22  &  10.26$\pm$0.21  &  10.46$\pm$0.15  & 0.63 & 0  \\ 
NGC 5846  &  E0-1 &        same   &  y  & 24.0 &  9.04$\pm$0.06  &  11.55$\pm$0.15  &  11.55$\pm$0.15  & 1.00 & 0  \\ 
NGC 6251  &  E    &        same   &  n  & 104.6 &  8.77$\pm$0.16  &  11.80$\pm$0.16  &  11.84$\pm$0.15  & 0.91 & 0  \\ 
NGC 6861  & SA0/ES,b    &  same   &  Y  & 27.0 &  9.30$\pm$0.08  &  11.07$\pm$0.19  &  11.15$\pm$0.18  & 0.83 & 0  \\ 
NGC 6926  &  SBbc  & $^{\star}$SABbc & Y  & 85.6 &  7.68$\pm$0.50  &   ...          &  11.13$\pm$0.14  & 0.00 & 0.087  \\
NGC 7052  &  E       &     same   &  n  & 61.9 &  9.35$\pm$0.05  &  11.46$\pm$0.13  &  11.46$\pm$0.13  & 1.00 & 0  \\
NGC 7332  &  SA0  &        SAB0    &  N  & 22.2 &  7.06$\pm$0.20  &  10.17$\pm$0.17  &  10.79$\pm$0.15  & 0.24 & 0.083 \\
NGC 7457  &   SA0   &      same   &  N  & 12.7 &  6.96$\pm$0.30  &   9.34$\pm$0.17  &  10.12$\pm$0.15  & 0.17 & 0  \\  
NGC 7582  &  SBab   &      same   &  Y  & 22.2 &  7.72$\pm$0.12  &  10.28$\pm$0.29  &  10.90$\pm$0.20  & 0.24 & 0.254 \\
NGC 7619  &  E      &      same   &  N  & 46.6 &  9.36$\pm$0.09  &  11.69$\pm$0.14  &  11.71$\pm$0.13  & 0.96 & 0  \\ 
NGC 7768  &  BCG E  &      same   &  n  & 108.2 &  9.09$\pm$0.15  &  11.90$\pm$0.16  &  11.90$\pm$0.16  & 1.00 & 0  \\ 
UGC 3789  &  SAab   &    SABab    &  Y  & 50.7 &  7.07$\pm$0.05  &  10.11$\pm$0.26  &  10.68$\pm$0.15  & 0.27 & 0.042 \\ 
\hline
\end{tabular}

Column~1 = Galaxy indentification.  
Columns~2 and 3 = Morphological type (old = entry in NED, primarily from the RC3; new = as informed  by the multicomponent decompositions). 
B = strong bar ($P\ge0.1$), AB = weak bar, A = no significant bar. 
m/SA0 = unbarred S0 merger remnant. 
Column~4 = Dust bin (Y = strong yes, y = weak and widespread, n = nuclear only, N = no obvious dust). 
Column~5 = Luminosity distance in Mpc \citep[][and references therein, unless otherwise noted in the text]{2019ApJ...887...10S, Graham:Sahu:22a}.
Column~6 = Logarithm of the black hole mass \citep[][and the reference chain therein]{Graham:Sahu:22a}. 
Columns~7 and 8 = Logarithm of the spheroid and galaxy stellar mass. 
Column~9 = Bulge-to-Total stellar mass ratio, $B/T$. 
Column~10 = Bar-to-Total stellar mass ratio, $P$.
Tablenotes: 
$^{\star}$ Very strong spiral.  
$^{**}$ Two bars are present in the decomposition (see Section~\ref{Sec-dubB}). 
$^{\#}$ Dust classification somewhat misleading because the galaxy is recognised as having experienced a major merger, which is the key underlying criterion typically signalled by a dust-rich character (see Section~\ref{Sec_dust}); these galaxies should arguably be moved to the dust $=$ Y bin, but this has not been done here.
A machine-readable version of this table is available online. 
\end{table*}

\begin{table*}
\centering
\caption{Parameters for 35 additional galaxies}\label{Table-extra}
\begin{tabular}{lllcrccccclc}
\hline
Galaxy    & \multicolumn{2}{c}{Morph.\ Type}  & Dust& Dist  & $\log(M_{\rm bh}/M_\odot)$ & $M_{K_s}$ & $M_{\rm \star}/L_{K_s}$ & $\log(M_{\rm \star,gal}/M_\odot)$ & \multicolumn{2}{c}{Bulge/Total} &  $P$  \\
Name      &    (old)  &     (new)             & bin & Mpc   &          dex       & [mag]    &                   &          dex              &     Obs.\ (band)      & $K_s$        &  Obs.         \\
(1)       &    (2)   &    (3)                 & (4) & (5)   &          (6)       & (7)      &            (8)    &           (9)            &    (10)               &  (11)       &  (12) \\
\hline
A1836-BCG  & BCG SA0  &   BCG E               &  n  & 158.0 &    09.59$\pm$0.06  & $--$26.40 &           1.37   &      12.00$\pm$0.15       & 1.00$^{\dag}$ ($K_s$) &  same   & 0.00 \\ 
Cygnus~A   &  BCG S?  &   BCG ES,e            &  Y  & 232.9 &    09.39$\pm$0.13  & $--$27.13 &     1.24$^{\P}$  &        12.17$\pm$0.15     & 0.95$^{\ddag}$ ($I$)  &  same   & 0.00  \\
ESO 558-G009 &   S   &      SAbc              &  Y  & 115.4 &    07.26$\pm$0.04  & $--$24.97 &     0.59$^{\P}$  &        11.06$\pm$0.15     & 0.07$^{\ddag}$ ($I$)  &  0.10   & 0.00  \\
IC 1481    &  S?     &        m/S             &  Y  &  89.9 &    07.15$\pm$0.13  & $--$24.21 &     0.98$^{\P}$  &      10.98$\pm$0.15       & 0.10$^{\P}$ (...)     &  same   &  0.00  \\
J0437+2456 &   ...   &       SABbc            &  Y  &  72.8 &    06.51$\pm$0.05  & $--$23.97 &     0.59$^{\P}$  &        10.65$\pm$0.15     & 0.09$^{\ddag}$ ($I$)  &  0.13   & 0.075$^{**}$  \\ 
Milky Way &   SBbc   &         same           &  Y  & 0.00786 &  06.60$\pm$0.02  &   ...    &       ...        &        10.78$\pm$0.15     & 0.15$^{\mathsection}$ &  0.15   &  0.15  \\
Mrk 1029  &   ...    &    SA0/a               &  Y  & 136.9 & 06.33$_{-0.13}^{+0.10}$ & $--$23.79 & 0.98$^{\P}$ &        10.81$\pm$0.15     & 0.18$^{\ddag}$ ($I$)  &  same   &  0.00  \\ 
NGC 0307  &   SA0    &      SAB0              &  N  &  52.8 &    08.34$\pm$0.13  & $--$24.00 &       1.06       &         10.92$\pm$0.15    & 0.47$^{\dag}$ ($r^{\prime}$) & 0.59  &  0.053 \\
NGC 0584  &   E4     &      ES,e              &  N  & 19.1  &     8.11$\pm$0.18  & $--$24.29 &       1.03       &       11.03$\pm$0.15      & 0.86$^{\|}$ ($L$)     &   same  &  0.00  \\
NGC 0613  &   SBbc   &   same                 &  Y  & 21.3  &     7.66$\pm$0.15  & $--$24.63 &       0.56       &      10.90$\pm$0.15       & 0.16$^{\P}$ ($L$)     &   same  &  0.15  \\ 
NGC 1068/M77 & SAb   & $^{\star}$SABb          &  Y  & 10.1  &     6.75$\pm$0.08  & $--$24.25 &     0.60$^{\P}$  &         10.78$\pm$0.15   & 0.31$^{\ddag}$ ($K_s$)&   same  &  0.033  \\ 
NGC 1365  &  SBb     &   same                 &  Y  & 17.8  &     6.60$\pm$0.30  & $--$24.89 &      0.57        &      11.01$\pm$0.15       & 0.25$^{\|}$  ($L$)    &   same  &  0.143  \\
NGC 1550  &  SA0     &     E                  &  N  & 51.6  &     9.57$\pm$0.06  & $--$25.07 &      1.14        &       11.38$\pm$0.15      & 1.00$^{\dag}$ ($K_s$) &   same  &  0.00   \\
NGC 1566  & SABbc    &   same                 &  Y  & 17.9  &     7.23$\pm$0.30  & $--$23.99 &      0.47        &       10.57$\pm$0.15      & 0.20$^{\P}$ ($L$)     &  same  &  0.026  \\
NGC 1672  &  SBb     &     same               &  Y  & 15.9  &     7.84$\pm$0.10  & $--$24.00 &      0.46        &      10.56$\pm$0.15       & 0.24$^{\|}$ ($L$)     &   same  &  0.121  \\
NGC 2748  &  SAbc    &    same                &  Y  & 25.1  & 7.68$_{-0.25}^{+0.17}$ & $--$22.54 &  0.62        &      10.39$\pm$0.15       & 0.00$^{\ddag}$ ($I$)  &   same  &  0.00  \\  
NGC 2784  &  SA0     &     same               &  n  &  9.6  &     8.00$\pm$0.31  & $--$23.66 &      1.11        &        10.81$\pm$0.15     & 0.23$^{\P}$ ($K_s$)   &  same  &  0.00  \\
NGC 3258  &   BCG E1 &     same               &  n  &  31.3 &     9.35$\pm$0.05  & $--$24.38 &      1.07        &        11.08$\pm$0.15     & 1.00$^{\P}$ (...)     &   same  &  0.00  \\
NGC 3393  &  SBa     &     same               &  Y  &  55.8 & 7.49$_{-0.16}^{+0.05}$ & $--$24.72 & 0.87$^{\P}$  &        11.13$\pm$0.15     & 0.17$^{\ddag}$ ($I$)  &   0.24  &  0.281 \\
NGC 3504  & SABab    &       SBab              &  Y  &  27.8 &     7.32$\pm$0.07  & $--$23.97 &       0.65       &         10.70$\pm$0.15   & 0.36$^{\|}$ ($L$)     &   same  &  0.164  \\
NGC 3640  &  E3      &     SA0                 &  N$^{\#}$ & 26.3 &   7.89$\pm$0.34  &  ...     &       0.94       &      10.57$\pm$0.15   & 0.19$^{\P}$ ($L$)     &   same  &  0.00   \\
NGC 3706  &  SA0     &     same               &  N  &  38.0 &     8.70$\pm$0.06  & $--$25.04 &        1.12      &       11.37$\pm$0.15      & 0.46$^{\P}$ ($L$)     &   same  &  0.00   \\
NGC 4281  &  S0      &     SA0               &  Y  &  24.4 &     8.73$\pm$0.08  & $--$24.02 &        1.08      &         10.94$\pm$0.15     & 0.50$^{\P}$ (...)     &   same  &  0.00  \\ 
NGC 4570  &  S0      &    SA0               &  N  & 17.1  &    7.83$\pm$0.14   & $--$23.49 &       1.03      &        10.71$\pm$0.15        & 0.10$^{\P}$ (...)     &   same  &  0.00  \\
NGC 4751  &  SA0     &     same               &  Y  & 26.9  &     9.15$\pm$0.05  & $--$23.96 &    0.98$^{\P}$  &         10.88$\pm$0.15     & 0.59$^{\dag}$ ($K_s$) &   same  &  0.00  \\
NGC 5102  &  SA0     &     same               &  y  &  4.0  &     6.06$\pm$0.38  & $--$21.11 &      0.57       &        9.50$\pm$0.15       & 0.26$^{\P}$ ($L$)     &    same &  0.00  \\
NGC 5206  &  SB0     &     dE                 &  N  &  3.2  &     5.76$\pm$0.36  & $--$19.08 &  0.92$^{\P}$    &       8.89$\pm$0.15        & 0.98$^{\P}$ ($K_s$)   &    same &  0.00  \\
NGC 5328  &   E1     &     same               &  N  &  64.1 &     9.67$\pm$0.15  & $--$25.86 &    1.22         &         11.73$\pm$0.15     & 1.00$^{\dag}$ ($K_s$) &    same &  0.00  \\
NGC 5516  &  SA0     &      E                 &  N  &  58.4 &     9.52$\pm$0.06  & $--$25.88 &  1.10$^{\P}$    &       11.69$\pm$0.15       & 1.00$^{\dag}$ ($K_s$) &   same &  0.00   \\
NGC 5765B &  SAab   &      SABab              &  Y  & 133.9 &     7.72$\pm$0.05  & $--$25.25 &  0.76$^{\P}$    &        11.28$\pm$0.15      & 0.08$^{\ddag}$ ($I$)  &   same &  0.091  \\
NGC 6086  &  BCG E   &    same                 &  N  & 138.0 &     9.57$\pm$0.17  & $--$26.08 &      1.20       &        11.81$\pm$0.15     & 1.00$^{\dag}$ ($r^{\prime}$) & same  &  0.00   \\
NGC 6264  &   Sb     &      SABb              &  Y  & 153.9 &     7.51$\pm$0.06  & $--$24.62 & 0.67$^{\P}$     &        10.97$\pm$0.15      & 0.09$^{\ddag}$ ($I$)  &  0.13    &  0.079 \\
NGC 6323  &  Sab     &      SABab                 &  Y  & 116.9 & 7.02$_{-0.14}^{+0.13}$ & $--$24.87 & 0.76$^{\P}$ &     11.13$\pm$0.15     & 0.07$^{\ddag}$ ($I$)  &  same     &  0.002 \\ 
NGC 7049  &  SA0     &     same               &  Y  &  29.9 &    8.51$\pm$0.12  & $--$25.16 &     1.24         &        11.46$\pm$0.15     & 0.54$^{\P}$  ($L$)    &  same    &  0.00   \\
UGC 6093  &  SABbc   &   $^{\star}$same             &  Y  & 152.8 & 7.41$_{-0.03}^{+0.04}$ & $--$25.08 &      0.73   &    11.19$\pm$0.15   & 0.12$^{\ddag}$ ($I$)  &  0.17    &  0.095 \\

\hline       
\end{tabular}

Similar to Table~\ref{Table-data} but for additional galaxies not in \citet{Graham:Sahu:22a}.
Column~7 = \textit{2MASS} $K_s$-band absolute magnitudes derived from the apparent magnitudes taken from NED, and corrected here for Galactic extinction, $5\log(1+z)$ mag of cosmological dimming, and missing light via Equation~\ref{Eqn1} for the few E, ES,e, and BCGs. 
Column~8 = ($B-V$)-dependent $K_s$-band $M_{\star}/L$ ratios (Equation~\ref{Eqn2}). $^{\P}$ indicates the $B-V$ colour is estimated in this work (see Section~\ref{Sec_non}).
Column~10 = Observed $B/T$ ratio (and passband within which it was measured). 
Galaxy decompositions provided by: 
$^{\dag}$ = \citet{2019ApJ...876..155S}; 
$^{\ddag}$  = \citet{2019ApJ...873...85D}; 
$^{\|}$ = S$^4$G;
$^{\mathsection}$ = \citep{2015ApJ...806...96L}; 
$^{\P}$ = this work (Appendix~\ref{App_fits}).
Column~11 shows the observed or expected \citep[][appendix~A]{2024MNRAS.531..230G} $K_s$-band $B/T$ ratio applied here.  
J0437+2456 = SDSS~J043703.67+245606.8, for which the larger (non-nuclear) bar was used to derive $P$. 
Abell~1836-BCG = MCG~-02-36-002. 
A machine-readable version of this table is available online.
\end{table*}

\begin{figure}
\begin{center}
\includegraphics[width=1.0\columnwidth]{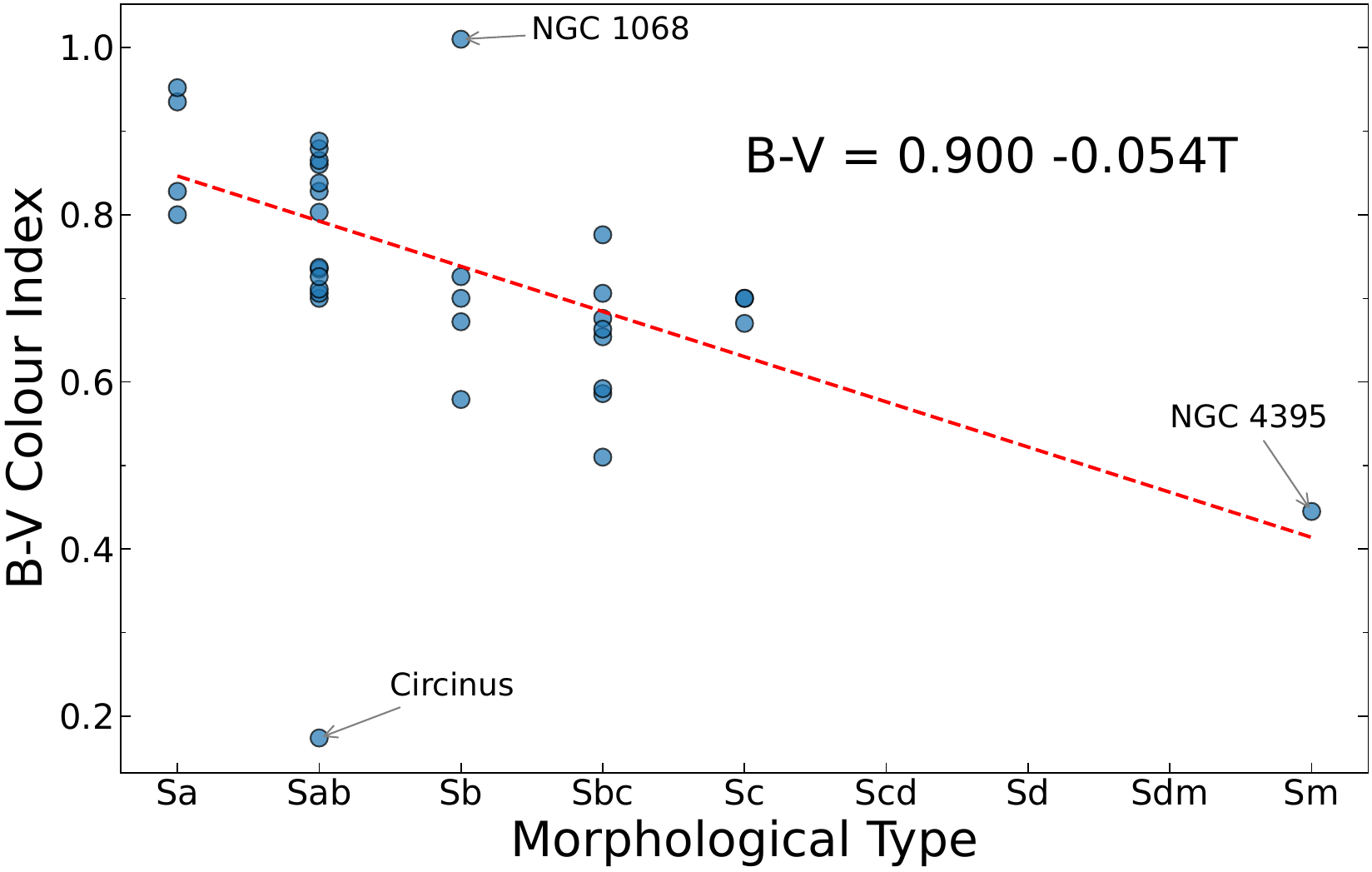}
\caption{ 
$(B-V)_{\rm Vega}$ colour (corrected for Galactic extinction) plotted against the numerical T type (such that Sa=1, Sab=2, Sb=3, etc.) for the 28 and 8 spiral galaxies in Tables~\ref{Table-data} and \ref{Table-extra}, respectively, with such available data. 
The line broadly matches the trend reported by \citet{1995PASP..107..945F} and was used to estimate the $B-V$ colours of six S galaxies in Table~\ref{Table-extra} for which no direct $B-V$ measurements were available.
}
\label{FigTBV}
\end{center}
\end{figure}

\subsubsection{Non-\textit{SST} data and \textit{2MASS} calibrations}\label{Sec_non}

What follows is a description of how additional galaxies with non-{\it SST} data have been included. 

(Galactic extinction)-corrected \textit{Two Micron All Sky Survey} \citep[\textit{2MASS}:][]{2000AJ....119.2498J} $K_s$-band apparent magnitudes were obtained via the NASA Extragalactic Database (NED)\footnote{\url{https://ned.ipac.caltech.edu/}} for the galaxies from \citet{Graham:Sahu:22a}. 
These exist for all of the galaxies listed in Table~\ref{Table-data}, except for NGC~404.  For NGC~2974, a bright foreground star has substantially contaminated this galaxy's measurement, so it too was excluded from the following derivation of the equation for the $K_s$-band stellar mass-to-light ratio. 

To account for missing $K_s$-band flux at large radii in high-Sérsic index galaxies, the galaxy magnitude correction from \citet{2013ApJ...768...76S}, of up to $\sim$0.2 mag, was applied. This absolute magnitude correction is such that:
\begin{equation}
K_{s,{\rm corr}} = 1.07 \times K_s + 1.527
\label{Eqn1}
\end{equation} 
for E, ES,e and E-BCGs.
The resulting $K_s$ and, when needed, $K_{s,\rm corr}$ stellar mass-to-light ratios required to give \textit{2MASS}-derived stellar masses consistent with those derived from the \textit{SST} data are shown in Figure~\ref{FigML} as a function of the extinction-corrected $(B-V)_{\rm Vega}$ colour taken from NED.

The dotted line in Figure~\ref{FigML} represents the relation in \citet[][their table~6]{2013MNRAS.430.2715I}, adjusted to the diet-Salpeter IMF of \citet{2001ApJ...550..212B}, and is such that $\log(M_\star / L_{K_s}) = 0.866(B-V)_{\rm Vega} - 0.851$. 
However, the solid line given by 
\begin{equation}
\log(M_\star / L_{K_s}) = 1.034 \times (B-V)_{\rm Vega} - 0.940
\label{Eqn2}
\end{equation}
provides a better fit to the data. It recovers the \textit{SST}-based stellar masses with an accuracy of $\sim$0.1~dex, except that a few bright E galaxies are offset by $\sim$0.1--0.2~dex, suggesting that the correction from \citet{2013ApJ...768...76S} to the \textit{2MASS} $K_s$-band magnitudes (Equation~\ref{Eqn1}) may have been conservative for some galaxies.

To increase the sample size, an additional 35 galaxies for which a 3.6~$\mu$m decomposition was not previously performed by past team members, but for which a black hole mass is recorded in \citet{2019ApJ...887...10S}, are given in Table~\ref{Table-extra}. 
\textit{2MASS} $K_s$-band galaxy magnitudes exist for all additional 36 galaxies listed in \citet{2019ApJ...887...10S}, except for NGC~5495, for which a bright star appears to have interfered with the flux, leading to the exclusion of this galaxy here.  
To correct for missing light at large radii among the sample in Table~\ref{Table-extra}, Equation~\ref{Eqn1} was applied to the magnitudes of four BCGs, three non-BCG E galaxies, and one non-BCG ES,e galaxy (NGC~584). 
Prior to calculating the absolute magnitude using the luminosity distances given in Table~\ref{Table-extra}, 
the magnitudes were corrected for dust extinction in our Galaxy and for $5\log(1+z)$ mag of cosmological dimming.
Equation~\ref{Eqn2} was applied to almost all of these additional galaxies. For the Seyfert~2 spiral NGC~1068,  heavy internal dust extinction and a red colour, $(B-V) > 1$, result in an overestimated mass when using Equation~\ref{Eqn2}. For NGC~1068, a fixed ratio of $M_\star/L_{K_s} = 0.6$ is adopted.  
The Milky Way galaxy's stellar mass was taken from \citet{2015ApJ...806...96L}, and NGC~3640's stellar mass was derived from a 3.6\,$\mu$m \textit{SST} image modelled in Appendix~\ref{App_fits}, as the \textit{2MASS} magnitude appears to be biased, perhaps by the neighbouring galaxy.

For 12 of these 35 galaxies, there was no reliable $B-V$ colour, and so the following estimated colour was assigned based on the results in Figures~\ref{FigML} and \ref{FigTBV}, with the latter showing the $B-V$ colour as a function of the spiral morphological type.  
For six spiral galaxies, the equation inset in Figure~\ref{FigTBV} was used, with the numeric morphological T index given by 1 $=$ Sa, 2 $=$ Sab, 3 $=$ Sc, and so on.  Based on the results in Figure~\ref{FigML}, a $B-V$ colour of 0.875 was used for one dust-poor S0 galaxy (dust bin = N; see Section~\ref{Sec_dust}), 0.9 for one spiral merger (Type = m/S) and the remaining two S0 galaxies, while 0.925 and 0.95 were used for one ES and one E galaxy, respectively.

The next step was to apply consistent bulge-to-total ($B/T$) ratios to obtain the stellar masses of the spheroids. The galaxies in Table~\ref{Table-extra} were modelled using the same decomposition methodology as applied to the \textit{SST} sample in Table~\ref{Table-data}, but the analysis was applied to \textit{2MASS} $K_s$-band images, {\it Sloan Digital Sky Survey} \citep[{\it SDSS:}][]{2000AJ....120.1579Y}\footnote{\url{https://www.sdss3.org/}} $r^{\prime}$ images (NGC~307 and NGC~6086), or \textit{HST}/F814W images (see Column~10 in Table~\ref{Table-extra}). The \textit{HST} images were typically used in the past because of the desire for better spatial resolution to quantify the bulge component of the more distant or later-type S galaxies.  The adopted works and filter are indicated in Table~\ref{Table-extra} (Column 10). 

$L$-band (3.6~$\mu$m) \textit{SST} image decompositions for four galaxies were taken from the S$^4$G,   
and eight new decompositions were performed in this work to acquire new or improved $B/T$ (and $P \equiv$ Bar/Total) ratios for all of the additional galaxies provided in Table~\ref{Table-extra}. For these eight, four \textit{SST} images were retrieved from the NASA/IPAC InfraRed Science Archive (IRSA)\footnote{\url{https://irsa.ipac.caltech.edu/data/SPITZER/S4G/}}, two additional \textit{SST} images were retrieved from the \textit{Spitzer Heritage Archive} \citep[SHA;][]{2010SPIE.7737E..16W}\footnote{\url{https://sha.ipac.caltech.edu/}}, and two \textit{2MASS} galaxy images were also retrieved from IRSA.\footnote{\url{https://irsa.ipac.caltech.edu/applications/2MASS/IM/interactive.html}}
Appendix~\ref{App_fits} details the derivation of the $B/T$ ratios and the $P$ ratios for two of these galaxies that are barred (NGC~613 and NGC~1566). 
The stellar masses derived from the \textit{SST} images agree to within 0.1~dex of the \textit{2MASS}-derived values, with the exception of NGC~3640.  As noted, for NGC~3640, the \textit{2MASS} $K_s$-band magnitude appears biased by a neighbouring galaxy, so the 3.6\,$\mu$m-derived mass measurement is used.

Finally, before proceeding, it is necessary to account for differences in $B/T$ across different wavelengths. 
The appendix of \citet{2024MNRAS.531..230G} quantifies the typical $B/T$ ratios in different filters as a function of the disc galaxy type. It is apparent that the ratios at optical wavelengths usually underestimate those in the $K$-band, assumed here to roughly match those in the $L$-band. The galaxy-morphology-dependent $B/T$ ratio corrections provided there are applied here. They are such that NGC~307 has had its observed $r^{\prime}$-band $B/T$ ratio increased by $1/0.8$, and the following five late-type S galaxies observed with the F814W filter (approximately the $I$-band) have had their $B/T$ ratios increased by $1/0.7$: ESO558-G009; J0437+2456; NGC~3393; NGC~6264; and UGC~6093. The five early-type S galaxies observed with the F814W filter do not require a correction, nor do the other galaxies' $B/T$ ratios in Table~\ref{Table-extra} (see Columns~10 and 11). 

\begin{figure*}
\begin{center}
\includegraphics[width=1.0\textwidth]{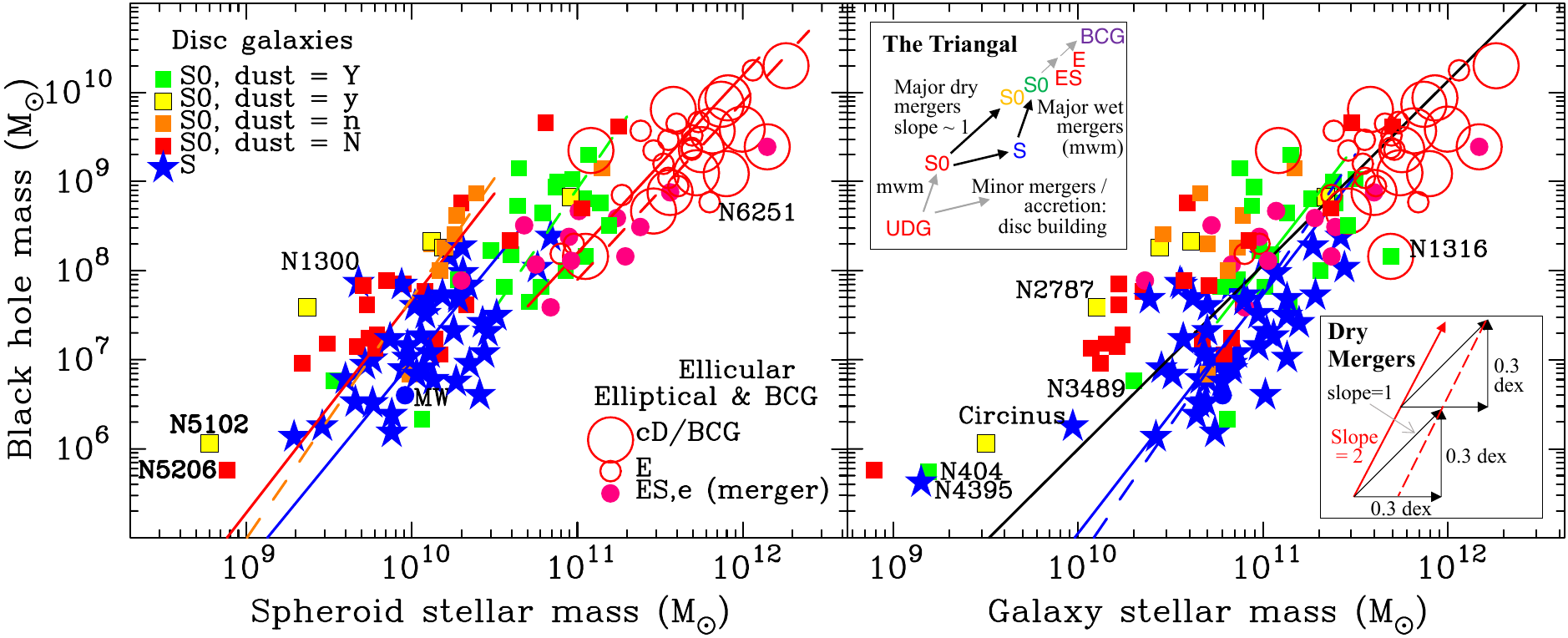}
\caption{ 
Left: Expanded $M_{\rm bh}$--$M_{\rm \star,sph}$ diagram building on that presented in \citet{Graham-triangal}, where the evolutionary pathways that led to the `Triangal' were first shown. Here, updated masses for some galaxies in Table~\ref{Table-data} are used, and the galaxies in Table~\ref{Table-extra} are added. 
The ES,e galaxy with the highest stellar mass is the distant BCG Cygnus~A.  The galaxy in the dust $=$ n bin with the highest stellar mass is the ES,b galaxy NGC~1332.
Right: Expanded $M_{\rm bh}$--$M_{\rm \star,gal}$ diagram building on that presented in \citet{Graham-triangal}, now showing the `Triangal' for the first time, and mirroring the `Triangal' shown in the galaxy (central surface brightness)-size diagram \citep{2025PASA...42..155G} for high and low surface brightness galaxies, including ultra-diffuse galaxies (UDGs), themselves a subset of the largely forgotten IC~3475 type galaxies.  
The blue dot denotes the Milky Way. 
The lines in both panels have come from \citet{Graham-triangal}.
 Major dry mergers preserve the $M_{\rm bh}/M_{\rm \star,gal}$ ratio and thus shift galaxies off the  $M_{\rm bh} \propto M_{\rm \star,gal}^2$ relation (black line) shown here for E, ES,e, and dust $=$ Y S0 galaxies. 
 For clarity, the offset dashed red line for the BCGs is only shown in the left panel.
}
\label{Fig-M-M}
\end{center}
\end{figure*}

The galaxies from Tables~\ref{Table-data} and \ref{Table-extra} are shown in Figure~\ref{Fig-M-M}, which displays the black hole mass against the stellar mass of the spheroid and galaxy for this enlarged sample.

\subsubsection{Distances}

The analysis presented here focuses on the local Universe (predominantly $z < 0.04$), 
where all 137 galaxies have direct (dynamical) black-hole mass measurements. 
For the entries in Table~\ref{Table-data}, distances have come from the compilation (with references) in \citet{Graham:Sahu:22a}. 
For the entries in Table~\ref{Table-extra}, distances have come from \citet{2019ApJ...887...10S}, with the following updates. 
A distance of 15.9$\pm$0.9~Mpc to NGC~1672 has been reported by \citet{2020AstBu..75..384T}, 
which also reports a distance of 14.9$\pm$1.0 to NGC~1566.
The distances for Cygnus~A, NGC~0613, NGC~2748, and NGC~3504 are the (Virgo $+$ GA $+$ Shapley)-corrected Hubble flow distances \citep{2000ApJ...529..786M} based on H$_0 = 73$ km s$^{-1}$ Mpc$^{-1}$, taken from NED.
For NGC~5102, the surface brightness fluctuation distance from \citet{2001ApJ...546..681T} has been used, 
and for
NGC~5206, a distance of 3.2~Mpc from \citet{2015ApJ...802L..25T} has been used.

\subsubsection{The squashed sample bias concern}

It is noted that this sample of galaxies with direct black hole mass measurements is not biased relative to the general population of galaxies of similar stellar mass. That suspicion, raised by \citet{2016MNRAS.460.3119S}, was resolved by \citet{2023MNRAS.518.1352S}, who demonstrated that inconsistent stellar masses had been used in \citet{2016MNRAS.460.3119S} between galaxies with and without directly measured black hole masses, which was one of the yet-to-be-checked possibilities raised by \citet{2016MNRAS.460.3119S}. A subsequent rebuttal by \citet{2025MNRAS.541.2070S}, based on a comparison in the (velocity dispersion, $\sigma$)--(luminosity, $L$) plane, claimed a residual bias persists at faint magnitudes. However, that analysis overlooked the bent nature of the $\sigma$--$L$ distribution for ETGs \citep[e.g.,][their figures 13--15]{1983ApJ...266...41D, 2005MNRAS.362..289M, 2019ApJ...887...10S}. The \textit{Spitzer Survey of Stellar Structure in Galaxies} \citep[S$^4$G:][]{2010PASP..122.1397S}\footnote{\url{https://irsa.ipac.caltech.edu/data/SPITZER/S4G/}} 
% \url{https://www.oulu.fi/astronomy/S4G_PIPELINE4/MAIN/}  
reference sample used by \citet{2025MNRAS.541.2070S} extends deep into the faint-luminosity regime, where ETGs follow an $L \propto \sigma^{2\text{---}2.5}$ distribution.  This pulled the single $\sigma$--$L$ line fit by \citet[][their figure~A1]{2025MNRAS.541.2070S} downward at faint magnitudes. In contrast, the sample with direct black hole mass measurements is anchored almost entirely at bright luminosities, where luminosity increases more steeply with velocity dispersion. Their extrapolation of the $\sigma$--$L$ line for the sample with direct black hole mass measurements to fainter luminosities explains the mismatch with their single-power-law relation for the S$^4$G sample.  Therefore, the conclusion of \citet{2023MNRAS.518.1352S} stands.

\subsection{Galaxy morphology}
\label{Sec_morph}

The existence of galaxy-morphology-specific $M_{\rm bh}$--$M_\star$ scaling relations \citep{Graham-triangal} established using the sample in Table~\ref{Table-data} necessitates an inspection of the galaxy types for the new sample (Table~\ref{Table-extra}).  
Most galaxy morphologies in Table~\ref{Table-data} follow the designations in \citet{Graham:Sahu:22a}. However, a review of these revealed a few refinements.  Changes and galaxies of interest are noted below. 

NGC~2974 (aka NGC~2652) and 
NGC~4594 (the Sombrero galaxy) were initially regarded as S galaxies by \citet{2019ApJ...873...85D} and \citet{Graham:Sahu:22a} before \citet{2023MNRAS.521.1023G} assigned them S0 status, in accord with many other works. 
The young spiral structure in NGC~2974 is evident in \textit{Galaxy Evolution Explorer} \citep[\textit{GALEX}:][]{2007ApJS..173..185G, 2007ApJS..173..682M, 2011MNRAS.411..311M} images and appears in the unsharp mask of an {\it HST/F555W} image \citep{2007MNRAS.376.1021J}. It has also been reported to contain a bar \citep{2017MNRAS.471.2187D}.
Here, NGC~2974 is (tentatively) designated S0/a given that its faint spiral and ring-like pattern \citep{2007MNRAS.376.1021J} is weak and atypical of standard spiral galaxies.
NGC~4594 is less clear-cut, with the \textit{SST} image arguably showing a more ring-like structure than a faint spiral. \citet{2012MNRAS.423..877G} discuss the dual S0/S nature of NGC~4594.  
Inspection of \textit{SST} 3.6~$\mu$m images (and GALEX/optical where noted) revealed a few other faint spiral-like features, as noted in Table~\ref{Table-data}. 
From Table~\ref{Table-extra}, Mrk~1029 is also identified as an S0/a galaxy. 
Cygnus~A was initially considered a potential S0/a candidate due to a spiral-like feature likely influenced by dust \citep{2019ApJ...873...85D}, but it was ultimately classified here as an ES,e galaxy.

\citet{2019ApJ...873...85D} report an inner disc in ESO~558-G009, creating an anti-truncated light profile. If this `inner disc' is instead considered a rotating bulge formed via a merger, the $B/T$ ratio would increase significantly. The spiral arms are rather faint for this galaxy's Sbc classification \citep{2017A&A...605A..84K}, and the dust appears largely confined to the inner region.  This massive system may be a misclassified S0/a galaxy, although its current spiral classification is retained for this work.

The reassigned morphological types for five of the nine ETGs modelled in \citet{2024MNRAS.535..299G} are adopted here. 
This includes the identification of an intermediate-scale disc in NGC~4291 and large-scale discs in NGC~3379, NGC~3091 and NGC~4649 (galaxies previously classified as E), and the reclassification of the ES galaxy\footnote{\citet{1966ApJ...146...28L} introduced the ES galaxy type for those with intermediate-scale discs.} NGC~4697 to a system with a large-scale disc. 
Following \citet{Graham:Sahu:22b},  ES galaxies---defined by intermediate-scale discs fully embedded within a spheroidal/dynamically-hot structure---are distinguished as either ES,b (more closely linked to S0 galaxies) or ES,e (more closely linked to E galaxies). 
The morphologies for the galaxies in Table~\ref{Table-extra} reflect the results of their decompositions, noting the presence or absence of discs, spirals, and bars. 
While some reclassifications differ from \citet{1991rc3..book.....D} (RC3) designations, they are consistently informed by multi-component decompositions, structural diagnostics, and recourse to kinematic data when available, ensuring that the adopted morphologies reflect physically distinct components rather than visual impressions alone.

Finally, \citet{2019ApJ...876..155S} are followed in classifying NGC~1550, NGC~5516, and A1836-BCG as E galaxies rather than as RC3 S0s. 
NGC~1550 is the Brightest Group Galaxy (BGG) of the NGC~1550 Group, a near `fossil group' with an X-ray halo and little rotation \citep{2009ApJ...691..971K, 2014MNRAS.441.2013L}.
NGC~5516 is also a BGG and possesses a stellar envelope \citep{2013AJ....146..160R}, similar to that seen around many BCGs such as the E/D galaxy A1836-BCG \citep{2009AJ....137.4795C, 2009ApJ...690..537D}, which may have led to its prior S0 misclassification.

\subsection{Dust bins --- a guide to the origin of the S0 galaxies}
\label{Sec_dust}

The $M_{\rm bh}$--$M_{\star}$ diagrams of 
\citet{2019ApJ...876..155S} solved one mystery\footnote{Most previous $M_{\rm bh}$--(luminosity, $L$) and $M_{\rm bh}$--$M_{\star}$ relations for ETGs had slopes around $\sim$1 to $\sim$1.3, and were inconsistent with the \citet{1969Natur.223..690L} ultra-massive black holes ($10^{10} < M_{\rm bh}/M_\odot \lesssim 10^{11}$). The steep slope of $\sim$2 reported by \citet{2019ApJ...876..155S} for E galaxies in the $M_{\rm bh}$--$M_{\star}$ diagrams rectified this. This improved consistency with the steep slopes in the $M_{\rm bh}$--$\sigma$ diagram of 7--8 for core-S\'ersic galaxies \citep{2013ApJ...764..151G, 2019ApJ...887...10S}, and 9--10 for massive BGGs/BCGs \citep{2018ApJ...852..131B}.} but uncovered another: S0 galaxies did not form a bridging population between E and S galaxies in the $M_{\rm bh}$--$M_{\rm \star,sph}$ diagram.
The solution emerged in \citet{2023MNRAS.521.1023G}, which revealed that dust-poor and dust-rich S0 galaxies occupy distinct regions of this diagram, differing in their mean $M_{\rm bh} / M_{\rm \star,sph}$ ratio by an order of magnitude. Dust-rich S0 galaxies bridge the S and E sequences, while dust-poor S0 galaxies are offset to lower stellar masses. \citet{Graham-triangal} and \citet{2025PASA...42...68G} refer to these low-mass, dust-poor S0 galaxies as `primeval' or `primordial'.
These distributions are shown in Figure~\ref{Fig-M-M}. 

The dust-poor/dust-rich distinction was based on the visual appearance of dust in optical images, leading to four dust classifications or "bins": `N' (no dust); `n' (nuclear dust only, $\lesssim$200~pc); `y' (weak, widespread dust); and `Y' (strong Yes, widespread dust). In the dust $=$ Y bin, dust tends to be distributed over the inner regions rather than spread equally across the galaxy. 
Although the dust classifications are observationally defined, they are not arbitrary. Previous studies have demonstrated that widespread dust in S0 galaxies is strongly associated with gas-rich merger activity, while its absence typically reflects either gas-poor formation pathways or subsequent removal processes. The dust bins used here should therefore be interpreted as empirical proxies for formation history, rather than purely descriptive labels.
 
Indeed, past studies \citep[see the references in][table~2]{2023MNRAS.521.1023G} revealed that most of the dust-rich (dust $=$ Y) S0 galaxies are products of major wet mergers. Over time, environmental processes or feedback may remove their dust, resulting in dust-poor S0 galaxies that were nonetheless built via major mergers. Despite this, while the dust bins are observational, they appear to proxy for formation history quite well (albeit imperfectly). Indeed, only a few massive S0 galaxies known to be merger remnants are currently dust-poor (dust $=$ N/n), perhaps due to sputtering from hot X-ray halos or ram pressure stripping. As noted in Table~\ref{Table-data}, 
NGC~3091, NGC~3379, NGC~4697, and NGC~4649 may belong to the merger-built sequence despite their current dust-poor appearance. As noted in \citep[][figure~1]{Graham-triangal}, they may also have arisen from dry major mergers of S0 galaxies at the top of the dust-poor S0 galaxy $M_{\rm bh}$--$M_{\rm \star,sph}$ sequence.
Similarly, NGC~3640 (Table~\ref{Table-extra}), reclassified here as S0, shows no obvious dust in \textit{HST} images. However, deep imaging reveals shells and distortions \citep{2009AJ....138.1417T, 2023A&A...670L..20R, 2024A&A...691A.104M} indicative of a tumultuous past, consistent with its anti-truncated disc profile (Appendix~\ref{App_fits}), as these are often found in major-merger-built S0 galaxies \citep{2014A&A...570A.103B, 2024MNRAS.535..299G}.
These five galaxies are flagged in the tables but not reassigned here.

The sample also includes four compact massive ES,b galaxies---NGC~1332 (dust $=$ n), NGC~3115 (dust $=$ Y), NGC~5845 (dust $=$ n), and NGC~6861 (dust $=$ Y)---which are grouped with the S0 galaxies following \citet{Graham:Sahu:22b}. These may represent relics of ancient major mergers between primeval S0 galaxies that were dustier in the past \citep{2011ApJ...730....4B}. 

\begin{figure}
\begin{center}
\includegraphics[width=1.0\columnwidth]{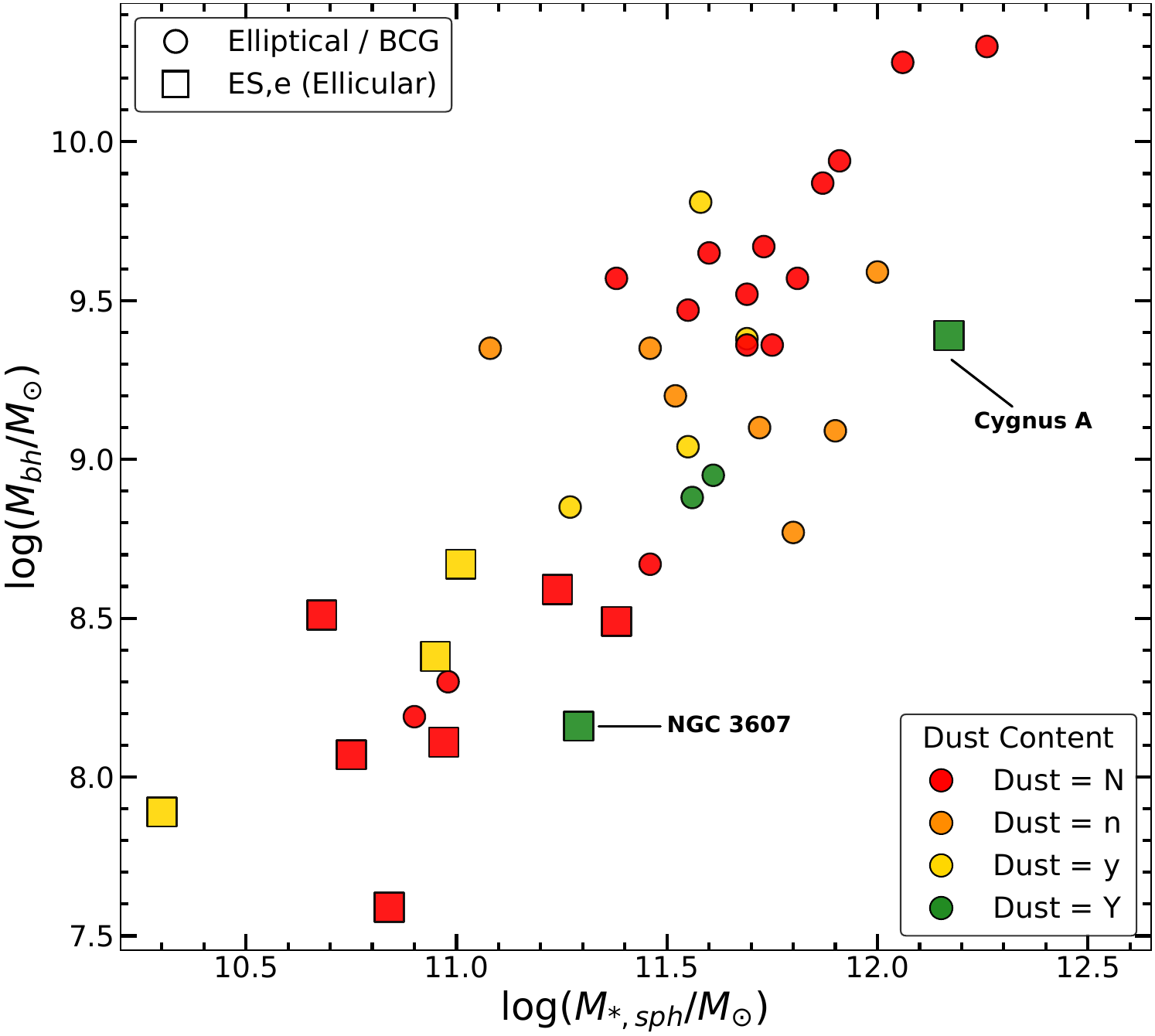}
\caption{ 
Similar to the left-hand panel of Figure~\ref{Fig-M-M}, but zoomed in and just displaying the E and ES,e galaxies. 
The ES,e tend to reside toward the lower mass range, while the E galaxies with nuclear dust discs tend to reside between these ES,e galaxies and the pure (discless) E galaxies, creating a `Disc Down-sizing' sequence.
}
\label{Fig_E-dust-bins}
\end{center}
\end{figure}

The dust `bin' for each S0 (and S0/a) galaxy is displayed in Figure~\ref{Fig-M-M}.
All of the S galaxies are dust-rich. This just leaves the E (and ES,e) galaxies, some of which are cD or BCGs. It was too crowded to show their dust `bin' in Figure~\ref{Fig-M-M}, and, therefore, it is shown in Figure~\ref{Fig_E-dust-bins}.  Patterns and observations will be discussed in Section~\ref{Sec_down_sizing}.

\begin{figure*}
\begin{center}
$
\begin{array}{ccc}
\includegraphics[trim=0.0cm 0cm 0.0cm 0cm, height=0.3\textwidth, angle=0]{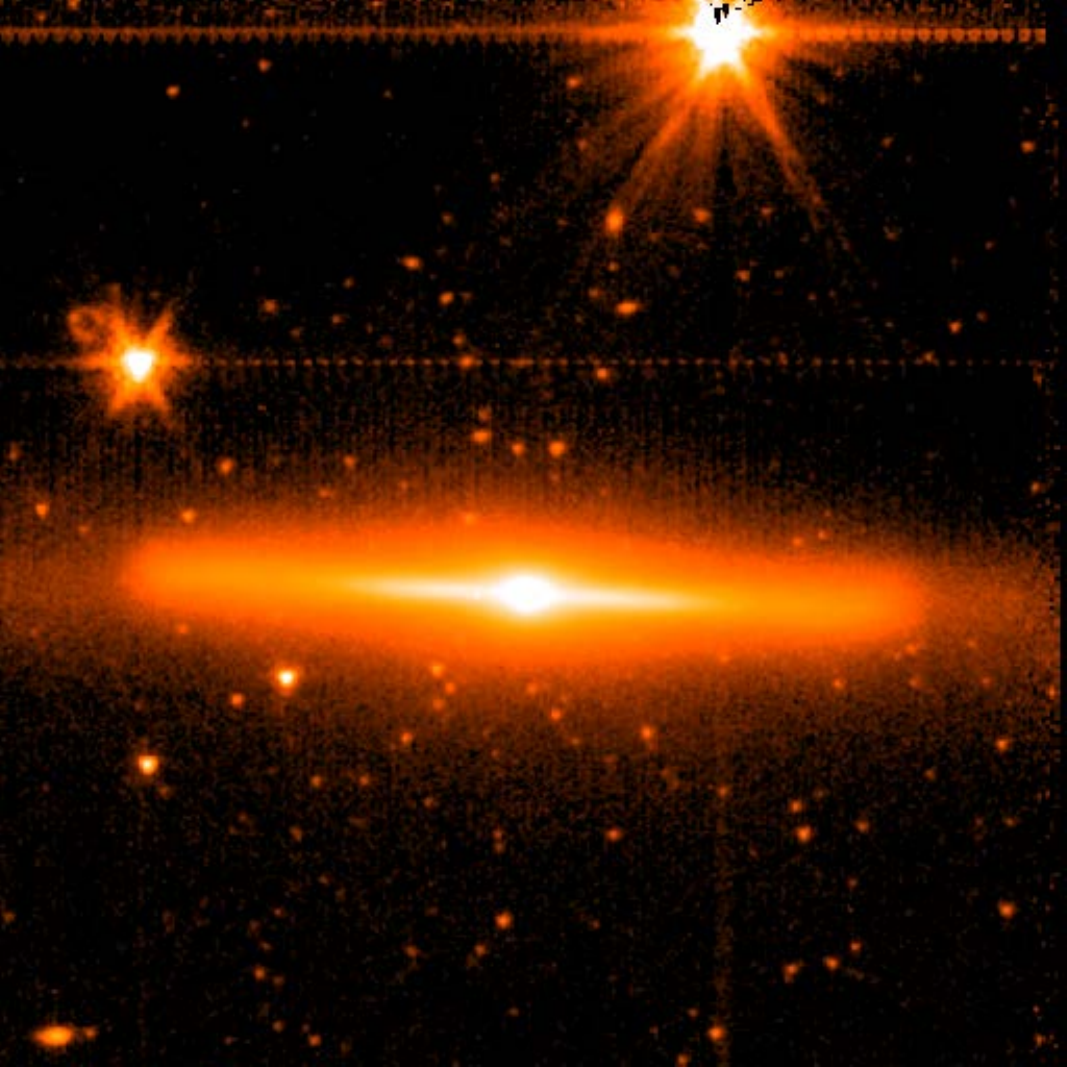} &
\includegraphics[trim=0.0cm 0cm 0.0cm 0cm, height=0.3\textwidth, angle=0]{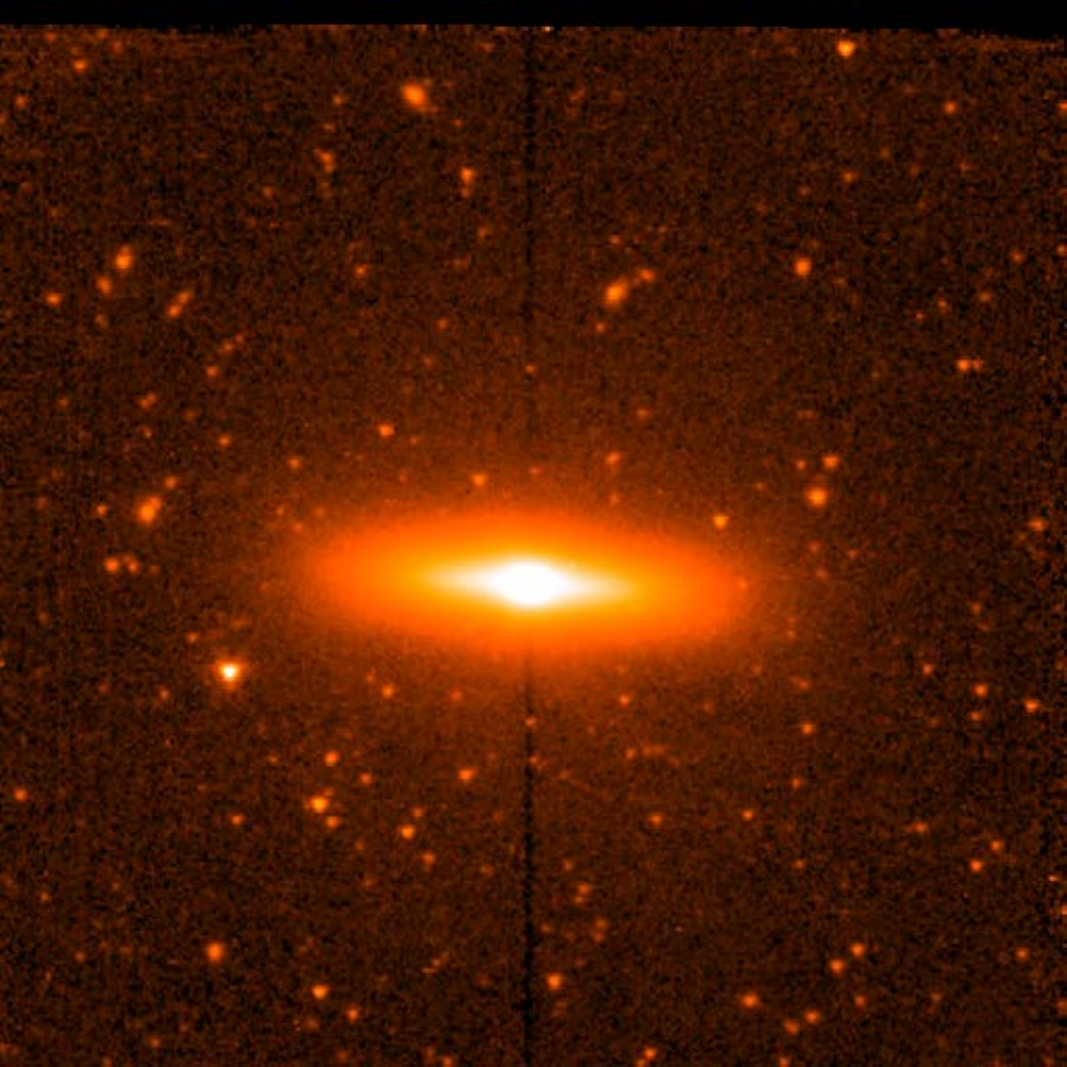} &
\includegraphics[trim=0.0cm 0cm 0.0cm 0cm, height=0.3\textwidth, angle=0]{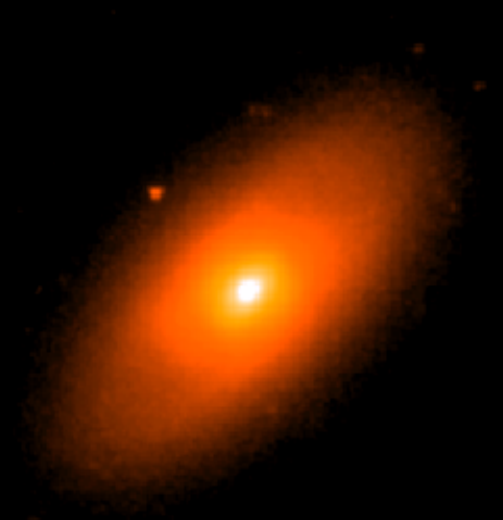} \\ 
\includegraphics[trim=0.0cm 0cm 0.0cm 0cm, height=0.3\textwidth, angle=0]{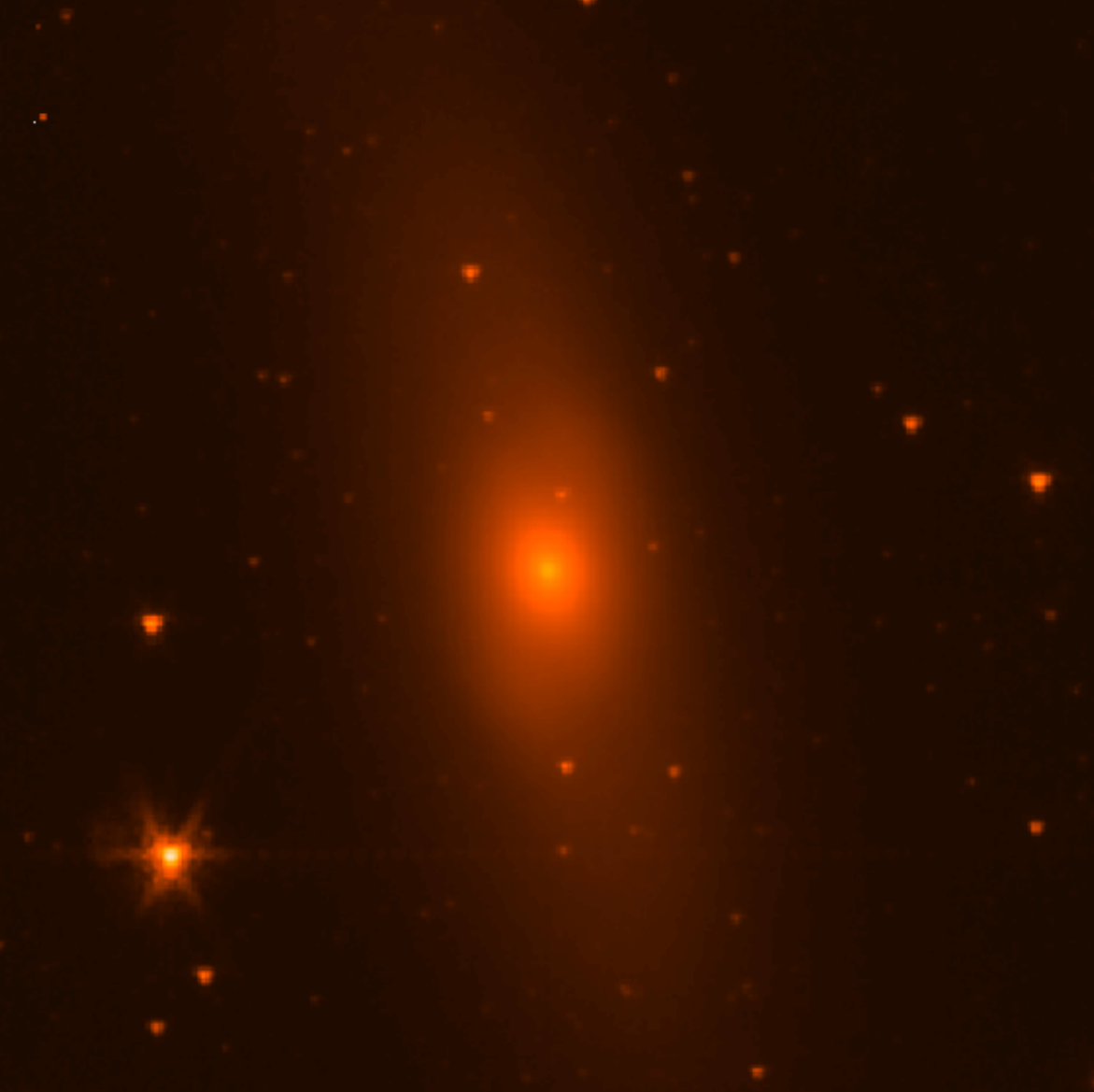} &
\includegraphics[trim=0.0cm 0cm 0.0cm 0cm, height=0.3\textwidth, angle=0]{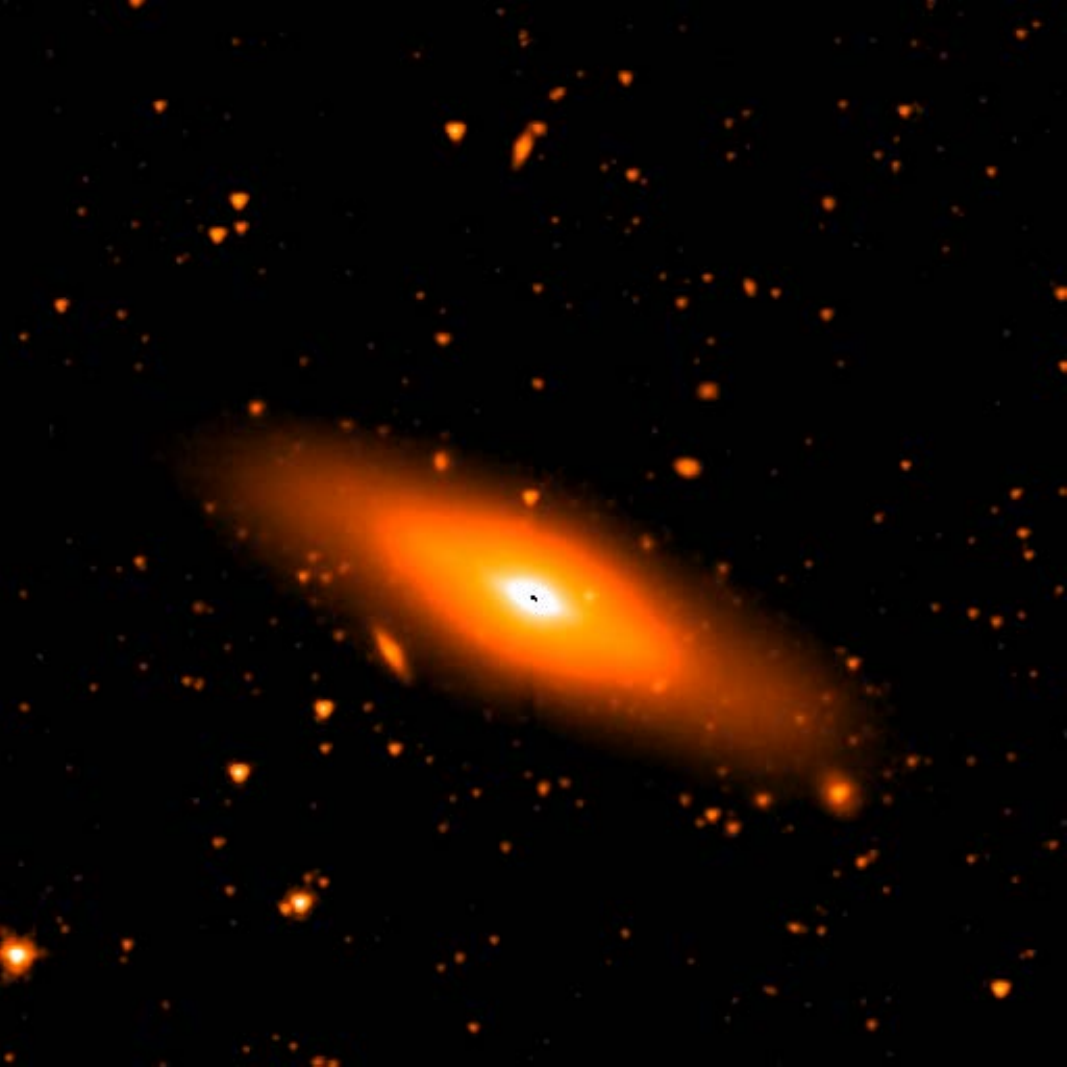} &
\includegraphics[trim=0.0cm 0cm 0.0cm 0cm, height=0.3\textwidth, angle=0]{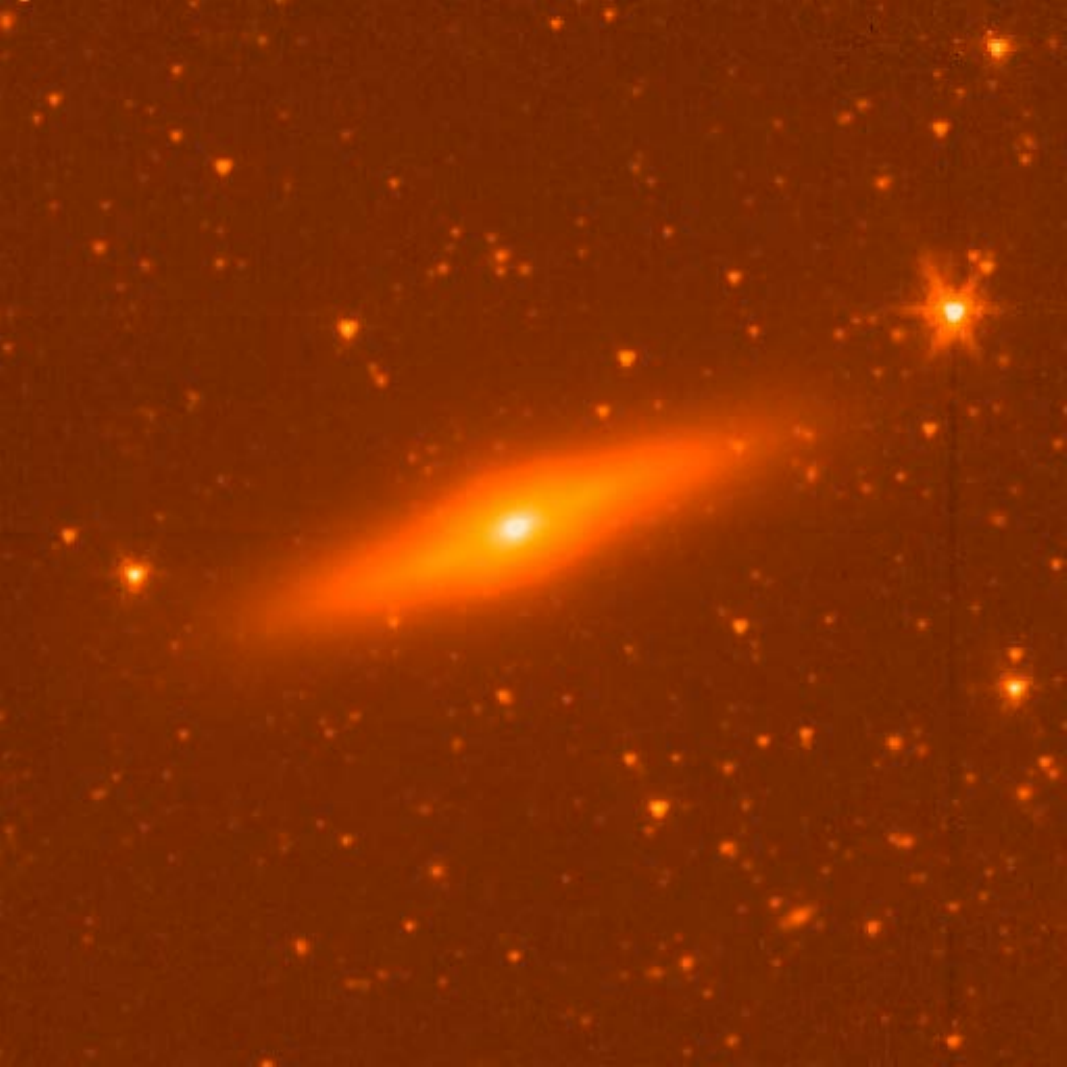} \\
\end{array}
$
\end{center}
\caption{\textit{SST} 3.6~$\mu$m images of six S0 galaxies with bars. 
Upper row, left to right:  NGC~4762 (SB0, dust $=$ N); NGC~4350 (SAB0, dust $=$ n); and NGC~3489 (SB0/a, dust $=$ Y).
Lower row, left to right: NGC~1023 (SAB0, dust $=$ N); NGC~2549 (SB0, dust $=$ N); and NGC~7332 (SAB0, dust $=$ N).  
}
\label{Fig-clean-S0}
\end{figure*}

\begin{figure}
\begin{center}
\includegraphics[width=1.0\columnwidth]{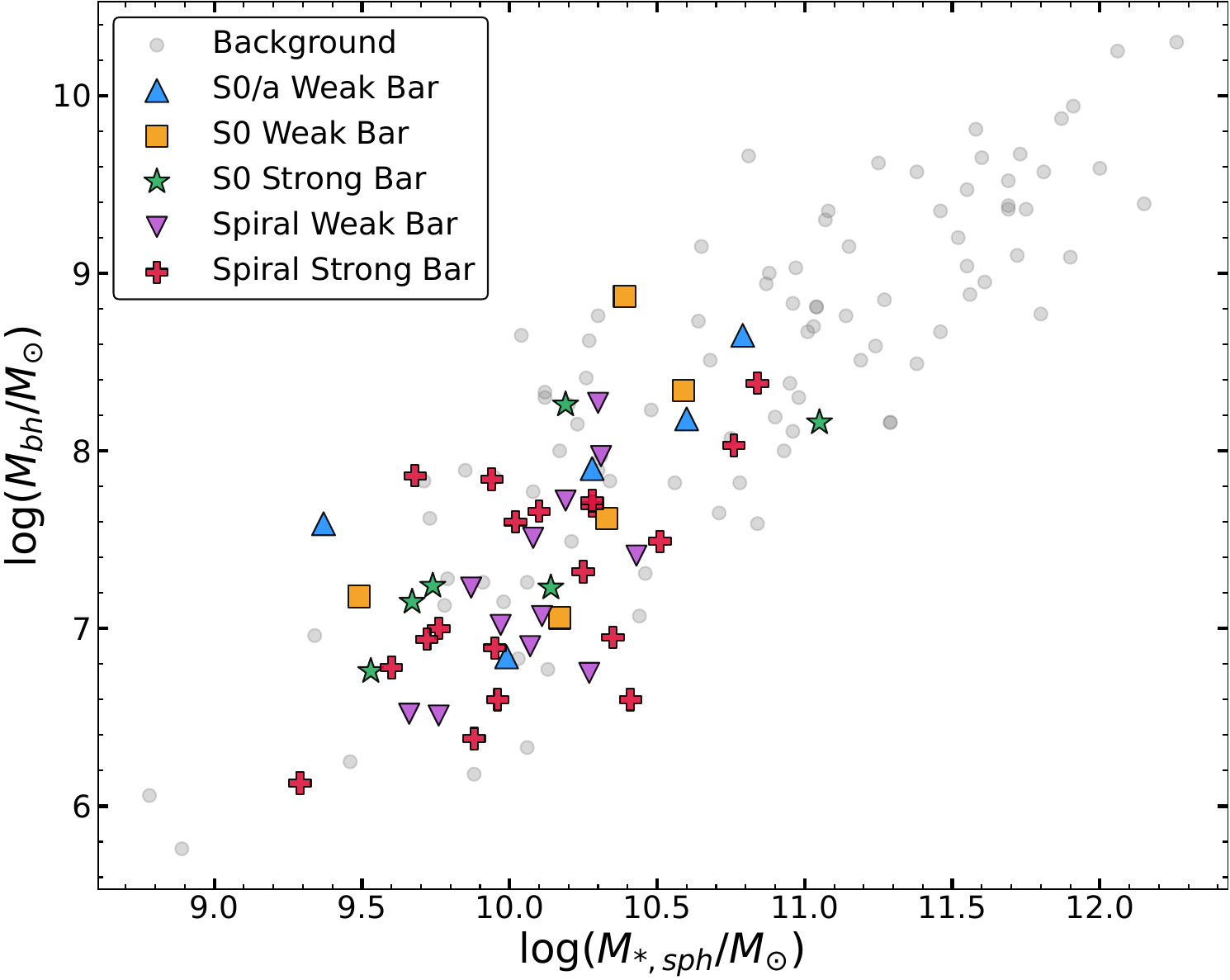}
\caption{
Variant of the left-hand panel of Figure~\ref{Fig-M-M} showing the location of galaxies with no (A), weak (AB), and strong (B) bars.
The S0/a galaxy NGC~3489 is shown here as a strongly barred S0; it is the only S0/a galaxy in the sample with a strong bar.
}
\label{Fig_colour_bars}
\end{center}
\end{figure}

\subsection{Weak and strong bars: Bar-to-total stellar mass ratios, {\it P}}
\label{Sec_weak_n_strong}

Putting some meat on the `Tuning Fork', \citet{1959HDP....53..275D} introduced a qualitative visual bar strength criterion in his classification volume/barrel, designating galaxies as A (absent), AB (weak/intermediate), or B (barred, strong). 
It appears that approximately one-quarter to one-third of stellar bars are missed in optical images but seen more clearly in the near-infrared \citep{1991Natur.353...48B, 2000AJ....119..536E, 2007ApJ...659.1176M, 2015ApJS..217...32B, 2024MNRAS.52711777M}. Most Table~\ref{Table-extra} reclassifications concern weak bars missed in the optical RC3 catalogue \citep{1991rc3..book.....D}, or to RC3-classified weak bars that appear strong in infrared images.\footnote{One exception may be NGC~4258 \citep{2019ApJ...873...85D}, where an anti-truncated exponential disc (not fit but possibly required) may have been conflated with the bar, potentially inflating the bar strength.}
Figure~\ref{Fig-clean-S0}  shows six examples of bars in S0 galaxies.

While bar strength can be quantified in relatively face-on galaxies via gravitational torque parameters, $Q$ \citep{1980ApJ...235..803S, 1981A&A....96..164C, 2001ApJ...550..243B}, and a raft of measures stemming from an isophotal Fourier analysis can quantify the X/(peanut shell)-shaped structures associated with buckled/unstable bars in relatively edge-on galaxies \citep{2016MNRAS.459.1276C}, a simpler metric is employed here. 
Specifically, the ratio of the luminosity (or mass) of the stars considered to be a {\it part} of the bar relative to the entire galaxy --- as informed by the galaxy decompositions --- gives rise to what could be termed a `party' parameter, $P$, reflecting `how many stars are in the bar'.\footnote{Alphabetically, the letter P precedes Q, and the $P$ parameter can be thought of as a `p'recursor or `p'rior metric to the more sophisticated evaluation of $Q$.}
Since a single colour-dependent $M/L$ ratio per galaxy is applied, $P$ is equivalent to the bar-to-total luminosity ratio from the decompositions. These $P$ parameters are provided in Tables~\ref{Table-data} and \ref{Table-extra}.\footnote{Note: Due to a missing factor of 2.5 in the output of the \textsc{Profiler} code \citep{2016PASA...33...62C} that was used for the bulk of the decompositions, the reported $\alpha$ values for the Ferrers function in the appendix figures of \citet{2019ApJ...873...85D} are 2.5 times too large.} In calculating $P$, barlenses\footnote{Barlenses
\citep[e.g.,][]{1978A&A....70...63V, 1983A&A...121..297D, 2002A&A...385....1S, 2011MNRAS.418.1452L}  
have been associated with bar buckling events that can produce boxy/X/(peanut shell)-shaped `pseudobulges'
\citep{1975IAUS...69..297B, 1975IAUS...69..349H, 1981A&A....96..164C, 2005MNRAS.358.1477A, 2016MNRAS.459.1276C}.}
 were excluded, while ansae were included.
Bars are designated as `strong' (B) if $P \ge 0.1$, and `weak' (AB) if $0 < P < 0.1$. This threshold (0.1) approximately corresponds to the peak in the bar strength distribution reported by \citet{2011MNRAS.415.3308G}.
Figure~\ref{Fig_colour_bars} shows the location of galaxies in the $M_{\rm bh}$--$M_{\rm \star,sph}$ diagram with a strong, weak, or no bar.

\begin{figure*}
\begin{center}
\includegraphics[width=0.85\textwidth]{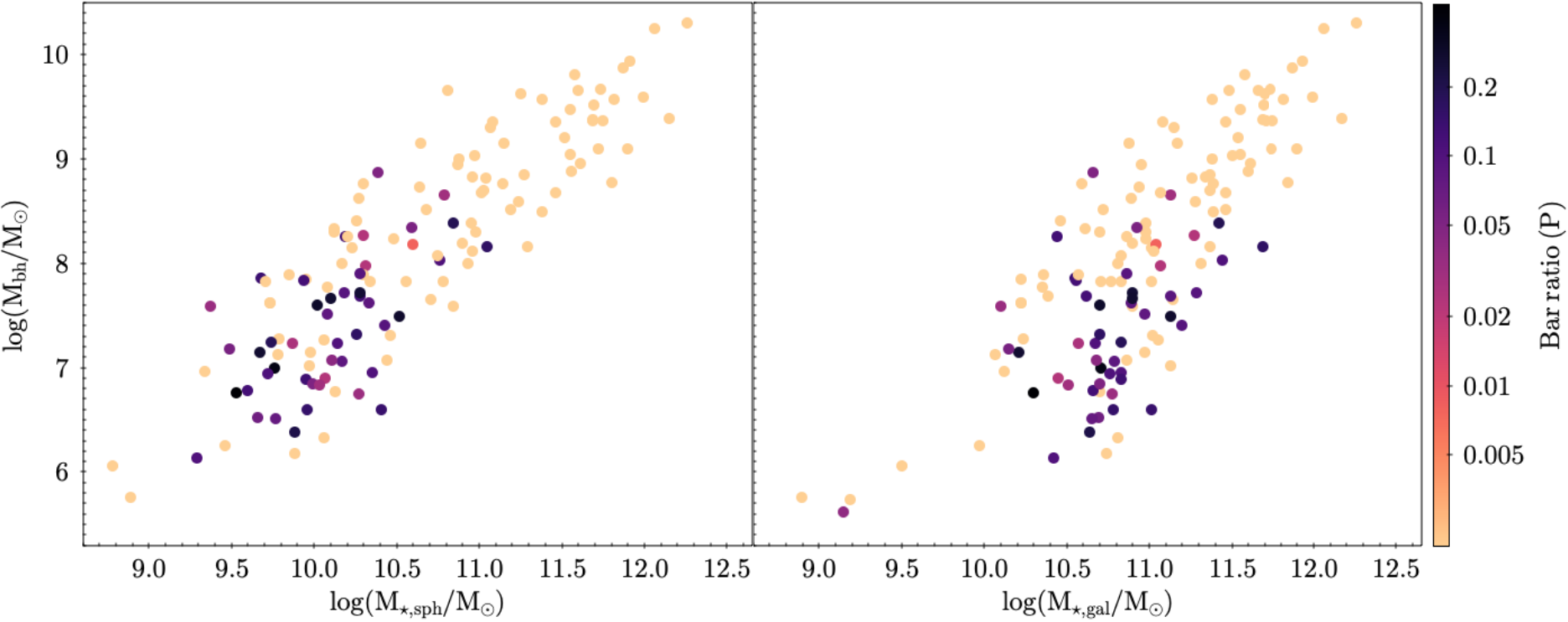}
\caption{ 
Variant of Figure~\ref{Fig-M-M} (and \ref{Fig_colour_bars}), but going beyond the classical weak/strong bar designation to now reveal the quantitative bar strength parameter, $P$.
}
\label{Fig_MMsph_bar}
\end{center}
\end{figure*}

\begin{figure*}
\begin{center}
\includegraphics[width=0.85\textwidth]{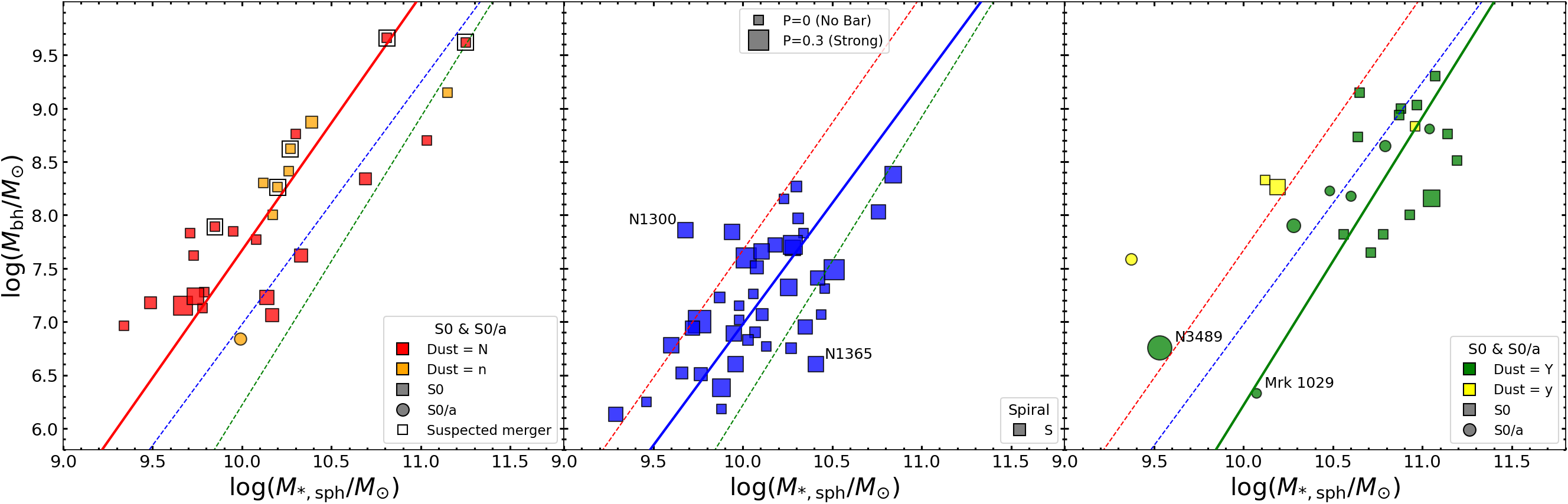}
\caption{Teasing out the detail from  a zoomed-in portion of the left-hand panel of Figure~\ref{Fig_MMsph_bar} for the disc galaxies.  The lines are those shown in Figure~\ref{Fig-M-M}.  The deficit of low-mass ($\lesssim 2\times10^9$ M$_\odot$) dust $=$ N S0 galaxies (dETGs) with $M_{\rm bh} \lesssim 10^7$ M$_\odot$  is thought to be an observational selection effect. The S0 galaxies with nuclear dust discs (dust $=$ n) tend to reside between the dust-poor (dust $=$ N) and dust-rich (dust $=$ Y) S0 galaxies. Massive dust-rich S0 galaxies are built from major wet mergers, and some of those S0 galaxies with dusty nuclear discs might have formed from major damp mergers.
}
\label{Fig_disc-bar}
\end{center}
\end{figure*}

Results for a few galaxies are worth noting. 
NGC~2549 is reported here to have a very strong bar \citep{2016ApJS..222...10S}, despite being classified as unbarred in the RC3, which is likely due to its edge-on orientation. 
Similarly, the edge-on galaxy NGC~253 (classified as AB in the RC3) is also re-reclassified here as having a very strong bar that is revealed in the \textit{SST} image along with a prominent ring at the bar's end. This is not unexpected because galaxies with thin discs, like NGC~253, are expected to form bars, modulo any stabilising effect from a dark matter halo. 
The galaxy NGC~4762 also has a particularly strong bar-plus-barlens in \citet{2019ApJ...876..155S}, 
with the bar-to-total ratio equal to 0.184 and the (bar plus barlens)-to-galaxy ratio equal to 0.496. 
Conversely, the quantitative measure of bar strength downgraded the RC3 classification from `B' to `AB' for six galaxies. This is largely due to the decomposition, which separates the bar from other components. For example, including the barlens with the bar in 
NGC~4371 increases the ratio from 0.051 to 0.201, and in 
NGC~2787 from 0.037 to 0.142, while in 
IC~2560 the ratio changes from 0.057 to 0.083, approaching the strong bar classification used here for Figure~\ref{Fig-clean-S0}.
In NGC~6926, including the inner disc/barlens(?) raises $P$ from 0.087 to 0.108; and in 
NGC~4596, from 0.089 to 0.167.
NGC~1023 ($P=0.083$) is the only galaxy where additional components are not present to have potentially bolstered the apparent bar strength in visual inspections. 

The quantification used here ensures that the bar designations reflect the physical bar component, distinct from barlenses and inner discs. 
It is also noted that while 
no galaxies with an AB designation in the RC3 were found to be barless in the decompositions, 
NGC~5206, classified as strongly barred in RC3, \textit{was} found to be barless (see Appendix~\ref{App_fits}, which assigns a nuclear disc ($h \approx 10$~pc) to this galaxy, as also done by \citet{2018ApJ...858..118N}). 

Bar-to-total stellar mass ratios, $P$, are absent for just three galaxies. 
For NGC~4945, the text pertaining to the decomposition of this galaxy in \citet{2016ApJS..222...10S} noted, but excluded from the fit, the presence of a (strong) bar. 
For NGC~4526 and NGC~4736, the text pertaining to the decomposition provided by \citet{2019ApJ...876..155S} and \citet{2019ApJ...873...85D}, respectively, noted, but excluded from the fit, a (weak) bar.
For the Milky Way, a {\it bar}-to-total ratio of 0.15 is used \citep{2007AJ....133..154L, 2015MNRAS.450.4050W}. 
Note that this is separate from the {\it bulge}-to-total ratio, which, coincidentally, is also around 0.15 \citep{2015ApJ...806...96L, 2015MNRAS.448..713P}.

Advancing beyond the strong, weak or absent bar designation of past classification schemas, 
Figure~\ref{Fig_MMsph_bar} displays a variant of Figures~\ref{Fig-M-M} and \ref{Fig_colour_bars}, revealing the $P$ values of the galaxies.  
Figure~\ref{Fig_disc-bar} goes further, revealing the bar strength for the different disc galaxy morphologies.

\subsubsection{Double barred galaxies}
\label{Sec-dubB}

Few galaxies in the sample are known to be double-barred.  
Following \citet{2010MNRAS.403..646N}, NGC~3368 was modelled in this way by \citet{2019ApJ...873...85D}, as was J0437+2456, in which the inner bar accounts for just 0.5 per cent of the light.
In contrast, NGC~3031 contains two bars that are so faint \citep{1995AJ....110.2102E, 2019ApJ...873...85D} that it continues to be labelled unbarred here. A similar situation applies to the S0 galaxy NGC~3998, which hosts a very faint single bar \citep{2011AJ....142..145G} and is often classified as unbarred.
Some confusion can  arise between rings and bars; NGC~1097 is considered here to have an inner/nuclear ring rather than an inner/nuclear bar \citep{2019ApJ...873...85D}, as is the case with NGC~2273 \citep{2004A&A...415..941E}. 

Ambiguity also exists regarding the outer primary bar in some allegedly double-barred systems. For NGC~1068 and NGC~4736, small inner bars were noted by \citet{2019ApJ...873...85D}, while \citet{2024MNRAS.528.3613E} considers any outer bar (not included in the decompositions used here) to be tentative or weak.
Furthermore, the secondary `nuclear bar' in double-barred systems often has a radial extent of only a few arcseconds.
For example, \citet{1997AJ....114.1771F} reveal a sub-arcsecond nuclear disc or bar in the SAB0 galaxy NGC~1023. Other examples include the strongly barred galaxies NGC~3504 \citep[$r_{\rm bar} \approx 2\farcs6 \approx 0.35$~kpc;][]{1993ApJ...418..687K, 2021MNRAS.504.3111W} and NGC~4303 \citep[$r_{\rm bar} \approx 1\farcs6 \approx 0.15$~kpc;][]{2000ApJ...529..845C}.

\begin{figure*}
\begin{center}
$
\begin{array}{ccc}
\includegraphics[trim=0.0cm 0cm 0.0cm 0cm, height=0.3\textwidth, angle=0]{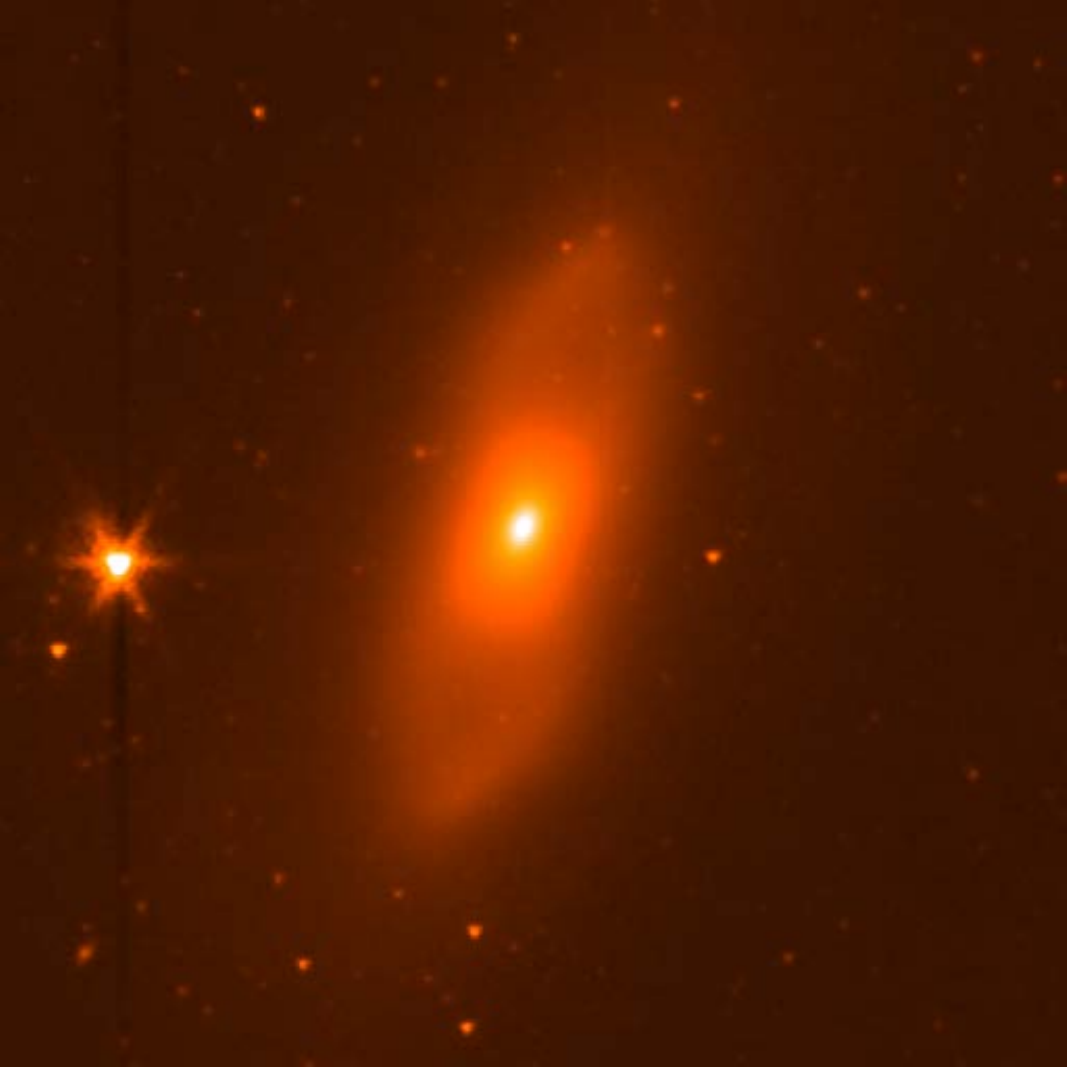} &
\includegraphics[trim=0.0cm 0cm 0.0cm 0cm, height=0.3\textwidth, angle=0]{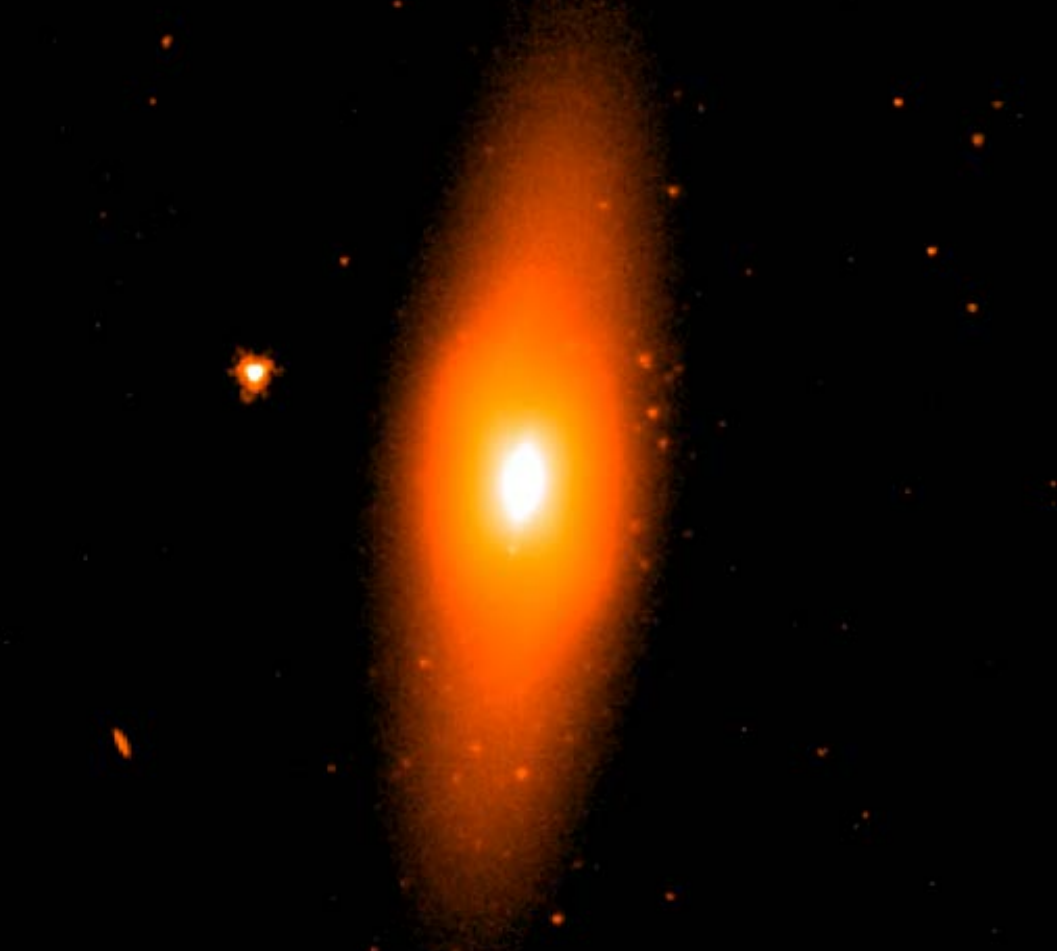} & 
\includegraphics[trim=0.0cm 0cm 0.0cm 0cm, height=0.3\textwidth, angle=0]{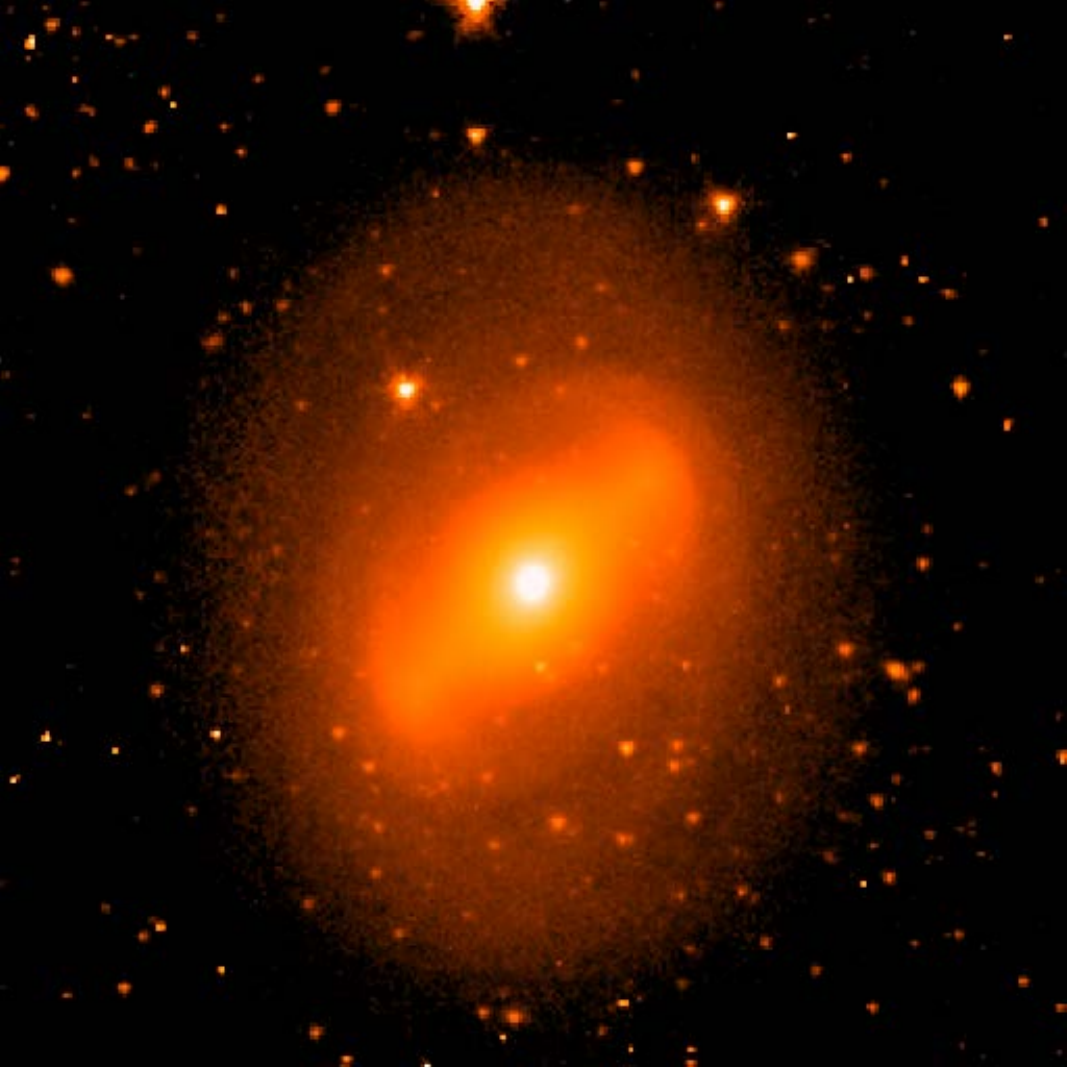} \\
\includegraphics[trim=0.0cm 0cm 0.0cm 0cm, height=0.3\textwidth, angle=0]{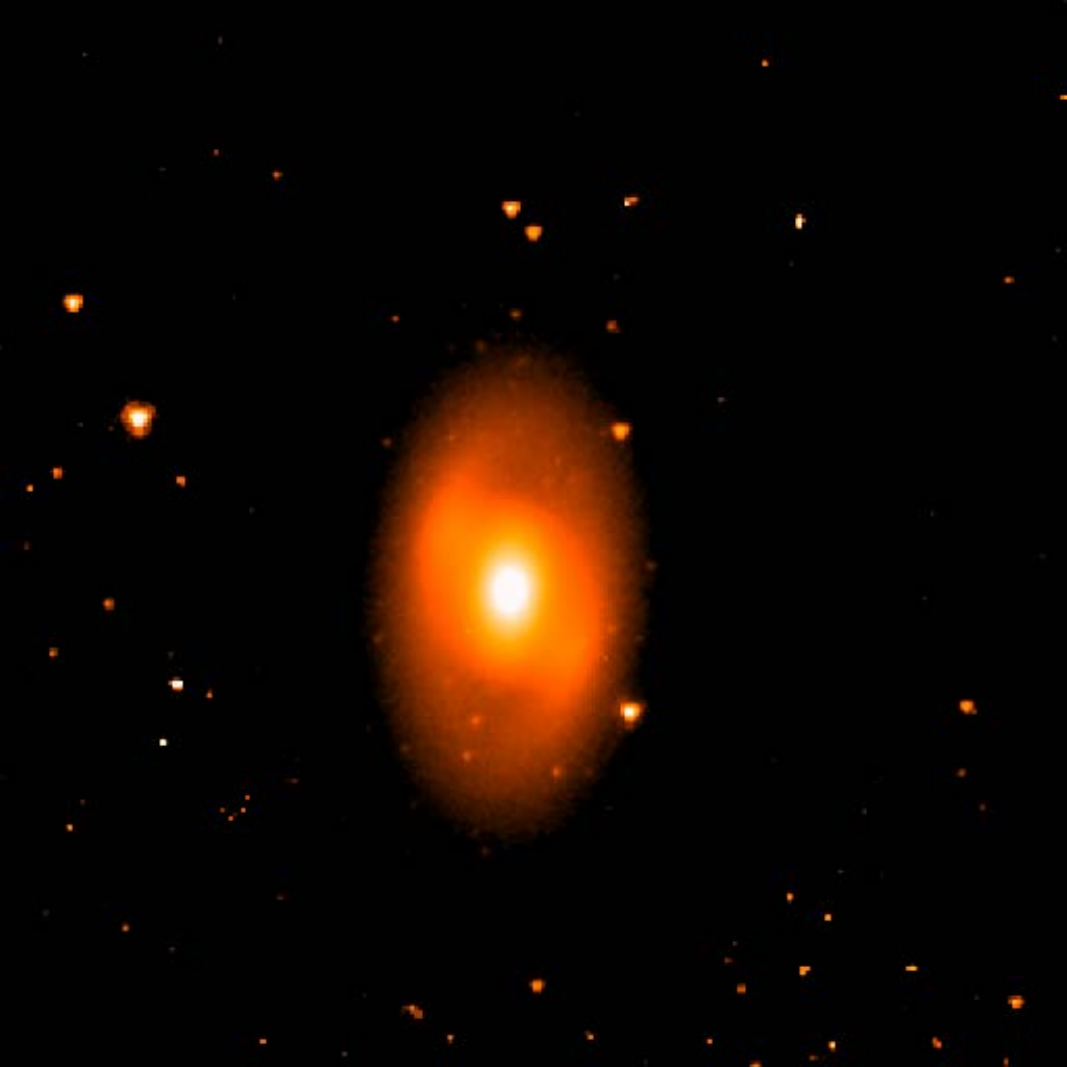} &
\includegraphics[trim=0.0cm 0cm 0.0cm 0cm, height=0.3\textwidth, angle=0]{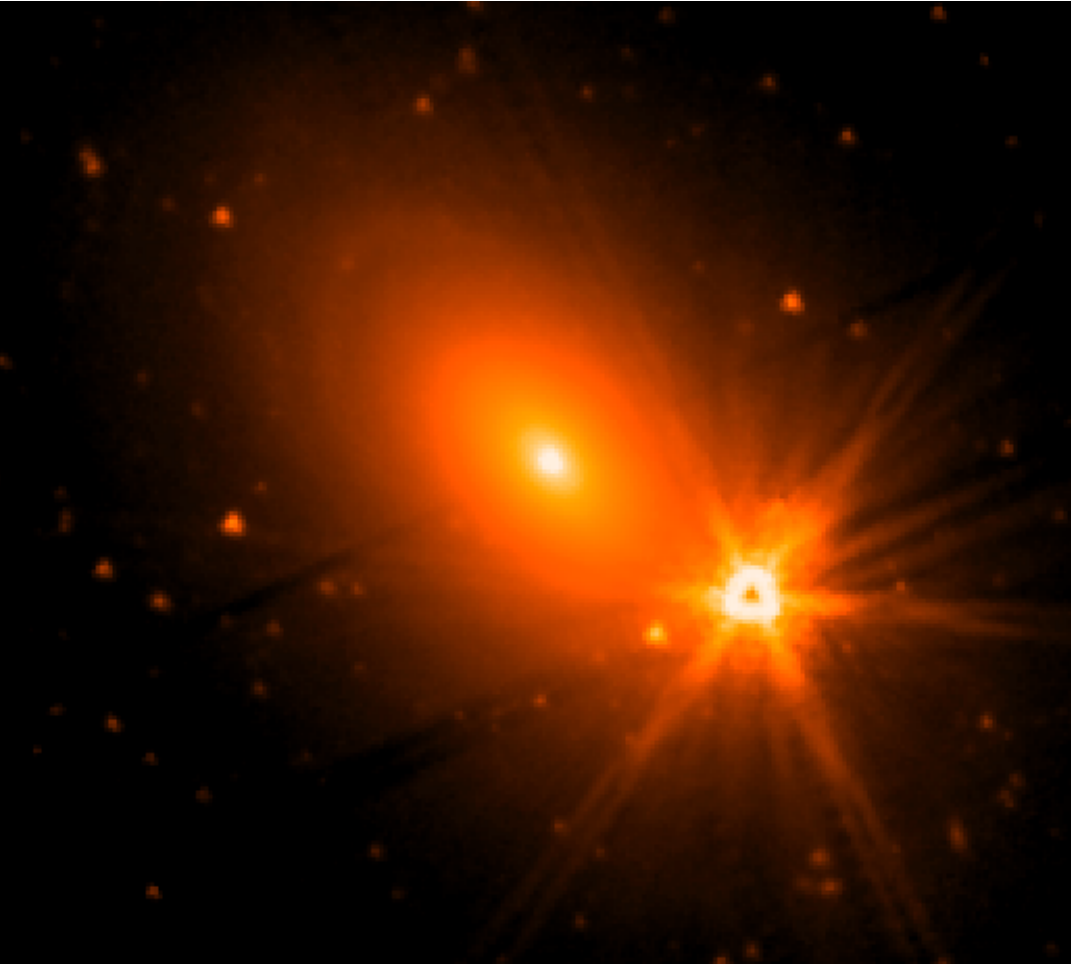} &
\includegraphics[trim=0.0cm 0cm 0.0cm 0cm, height=0.3\textwidth, angle=0]{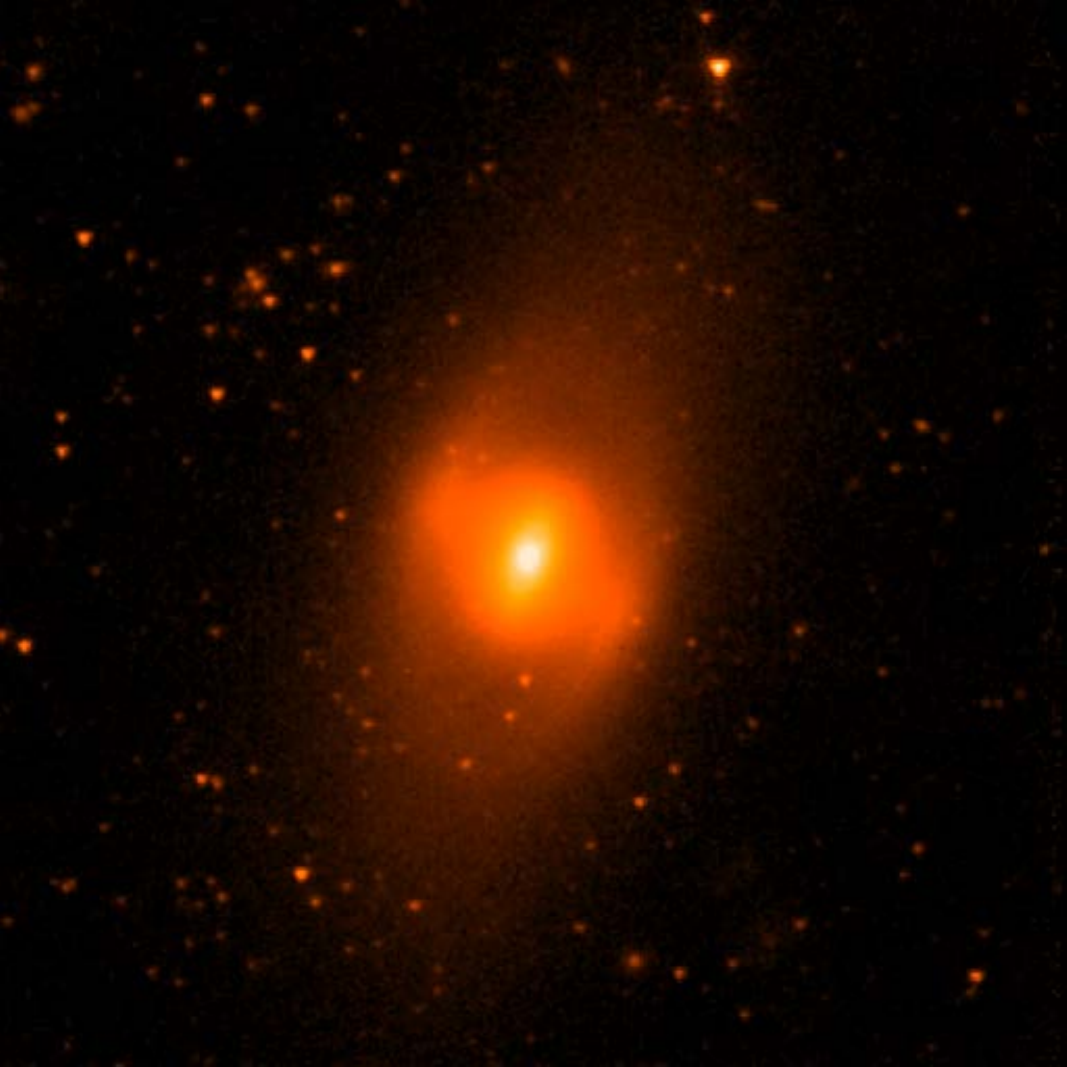} \\
\end{array}
$
\end{center}
\caption{\textit{SST} 3.6~$\mu$m images of six (mostly dust-rich) S0/a galaxies. 
The image stretch is optimised to show the weak spiral, rather than the full extent of the galaxies.
Upper row, left to right: 
NGC~4429 (dust $=$ Y);                % SA0^+(r)
NGC~4526 (aka NGC~4560, dust $=$ Y);  % SAB0^0?(s)
and NGC~4596 (dust $=$ Y).            % SB0^+(r)
Lower row, left to right: 
NGC~2787 (dust $=$ y);                % SB0^+(r)
NGC~2974 (dust $=$ Y); and            % E4 (S0/a in 1995MNRAS.274.1107N)
NGC~4371 (dust $=$ n).                % SB0^+(r)
The spiral-like feature in NGC~4371 
appears like an incomplete ring in the {\it SDSS} Data Release 14 image. 
% https://skyserver.sdss.org/dr14/en/tools/chart/navi.aspx
NGC~4526 appears different from the others in that its weak spiral pattern is {\it within} the lentil-shaped feature, perhaps associated with an X-shaped bulge, and thus making NGC~4526 an S0(s) galaxy.
}
\label{Fig-dusty-S0}
\end{figure*}

Such small nuclear bars are easily missed in decomposition analyses of limited-resolution \textit{SST} and \textit{2MASS} images. Consequently, the current census of \textit{bona fide} double-barred galaxies is likely incomplete.
The census of triple-barred galaxies is even more incomplete.

\subsection{Weak, normal, and strong spirals: S0/a, S, and \texorpdfstring{S$^{+}$}{S+}}
\label{Sec_S0a}

Following \citet{1976ApJ...206..883V}, three disc types are recognised, corresponding to the three prongs of his Trident: the normal S galaxies and the S0/a and S0 galaxies \citep{1920MNRAS..80..746R}.   A fourth type, S$^+$, is introduced here  for those  disc galaxies with very strong spiral structure. In such galaxies, the disc plays second fiddle to the prominent spiral structure and, in extreme cases, may appear to be barely there. 
Although the RC3 \citep{1991rc3..book.....D}, accessed via NED, did not list any of the sample galaxies as S0/a, inspection of \textit{SST} 3.6~$\mu$m images reveals weak spiral-like structures in seven S0 galaxies, with two additional systems (NGC~4594 and Mrk~1029) exhibiting more uncertain or ambiguous spiral-like features already reported in the literature \citep{2012MNRAS.423..877G, 2017MNRAS.471.2187D}. Six of these seven are displayed in Figure~\ref{Fig-dusty-S0}, and the seventh, NGC~3489, was shown in Figure~\ref{Fig-clean-S0}. 

Specific features include:
\begin{itemize}
    \item \textbf{NGC~4429} \citep[dust $=$ Y. S0/Sa pec:][]{1981RSA...C...0000S}. Displays an incomplete figure-of-eight pattern --- characteristic of a dissolved bar that had near-ring-like ansae and an inner X-shaped structure --- which can resemble spiral arms \citep{2015ApJS..216....9C}.
    \item \textbf{NGC~4526} (dust $=$ Y). Possesses an X-shaped bulge and a reported faint spiral \citep{1979ApJS...41..435B, 2022A&A...659A..46B}.  An impressive low-surface-brightness envelope around NGC~4526 has recently been revealed by the Simonyi Survey Telescope at the Vera Rubin Observatory. 
    \item \textbf{NGC~4596}  \citep[dust $=$ Y. SBa very early:][]{1981RSA...C...0000S}. Shows spiral arm segments associated with an incomplete ring system at the ends of its bar, somewhat resembling the face-on profile of a Star Wars `Tie Interceptor'. As with the previous two galaxies, the dusty disc region of this galaxy, which is well-known and visible in optical passbands, is well inside the domain of the weak spiral-like features \citep{1999MNRAS.306..926G}. 
    \item \textbf{NGC~2787} \citep[dust $=$ y. SB0/a:][]{1981RSA...C...0000S}. Contains misaligned dust rings and faint irregular dust lanes stretching across the spiral arm structure, encircled by a misaligned H\,{\footnotesize I} ring \citep{1987A&A...175....4S}.
    \item \textbf{NGC~2974} \citep[dust $=$ Y. S0/a:][]{1995MNRAS.274.1107N}. Features a star-forming ring and rotating H\,{\footnotesize I} ring of $\sim$$0.7\times10^9\,M_\odot$ \citep{1988ApJ...330..684K, 2008MNRAS.383.1343W, 2011ApJ...736..154M}. Merger-induced shells have been detected in deep imaging \citep{2009AJ....138.1417T}.  While a faint inner bar may be present \citep{2007MNRAS.376.1021J},  2$\arcsec$ gaseous spiral arms \citep{2003MNRAS.345.1297E} and dust extending across the galaxy are noted.
    \item \textbf{NGC~4371} (dust $=$ n). Contains a nuclear dust ring within a disc older than 7~Gyr \citep{2015A&A...584A..90G}. NGC~4371 \citep{2015ApJS..217...32B, 2016MNRAS.462.3800D} is the only galaxy in this subset where a weak spiral may not have been previously reported.   
Its spiral features might be slightly different from those of other S0/a galaxies, emanating from a ring and being strongest near the ring.  Although this may also be the case with NGC~3489 (dust $=$ Y), weakening the argument for any clear, distinguishing appearance in the non-dust-rich (dust $=$ n) S0/a galaxies.
    \item \textbf{NGC~3489} \citep[dust $=$ Y. S0/Sa:][]{1981RSA...C...0000S}. A Seyfert galaxy with a strong bar and weak spiral arms or a partial ring (Figure~\ref{Fig-clean-S0}).  The weak spiral is also visible in optical \textit{HST} images, where a strongly lop-sided dust disc is seen.
\end{itemize}

Curiously, all but perhaps one of these seven systems (NGC~2974) also possess a weak stellar bar. Except for NGC~2974, these galaxies were previously recognised \citep{1991rc3..book.....D, 2007MNRAS.376.1021J} as having inner rings or, in the case of NGC~4526 and NGC~3489 (plus NGC~4594 noted below), an inner spiral pattern.\footnote{Based on the de Vaucouleurs classification system (the CVRHS), the inner spiral (s) designation reflects the \textit{Variety} (inner structural details) while the S0/a designation reflects the \textit{Stage} (global evolution).}

The two additional S0/a candidate galaxies are less certain. Sometimes regarded as S galaxies \citep{2017MNRAS.471.2187D}, they are \textbf{NGC~4594} (dust $=$ Y, Sombrero galaxy) --- a merger product with a dusty annulus and questionable spiral --- and \textbf{Mrk~1029} \citep[dust $=$ Y. S:][]{2017MNRAS.471.2187D} with shells and what might be considered tidal debris/arms but not clear spiral structure.

Finally, six spiral galaxies with particularly strong spiral patterns relative to their host disc have been identified: NGC~1097; NGC~4303; NGC~1068; NGC~4501; NGC~6926 (disturbed, suspected inner disc rather than bulge); and UGC~6093. 

\begin{figure*}
\begin{center}
\includegraphics[width=0.85\textwidth]{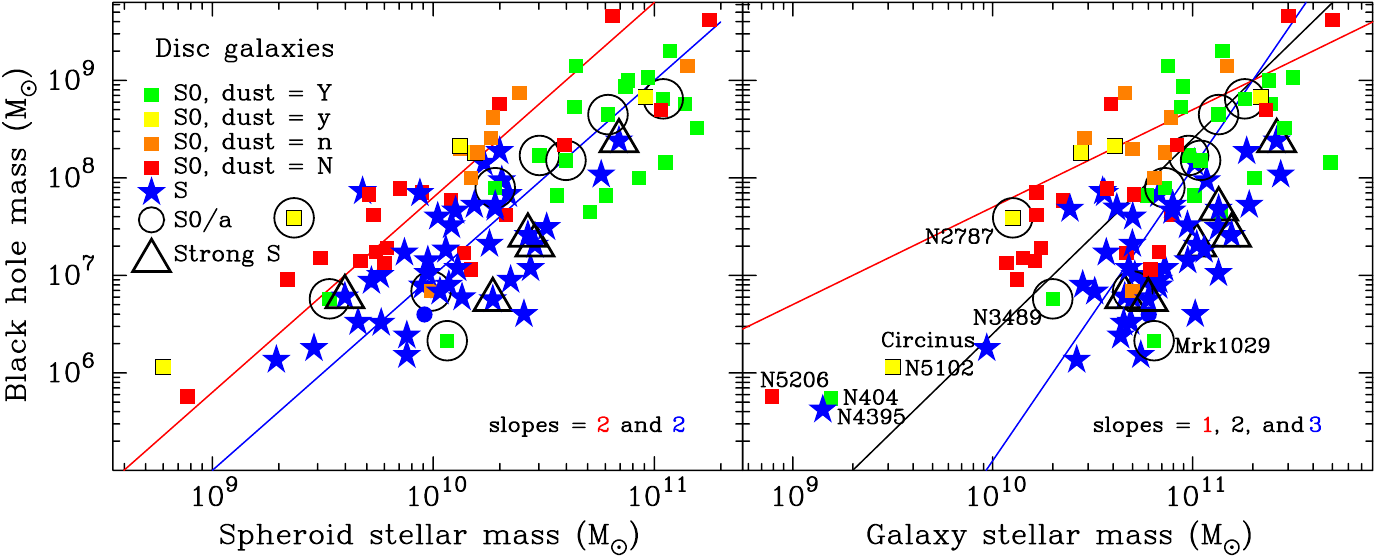}
\caption{ 
Similar to Figure~\ref{Fig-M-M}, but zoomed in and just displaying the disc galaxies.  Those with particularly weak (S0/a) and strong (S$^+$) spiral patterns are circled and enclosed with a triangle, respectively. Two ambiguous S0/a galaxies are included: Sombrero with the highest black hole mass of the nine S0/a galaxies indicated here, and Mrk~1029 with the lowest. NGC~4303 is a galaxy with a strong spiral structure and the smallest stellar spheroid mass. 
}
\label{Fig_MM_S0a}
\end{center}
\end{figure*}

Figure~\ref{Fig_MM_S0a} identifies which galaxies have particularly weak or strong spiral arms.
These identifications are utilised in the following analysis.

\section{Results}
\label{Sec_Anal}

\subsection{Weak and strong spiral patterns in the (black hole)-stellar mass diagrams}
\label{Sec_spiral}

The distribution of the S0/a and candidate S0/a galaxies in the $M_{\rm bh}$--$M_{\rm \star,sph}$ diagram (Figure~\ref{Fig_MM_S0a}) reveals that they do not form a single monolithic class. Instead, they appear to be associated with three distinct physical states of S0 galaxies noted in \citet{2024MNRAS.531..230G}, and largely correlated with their dust content and $B/T$ ratios.

\begin{enumerate}
    \item \textbf{Faded Spirals:} NGC~4371 is the only S0/a galaxy with a nuclear dust disc (dust $=$ n). It resides in the region of the diagram associated with S galaxies, distinct from the merger-built sequence of S0 galaxies. Furthermore, its large disc scalelength, $h\approx$6--7~kpc \citep{2019ApJ...876..155S}, is more in line with S galaxies than dust-poor S0 galaxies \citep[][figure~12]{2025PASA...42..155G}. 
    
    \item \textbf{Primeval/Accreting Systems:} NGC~2787 (dust $=$ y) resides where the primeval (dust-poor) S0 galaxies are located in the $M_{\rm bh}$--$M_{\rm \star,sph}$ diagram. It displays faint dust beyond the nucleus and possesses a small disc scalelength of $h=1.6$~kpc \citep{Graham:Sahu:22b}. 
    
    \item \textbf{Major Merger Remnants:} The bulk of the remaining S0/a galaxies occupying the upper end of the distribution (NGC~2974, NGC~4429, NGC~4526, NGC~4594, and NGC~4596) are dust-rich (dust $=$ Y) and consistent with massive S0 systems. Two additional low-$B/T$, dust-rich systems --- NGC~3489 \citep[$h\approx1.9$~kpc:][]{2016ApJS..222...10S} and Mrk~1029 \citep[$h\approx3.6$~kpc:][]{2019ApJ...873...85D} --- reside in the lower-left of the diagram. Presumably, these two are on their way to becoming S galaxies (if there is sufficient accretion), or Mrk~1029 could be a fading S galaxy based on its large disc size.
\end{enumerate}

A pattern with bulge prominence is also noted across these groupings. The four S0/a or candidate S0/a galaxies in the lower-left of Figure~\ref{Fig_MM_S0a} have $B/T=0.17$--$0.19$, while the five in the upper-right have higher values of $B/T=$ 0.26, 0.32, 0.36, 0.46, and 0.60. 
As noted, if the two dust-rich S0/a galaxies in the lower-left (Mrk~1029 and NGC~3489) had already experienced a major wet merger, then one might expect their $B/T$ ratios to be greater. Their smaller ratios lends weight to the migratory and faded scenario just mentioned.

While the number of galaxies with particularly strong spiral structure is modest (six), their tendency to reside toward the high-$M_{\star,\rm gal}$ side of the distribution suggests a link between spiral prominence and disc growth. 
This is consistent with the theory that dynamically colder discs, which are more susceptible to spiral formation, are more prevalent in systems where the disc has grown significantly via gas accretion \citep{2013ApJ...766...34D}.  As the disc grows and cools to the midplane, it becomes flatter \citep{1983MNRAS.202.1025G} and more prone to strong spiral instabilities. A strong spiral structure is therefore associated with dynamically cold, massive stellar discs, although spiral amplitudes can also be  enhanced by tidal forcing, interactions, and halo wake coupling, effects that may introduce increased scatter rather than the systematic trends observed here. 
However, given the limited sample size, this result should be regarded as indicative rather than definitive.

\subsection{Bar strength across the \texorpdfstring{$M_{\rm bh}$--$M_{\rm \star,sph}$}{Mbh--Mstar,sph} diagrams}
\label{Sec_bars}

\subsubsection{Weak and strong bars in spiral galaxies}

Tables~\ref{Table-data} and \ref{Table-extra} contain 44 spiral galaxies and one spiral-merger (IC~1481), plus another seven likely S0/a galaxies and an additional two candidate S0/a galaxies, discussed further in the next section.
Among the S and m/S galaxies, 18 are identified as having a strong bar, 13 have a weak bar, and 14 are unbarred.

Spiral galaxies with weak and strong bars appear intermingled in the $M_{\rm bh}$--$M_{\rm \star,sph}$ diagram (Figures~\ref{Fig_colour_bars}--\ref{Fig_disc-bar}); their bar strength does not appear to dictate their cumulative evolutionary history.
Grouping S galaxies with and without bars, and grouping S galaxies with weak or no bars versus those with strong bars, similarly revealed no pattern. Furthermore, based on the (few) known examples of double-barred galaxies in this sample, they also do not appear to occupy any special location in the $M_{\rm bh}$--$M_{\rm \star,sph}$ diagram.

The absence of a preferred location for barred galaxies in these diagrams suggests that bar strength is not tightly coupled to the cumulative growth traced by $M_{\rm bh}$ and $M_{\star}$. This supports the view that bars are primarily manifestations of secular evolution, which can arise, dissolve, and recur over timescales that do not map cleanly onto a galaxy's overall assembly history.

\subsubsection{Weak and strong bars in S0 (and S0/a) galaxies}

\textbf{S0/a Galaxies:} 
As noted in Section~\ref{Sec_spiral}, five of the six S0/a galaxies displayed in Figure~\ref{Fig-dusty-S0} contain a weak bar\footnote{It is noted that NGC~4371 could be considered as having a strong bar if the barlens component is added to the bar flux, taking $P$ from 0.051 to 0.201.}, and the sixth galaxy (NGC~2974) may too, according to the references given there.  
The seventh S0/a galaxy, NGC~3489 (Figure~\ref{Fig-clean-S0}), has a strong bar according to the decomposition in \citet{2016ApJS..222...10S}, while the two ambiguous S0/a candidates (the Sombrero galaxy and Mrk~1029) have no sign of a bar. 

A possible pattern with dust content is evident among these S0/a galaxies. Aside from NGC~4371 (dust $=$ n) and NGC~2787 (dust $=$ y), the other five S0/a galaxies (NGC~2974, NGC~3489, NGC~4429, NGC~4526, and NGC~4596) are dust-rich (dust $=$ Y), as are the two possible S0/a galaxies (NGC~4594 and Mrk~1029).

\textbf{Turning now to the S0 Galaxies:} 
None of the four ES,b galaxies are unbarred. 
Among the 40 S0 (not to be confused here with the nine S0/a) galaxies, five display a weak stellar bar ($P < 0.1$). 
In contrast to the S0/a galaxies, these five appear to belong to a dust-poor population. Aside from NGC~4350 (dust $=$ n), the other four S0/a galaxies with weak bars (NGC~0307, NGC~1023, NGC~2778, NGC~7332) do not show visible dust and belong to the dust $=$ N bin.  

Six S0 galaxies have strong bars ($P\ge0.1$), and they display a split in dust properties. Three (NGC~2549, NGC~3384, and NGC~4762)\footnote{With $P=0.10$, NGC~3384 is on the weak/strong bar cusp.} do not show visible signs of dust (dust $=$ N), while the other three show widespread dust: NGC~1316 and NGC~3489 are dust-rich (dust $=$ Y), and NGC~4026 appears in the dust $=$ y bin.\footnote{The bar-like feature visible in the 3.6~$\mu$m image of NGC~1316 is unusual and the bar classification is tentative.}
Although NGC~1023 \citep{2012MNRAS.423.2957B}, NGC~4762 \citep{1994A&A...286L...5W}, and NGC~7332 \citep{2004MNRAS.350...35F} are noted in \citet[][his table~2]{2023MNRAS.521.1023G} as possible mergers without a dust-rich character, they may have been disturbed from an encounter or minor merger. As such, they are not reassigned into the dust $=$ Y bin, which applies to wet major mergers.

Five barred S0 galaxies plus the S0/a galaxy NGC~3489 are displayed in Figure~\ref{Fig-clean-S0}, providing a comparison with the predominantly weakly-barred S0/a galaxies (Figure~\ref{Fig-dusty-S0}), which obviously require a more face-on orientation for identification of their faint spiral pattern.

For the fuller S0 sample, a mass dependence is observed in the bar properties. Although some strong bars are found in low-mass dust-poor S0 galaxies, they are notably rare in the high-mass dust-rich population (see Figure~\ref{Fig_disc-bar}). The only high-mass, dust-rich merger-built S0 galaxy with a (debatable) strong bar is Fornax~A (NGC~1316). Finally, it is noted that the dozen S0 galaxies at the top-right of the S0 distribution in the $M_{\rm bh}$--$M_{\rm \star,sph}$ diagram (Figure~\ref{Fig_disc-bar}) do not appear to contain a (strong or weak) bar.

Although alternative measures of bar strength based on gravitational torques or Fourier modes may capture different aspects of bar dynamics, and bar length could track the age, the absence of a systematic separation of barred galaxies in the $M_{\rm bh}$--$M_{\star}$ planes suggests that the core (first-order) result seen here is not sensitive to the specific bar metric adopted.

\section{Discussion and Conclusions}
\label{Sec_d_and_c}

\subsection{The progression, and predictive power, of the (black hole)-stellar mass scaling relations}
\label{Sec_prog}

The analysis of bar and spiral strength presented here relies on interpreting galaxy positions within the $M_{\rm bh}$--$M_{\star}$ plane. 
To do this correctly, one must first address the historical context of the plane's scaling relations. 
Throughout this work, care is taken to distinguish between correlations that reflect underlying physical connections and those arising from coincident structural properties, with interpretations guided by both the distributions in parameter space and consistency with established formation mechanisms (Section~\ref{Sec_Modern}). 

For years, the community operated under the assumption of a single, near-linear relation \citep[e.g.,][]{1998AJ....115.2285M, 2005SSRv..116..523F, 2007MNRAS.379..711G}. 
The shift toward the steep, quadratic-like relations seen here with the expanded data sample (Figure~\ref{Fig-M-M}) was not arbitrary; it was driven partly by the necessity to reconcile the $M_{\rm bh}$--$M_{\star}$ relation with the $M_{\rm bh}$--$\sigma$ and bent $M_{\star}$--$\sigma$ relation, as explained in \citet{2012ApJ...746..113G}. 
While initial updates confirmed the super-quadratic behaviour \citep{2013ApJ...764..151G, 2013ApJ...768...76S}, subsequent refinements improved the picture. 
For instance, \citet{2019ApJ...876..155S} identified offset relations between E and S0 galaxies, while \citet{2019ApJ...873...85D} refined the slope for S~galaxies.  
\citet{2023MNRAS.521.1023G} separated the S0 population based on formation history---distinguishing dust-rich (merger-built) from dust-poor (primeval/stripped) systems.
The historical application of single, monolithic power laws to mixed-morphology samples is no longer supported by the data.

Confidence in these morphology-dependent, quadratic relations shown in Figure~\ref{Fig-M-M} is bolstered by their ability to shed light on several key astrophysical phenomena and resolve tensions that the near-linear monolithic relation cannot.  
First, they predict the existence of ultra-massive black holes ($10^{10} < M_{\rm bh}/M_\odot \lesssim 10^{11}$) without requiring unphysically massive host galaxies --- a limitation inherent in the near-linear approximation to the ETGs.  This point is worth emphasizing.  
By explicitly tracing the steep distributions of wet- and dry-merger-built systems in the $M_{\rm bh}$--$M_{\star}$ diagrams, the framework that led to the `Triangal' successfully recovers these extreme black hole masses. 
Using the relations from \citet{Graham-triangal}, UMBHs are expected in E/ES,e galaxies with stellar masses greater than $8.1^{+1.7}_{-1.4}\times10^{11}$ M$_\odot$ (based on the stellar mass-to-light ratios used there and here).
 This refinement at high masses is particularly crucial for gravitational wave astronomy. Predictions of the stochastic gravitational wave background, which is highly sensitive to the ultra-massive black hole binary population, can be significantly improved by using these quadratic scaling relations first established for E galaxies within OzGrav\footnote{OzGrav is the Australian Research Council's Centre of Excellence for Gravitational Wave Discovery.} by \citet{2019ApJ...876..155S} and subsequently expanded through the `Triangal' schema \citep{Graham-triangal}.

Second, the morphology-dependent approach elegantly resolves the historical `bridging' anomaly. Rather than appearing anomalously positioned between E and S0 galaxies in earlier diagrams, spiral galaxies are correctly placed between the primeval (dust-poor) and merger-built (dust-rich) S0 sequences, exactly as expected from their accretion and merger histories.
 Third, these relations successfully map galaxy morphology to other fundamental properties. For instance, they link the structural formation of dust-rich S0 galaxies to the `green mountain' in the colour--magnitude diagram \citep{2024MNRAS.531..230G} and demonstrate why morphology, rather than black hole mass, is the primary driver of specific star formation rates \citep{2024MNRAS.535..299G}. It is onto this physically grounded framework that the bar and spiral strengths are now mapped, allowing their evolutionary significance to be tested.

\subsection{Stellar bars}
\label{Sec_stellar_bar}

The presence of bar-like structures in spiral galaxies was first noted by \citet{1915HelOB..15..129K} and, more convincingly, by \citet{1918PLicO..13....9C}. 
\citet{1926ApJ....64..321H} used the presence/absence of bars to create the parallel sequences of spiral galaxies in what would become known as the Tuning Fork, and \citet{1959HDP....53..275D} added the weak bar as an intermediate family. 
However, an analysis of the dependence of bar frequency on galaxy properties is limited if it does not account for bar \textit{strength}, rather than treating it simply as a binary (present or absent) feature. 
It is, therefore, unusual that bulge strength is routinely quantified while bar strength is frequently treated as a binary entity, despite the bars in some galaxies (especially bulgeless spirals) containing more light than the bulge component. 

Stellar bars are generally understood to form via a global instability in rotationally supported, self-gravitating discs \citep{1973ApJ...186..467O, 2003MNRAS.341.1179A} --- modulo bar-suppression from a dark matter halo --- or via tidal perturbations from neighbouring galaxies \citep{1982MNRAS.199.1069E, 2018MNRAS.473.2608Z, 2022MNRAS.510.4394H, 2025MNRAS.538.1587F}.  While some bars might be transient features subject to destruction or inhibition  by central mass concentrations \citep[e.g.,][]{2018ApJ...858...24S}, the current consensus --- supported by recent $N$-body and cosmological simulations --- is that bars are robust, long-lived structures that persist for secular timescales \citep{2004ApJ...604..614S, 2017MNRAS.469.1054A, 2022MNRAS.512.5339R, 2022MNRAS.515.1524Z, 2025MNRAS.538.1587F} until a substantial encounter with another galaxy. This longevity might be observationally evidenced by the presence of "fossil" bars in passive lenticular galaxies, such as NGC 4371, where the bar is reported to have survived for $\sim$10~Gyr \citep{2015AandA...584A..90G}. 
The relative lack of bars in Sc/Sd galaxies is consistent with them being dynamically young discs formed in/around S0 systems. At the same time, more massive discs (prevalent in Sa galaxies) can develop bars more easily \citep[e.g.][]{2003MNRAS.341.1179A, 2016ApJ...831...65B, 2020MNRAS.491.2547R}. 

It might be argued that the lack of segregation between barred and unbarred galaxies in the $M_{\rm bh}$--$M_{\rm \star,sph}$ diagram is a consequence of the decomposition methodology, which explicitly separates out bar-related structures (e.g., barlenses, X-shaped features, and inner discs) from the classical spheroid. If bars simply redistribute existing stellar mass from the disc to such `pseudobulge' components without altering the classical spheroid, horizontal movement in the $M_{\rm bh}$--$M_{\rm \star,sph}$ plane would not be expected. However, this interpretation ignores the vertical axis. Theoretical models have long invoked bars as a mechanism for driving gas inflows to fuel AGN and grow supermassive black holes \citep[e.g.,][]{1989Natur.338...45S, 2023ApJ...958...77L, 2024MNRAS.527.3366K}. If such bar-driven secular growth were a dominant mode of black hole assembly, one might expect galaxies that have experienced strong bars to exhibit a systematic offset toward higher $M_{\rm bh}$ values relative to galaxies of similar stellar mass. The absence of such an offset in Figures~\ref{Fig_colour_bars}--\ref{Fig_disc-bar} suggests that bar-driven inflows do not leave a strong imprint on the cumulative growth of the black hole. 
This null result may indicate that, even if bars efficiently funnel gas to the central regions, the material does not make it to the accretion disc. 
However, the present-day bar classification may be incomplete, as bars can form, evolve, and dissolve over a galaxy's lifetime. The present results, therefore, do not preclude episodic bar-driven accretion, potentially at earlier epochs.

It is stressed that the absence of a preferred location for barred galaxies in the $M_{\rm bh}$--$M_{\rm \star,sph}$ plane does not imply that bars are dynamically unimportant. 
Rather, it indicates that their presence and strength are not uniquely predictive of a galaxy's location in parameter spaces that encode its integrated growth history. 
On the contrary, stellar bars are well known to redistribute angular 
momentum, funnel gas, and drive secular evolution within discs \citep{2003MNRAS.341.1179A, 2013MNRAS.429.1949A}. 
Rather, the result presented here indicates that 
bar strength does not uniquely encode a galaxy’s cumulative evolutionary state 
once anchored to integrated measures of mass growth such as 
$M_{\rm bh}$ and $M_{\star}$. Bars could form, weaken, dissolve, and reform over a 
galaxy’s lifetime, responding to changes in gas fraction, disc stability, external tidal perturbations and 
halo structure, without leaving a monotonic imprint on the 
(black hole)--stellar mass distribution.

\subsection{More on the tripartite nature of S0/a galaxies}
\label{Sec_tripartite}

The \citet{1976ApJ...206..883V} Trident classification was motivated largely by the hypothesis that S0 galaxies represent gas-stripped, faded spiral galaxies \citep{1951ApJ...113..413S}. In this scenario, the S0/a morphology represents a transition phase of gas removal, driven by mechanisms such as ram-pressure stripping within the hot X-ray halos of galaxy groups and clusters \citep{1972ApJ...176....1G}. Alternatively, S0 and S0/a morphologies can result from the merger of spiral galaxies, where the collision destroys or erodes the spiral structure and dynamically heats the disc \citep{1996ApJ...471..115B, 1998ApJ...502L.133B}.

However, while these ``fading'' and ``merging'' channels for S0 and S0/a galaxy formation are well-established in the literature \citep[e.g.,][]{2014MNRAS.440..889S, 2021MNRAS.508..895D}, they ignore a third formation channel: primeval evolution.  These are the true `early type' galaxies.  Many such S0 galaxies may have quenched long ago without ever developing a spiral pattern. Within the `Triangal' framework, some S0/a galaxies may not be fading spirals, but rather primeval S0 galaxies in the process of \textit{becoming} spirals via gas accretion and disc growth. While this pathway was likely more common in the gas-rich high-redshift Universe, it persists today in the realm of dwarf galaxies,\footnote{This is because the accretion of a $10^8$ M$_\odot$ H{\footnotesize I} gas cloud will have a greater impact on a dwarf galaxy than it will on a Milky Way mass galaxy; and $10^9$ M$_\odot$ H{\footnotesize I} gas clouds are not around today to build-up non-dwarf spiral galaxies.} evidenced by H{\footnotesize I}-bearing ultra-diffuse galaxies with emerging spiral patterns \citep{1987AJ.....94...23B, 2017ApJ...842..133L, 2017MNRAS.468.4039R, 2017ApJ...846...26S, 2019MNRAS.484.4865P, 2021MNRAS.506.5494P}.

It has been argued that spiral patterns may be transient \citep{2011MNRAS.410.1637S, 2012MNRAS.421.1529G}, though others contend they are long-lived \citep{2013ApJ...766...34D, 2019MNRAS.489..116S}. If spiral patterns were to disappear and reappear routinely, causing galaxies to oscillate between S and S0 morphologies, one would expect a greater overlap between the two populations in the $M_{\rm bh}$--$M_{\star}$ diagram than is observed. Furthermore, in the absence of substantial changes to the disc and bulge, the spiral geometry is tied to the density of the disc and the prominence of the bulge (and thus, indirectly, the black hole mass), limiting the degree to which a spiral can simply fade back and forth without structural change.

The $M_{\rm bh}$--$M_{\rm \star,sph}$ diagram was employed to discriminate between these evolutionary pathways. As already demonstrated in Section~\ref{Sec_spiral}, the distribution of S0/a galaxies in the sample reveals that they do not form a single, continuous population, but rather occupy three distinct loci corresponding to the three formation histories described above.  This observation is discussed further here.

\begin{enumerate}
    \item \textbf{The Major Merger Remnants (Dust-Rich):} As briefly noted in Section~\ref{Sec_spiral}~(iii), the majority of the S0/a sample (NGC~2974, NGC~4429, NGC~4526, and NGC~4596) reside on the high-mass end of the distribution of S0 galaxies in the $M_{\rm bh}$--$M_{\star}$ diagrams (Figure~\ref{Fig_MM_S0a}), aligned with the sequence of massive, dusty S0 galaxies known to have formed from major mergers \citep[][table~2 and references therein]{2023MNRAS.521.1023G}. Crucially, all of these are classified as dust-rich (dust $=$ Y). Their elevated black hole masses and substantial dust content are consistent with origins in major wet mergers, where gas inflows fueled black hole growth while leaving a dusty, dynamically heated disc with reforming or surviving weak spiral features. Presumably, the wet merger products did not experience a sufficiently violent collision to either fully erase a pre-existing bar\footnote{A major merger is expected to significantly disrupt, or at a minimum temporarily compromise, the structural integrity of an existing bar due to its delicate orbital structure, where stars are confined within coherent orbital families across large spatial and temporal scales.} or produce a dynamically hot disc capable of suppressing the re-emergence of a bar, as three of these four galaxies contain weak bars. It is noted that the Sombrero galaxy (NGC~4594, with an ambiguous S0/a classification) also resides in this high-mass, dust-rich regime, reinforcing its identification as a massive merger remnant rather than a standard spiral. This identification challenges the view that galaxies in the `green valley' are primarily spirals undergoing quenching \citep{2014MNRAS.440..889S}; instead, the high-mass, dust-rich population appears to be merger-dominated yet still disc-dominated.
    
    \item \textbf{The Faded Spiral (Dust-Poor):} NGC~4371 represents the classic fading pathway. It overlaps with the cloud of spiral galaxies in the $M_{\rm bh}$--$M_{\star}$ diagram. Its classification as having only nuclear dust (dust $=$ n) indicates that it has been stripped of much of the interstellar medium that helps sustain prominent spiral structure through ongoing star formation and to maintain a strong (grand design) spiral pattern \citep{2015MNRAS.451.1350G}.  
    It has not experienced the merger-driven growth of the first group; it is simply a spiral galaxy that has ceased star formation.
    
    \item \textbf{The Primeval/Accreting System (Dust-Widespread):} NGC~2787 appears to represent the third, often overlooked channel. It resides at the low-mass end of the S0 sequence ($\log M_{\rm \star,sph} \approx 9.4$ dex), near the dust-poor S0 galaxies lacking evidence of having formed from a major merger event \citep[][table~2 and references therein]{2023MNRAS.521.1023G}. It contains weak widespread dust (dust $=$ y), consistent with an initially dust-poor S0 that has accreted gas and is currently developing the weak spiral features that warrant the S0/a designation.
\end{enumerate}

In addition to NGC~4594 (Sombrero), Mrk~1029 is the other tentative/candidate S0/a galaxy. 
Mrk~1029 (dust $=$ Y) is identified as a notable outlier, residing firmly within the spiral galaxy cloud. As a dust-rich merger remnant hosting a Seyfert nucleus, it may represent a system caught in the immediate aftermath of a merger, before significant active galactic nucleus (AGN) growth drives it up onto the wet-merger-built S0 sequence in the $M_{\rm bh}$--$M_{\rm \star,sph}$ diagram. Alternatively, it may simply represent a rarer, gas-rich dwarf-mass merger.

Finally, situated in the spiral galaxy regime of the $M_{\rm bh}$--$M_{\rm \star,sph}$ diagram, the low-mass dust-rich S0/a galaxy NGC~3489 may be primed to (try and) metamorphose into a bona fide spiral galaxy should it have or acquire sufficient gas to enter the star-forming main sequence.

While these three categories are presented as distinct formation pathways, they likely represent idealised endpoints of a broader continuum. Nonetheless, their separation in the $M_{\rm bh}$–$M_{\star}$ diagrams indicates that they capture meaningful differences in assembly history.

In summary, the morphological classification ``S0/a'' is degenerate. It captures galaxies from three physically distinct evolutionary tracks: major wet mergers with damaged/re-emerging spirals; faded spirals with dissolving patterns; and primeval S0 galaxies undergoing incipient disc /spiral growth. This mirrors the three distinct S0 galaxy types reported by \citet{2024MNRAS.531..230G}: early-lenticulars (i.e., low-mass dust-poor S0), mid-lenticulars that are faded spirals, and late-lenticulars (i.e., wet-merger-built dust-rich S0 galaxies).

\subsection{The `Disc Down-sizing' sequence and the `Dust Attrition' sequence}
\label{Sec_down_sizing}

The `Disc Down-sizing' sequence pertains to the reduction in disc size through sustained pressure from hierarchical merging and the associated entropy increase of the stars in a galaxy. 

Figure~\ref{Fig_E-dust-bins} revealed a novel structural progression at the high-mass end of the early-type galaxy sequence, as foretold by \citet{2024MNRAS.535..299G}. While it is established that major wet mergers can generate dust-rich S0 galaxies with large-scale discs, the subsequent evolution of these systems via increasingly dry mergers appears to systematically reduce the scale of the surviving disc component.

A clear mass-dependent separation between the S0 galaxies, the ellicular (ES,e) galaxies with their intermediate-scale stellar discs, and the elliptical (E) galaxies with nuclear dust discs or rings (dust $=$ n) is observed. The ES,e galaxies predominantly occupy the lower-mass regime of the $M_{\rm bh}$--$M_{\rm \star,sph}$ diagram ($M_{\rm \star,sph} \lesssim 10^{11.4} M_{\odot}$). In contrast, the merger-built E galaxies with only nuclear dust discs  tend to have higher black hole and spheroid masses.  The pure (discless) E galaxies tend to reside at yet higher masses in the diagram.
That is, a new and detailed structural evolutionary sequence---from large-scale disc to intermediate-scale disc to nuclear disc to no disc---that tracks with increasing mass is revealed, and coined the `Disc Down-sizing' sequence.

The `Disc Down-sizing' sequence strongly suggests an evolutionary continuum driven by hierarchical merging. As ES,e galaxies (or dust-rich S0s) undergo further mergers, the violent relaxation and dynamical heating destroy the stellar disc, while the remaining gas and dust have their orbital energy reduced and are funneled inward. The resulting distinct population of high-mass E galaxies with nuclear dust discs likely represents the final stage of this disc-erosion  sequence, where the once large-scale disc has been erased, leaving only a compact nuclear remnant as a fossil record of the galaxy's dissipative history.

A second pattern was observed at the top of the (nowadays) dust-poor (primeval) S0 galaxy sequence (Figures~\ref{Fig-M-M} and \ref{Fig_disc-bar}), possibly also signalling past merger events for those galaxies.
Here, a `Dust Attrition' or `Dust Retention' sequence/pattern is observed such that S0 galaxies with nuclear dust discs (dust $=$ n) --- which includes the compact massive ES,b galaxies --- tend to reside between the dust-poor (dust $=$ N) and dust-rich (dust $=$ Y) S0 galaxies in the $M_{\rm bh}$--$M_{\rm \star,gal}$ diagram.  
The lower-mass S0 galaxies are less resilient to retaining their gas and dust due to various removal (and accretion-inhibiting) scenarios, such as ram pressure stripping and strangulation \citep{1972ApJ...176....1G, 2008ApJ...672L.103K, 2009MNRAS.399.2221B}.
In addition, while dust-rich S0 galaxies form from major wet mergers, the S0 and E galaxies with a dusty nuclear disc may have formed from `damp' mergers.\footnote{The expression `damp merger' was introduced by \citet{2007ApJ...659..188F} to describe a halfway station between dry and wet mergers.} where a reduced gas fraction formed a nuclear disc.
This scenario does not rule out the possibility that larger-scale dust/gas discs in lower-mass (now dust-poor) S0 galaxies once existed and have since been removed or exhausted.

Finally, Figure~\ref{Fig_E-dust-bins} reveals the distribution of dust among E and ES,e galaxies, demonstrating that not all are products of dry mergers. Notably, a tendency for wet-merger remnants to reside on the right-hand side of the distribution is apparent. Specifically, the four dust-rich galaxies highlighted in Figure~\ref{Fig_E-dust-bins} are the BCGs NGC~1275 (ES,e) and Cygnus~A (ES,e), the giant Virgo cluster E galaxy NGC~4374 (M84), and NGC~3607 (ES,e and BGG of the Leo~II Group). 
It is therefore speculated that these galaxies may be experiencing fueling of their central black holes, which could be driving them higher up the E/ES,e distribution to larger $M_{\rm bh}/M_{\star}$ ratios.  To provide further context, 
NGC~1275 is an FR I (Fanaroff-Riley Class I) galaxy \citep{1983MNRAS.204..151L}; 
Cygnus~A has a radio jet \citep[e.g.,][]{1992A&AS...95..249L}; and 
M84 and NGC~3607 are Seyfert~2 galaxies \citep{2006A&A...455..773V}.
A more quantitative analysis of how AGN activity may relate to location in the in the $M_{\rm bh}$--$M_{\star}$ diagrams will be left for future work. 
For BCGs like NGC~1275, gas and metals might also be accreted from cooling flows in the cluster core \citep[e.g.,][]{2023MNRAS.521.1794F}, a process that may partly mimic a wet merger remnant. Consequently, 
the presence of dust and gas may facilitate the formation of an inner disc, which could potentially transition some E galaxies into the ES,e type. Supporting this idea, the low-level star formation observed in M84 may be a signature of this process \citep{2012AAS...21910203F}.

\subsection{Resolution: Tuning Fork, Trident, or Triangal?}
\label{Sec_Resolution}

\begin{figure}
\begin{center}
\includegraphics[trim=0.0cm 0cm 0.0cm 0cm, width=1.0\columnwidth, angle=0]{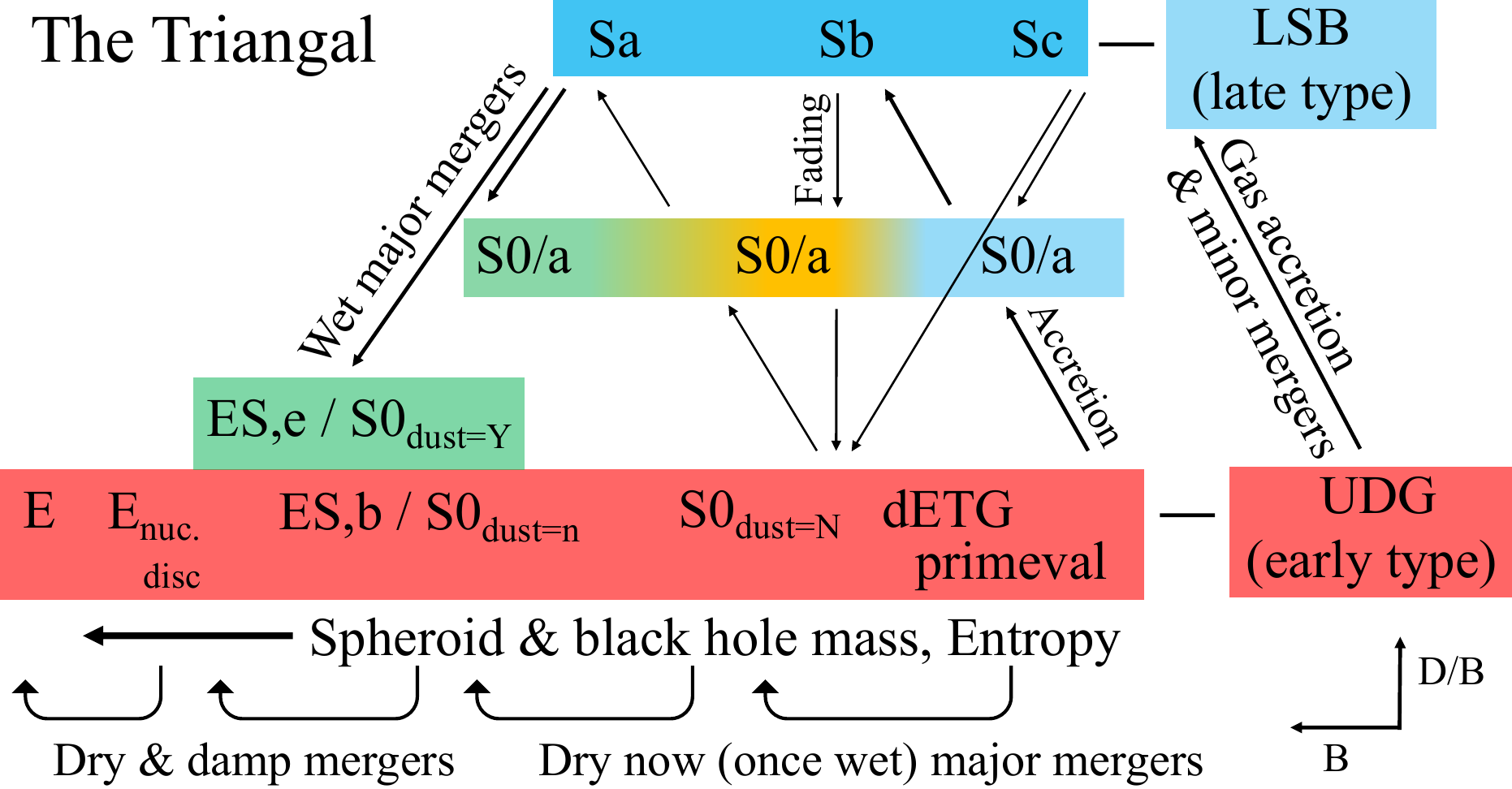}
\caption{The `Triangal' is an adaptation of the Trident and ATLAS$^{3D}$ Comb, providing an evolutionary schema for galaxy evolution.  Bars, and probably rings, can come and go in a galaxy's disc without substantially affecting the bulge or black hole mass. As indicated by the $M_{\rm bh}$--$M_{\rm \star,sph}$ diagram, such secular evolution does not play a dominant role in galaxy speciation. While the current sample does not contain low surface brightness galaxies --- hence the use of separators in this schematic --- the UDG-to-LSB galaxy connection is evident in the size-(central surface brightness) diagram shown by \citet[][figure~12]{2025PASA...42..155G}, and the UDG-dETG connection is explained there using various structural parameters.  For simplicity, Sd and Sm galaxies are not shown here, nor are the Irregular discs with their lumpy star formation.
}
\label{Triangal-Trident}
\end{center}
\end{figure}

The analysis presented here allows us to adjudicate the `clash' between the historical classification schemas. \citet{1926ApJ....64..321H} grafting of the then-recently discovered barred spirals onto the Jeans-Reynolds sequence appears to have captured a feature related to the elongation of disc orbits---a secular and often transient phenomenon---rather than an evolutionary stepping stone in the growth and speciation of galaxies. Galaxies spanning a wide range of bar strengths do not, to first order, occupy 
distinct or preferential regions of the $M_{\rm bh}$--$M_{\star}$ distributions, 
in contrast to trends associated with spiral strength or dust content. Consequently, the Tuning Fork's \citep{1928asco.book.....J, 1936rene.book.....H} primary bifurcation based on bar presence offers limited insight into the hierarchical assembly of mass.

On the other hand, the Trident \citep{1976ApJ...206..883V} and the ATLAS$^{3D}$ Comb \citep{2011MNRAS.416.1680C}, which use spiral strength (or specific angular momentum) rather than bar strength, capture a midway point for some galaxies transitioning between spiral and lenticular galaxy types. The Comb kinematically identified a sequence of fast rotators with decreasing bulge-to-total ratios, linking star-forming spirals to lenticulars. However, these schemas generally imply a gradual transition. The `Triangal' reveals that the evolutionary pathways are not necessarily gradual. For instance, major wet mergers can abruptly transform spirals into dust-rich S0s, effectively bypassing the weak-spiral phase of the Trident and Comb. \citet{Graham:Sahu:22a} referred to this merger-driven transformation as `punctuated equilibrium'.

In contrast to the predominantly secular evolution of bars, quasars, and ultra-luminous infrared galaxies \citep[ULIRGs; e.g.,][]{1996ARA&A..34..749S} are expected to occur within the population of systems climbing the `green mountain'\footnote{Upon applying dust-corrections, many dust-rich S0 galaxies move off the `red sequence' and into the `green valley' \citep{2018MNRAS.481.1183E, 2024MNRAS.531..230G}. This greater density makes the `green valley' something of a `green mountain.} of dust-rich, merger-built S0 galaxies. During the early phases of the merger-building events, gas dissipation leads to spatial offsets between the gaseous and stellar components; the distorted stellar potential then exerts strong gravitational torques on the gas, efficiently removing its angular momentum and driving it inward \citep{1991ApJ...370L..65B}.  This provides further impetus to pay attention to the dusty, or not, nature of (normal-to-massive) S0 galaxies, in addition to stellar features such as bars and rings.

The `Triangal' galaxy evolution schema --- a synthesis schematic informed by the preceding empirical results --- is visualised in Figure~\ref{Triangal-Trident}. It addresses these shortcomings by incorporating these merger-driven jumps alongside accretion-driven growth. It also explicitly accounts for the three physically distinct types of S0 galaxy identified here: primeval systems, the true `early-type' galaxies \citep{2023MNRAS.521.1023G}; faded spirals resulting from environmental stripping \citep{1972ApJ...176....1G}; and dust-rich remnants of major wet mergers \citep{1977egsp.conf..401T, 1979A&A....76...75R, 1981MNRAS.197..179G, 1983MNRAS.205.1009N}. It is grounded in the $M_{\rm bh}$--$M_{\rm \star,sph}$ diagram, which discovered here that ES,e galaxies (morphologically similar to the Comb's E(d) class) with discs fully embedded in the main spheroidal component, form part of a `Disc Down-sizing' sequence that bridges dust-rich S0 galaxies and E galaxies containing nuclear dust discs.  This sequence links to the (pure) E galaxies and reinforces the entropy-driven nature of the $M_{\rm bh}$--$M_{\rm \star,sph}$ diagram. 

The axes in the lower-right of Figure~\ref{Triangal-Trident} illustrate the primary growth vectors: gas accretion drives disc growth, while minor mergers contribute to both disc and bulge assembly. 
A well-known example consistent with this pathway is the Milky Way, which may have originated from a relatively simple, early disc system and subsequently grown through gas accretion and minor mergers, with its present-day spiral structure emerging later as a consequence of this evolution \citep{Graham-triangal}.  
Major mergers, conversely, redistribute disc stars into the merger remnant's bulge. 
While gas removal and subsequent fading reduce the optical brightness of discs, the removal of glowing dust may dim the 3.6\,$\mu$m disc magnitude, and thus the $D/B$ ratio. 
The $D/B$ axis shown in the lower-right of Figure~\ref{Triangal-Trident}, however, is not a defining characteristic of the schema and is only an approximation crudely borne out in the data \citep[][figure~A2]{Graham-triangal}.  
The $B/T$ ratio is a secondary parameter compared to the bulge mass itself \citep{1977PNAS...74.1767O, 2008MNRAS.388.1708G}, a view further supported by the quadratic $M_{\rm bh}$--$M_{\rm \star,sph}$ and cubic $M_{\rm bh}$--$M_{\rm \star,gal}$ scaling relations for late-type and dust-rich early-type galaxies. 
Variance about these axes is expected from secondary factors; for instance, massive dark matter halos can stabilize low-$B/T$ S0 discs against spiral formation \citep{1981seng.proc..111T, 1984ApJ...282...61S, 1987A&A...179...23A}.  

This framework has already delivered considerable insight beyond morphology. It explains the existence of ultra-massive black holes and the `green mountain' of dust-rich lenticulars in the dust-corrected colour--magnitude diagram \citep{2005ApJ...619L.111Y, 2018MNRAS.481.1183E, 2024MNRAS.531..230G}. 
It results in a greater awareness of the importance of major galaxy mergers \citep{Graham:Sahu:22a, 2023MNRAS.518.6293G} relative to suspected AGN feedback. 
It reveals that galaxy morphology (reflecting accretion/merger history) better predicts the specific star formation rate than black hole mass \citep{2024MNRAS.52710059G}.  By encompassing the accretion-dominated and merger-dominated regimes, the `Triangal' provides a consistent evolutionary context for the diversity of galaxy forms.

\subsection{Future work}
\label{Sec_future}

\subsubsection{Rings and nuclear bars}
\label{Sec_Rings}

As discussed in Appendix~\ref{AppdxA}, \citet{1920MNRAS..80..746R} designated galaxies with outer rings  as `Stage~IV' systems in his sequence.  Attention to rings was not included in the Tuning Fork, but they were reintroduced by \citet{1959HDP....53..275D} in his barrel-like scheme, which additionally introduced inner rings (and inner spirals). 
The painstakingly catalogued plumage of the RC3 and the \textit{Comprehensive de Vaucouleurs revised Hubble-Sandage} system \citep[CVRHS;][]{2007dvag.book.....B} reveal a wealth of detail not considered here. Like bars, many of these features may come and go (or linger), changing like the flowering and fruiting of plant species with the seasons, i.e., when the environmental (and internal) conditions are right. These flashy visual changes may not, however, significantly alter the bulge or total stellar mass; that is, they may not signal speciation---the kind of galaxy evolution emphasised here.

Nonetheless, there is insight to be gained upon closer investigation. Indeed, bars alter the gravitational field in galaxies, directing gas inwards and potentially aiding black hole growth and Seyfert activity for S galaxies. However, decades of debate reflect that this is not a dominant process continuously fuelling central black holes \citep{1997ApJS..112..315H, 1997ApJ...482L.135M, 2015MNRAS.447..506C}. It has also not led to barred galaxies having overly large black holes, as evidenced by the absence of barred spiral galaxies clustered on the high-black-hole-side of the S galaxy distributions in the $M_{\rm bh}$--$M_\star$ diagrams presented here.

Rather than fueling the black hole, much of the inward-driven gas may instead form a ring at the Inner Lindblad Resonance \citep[ILR; e.g.,][]{1984PhR...114..319A}, unless a secondary `nuclear' bar transports the gas further down to the black hole's accretion disc \citep{1989Natur.338...45S, 2002ApJ...567...97L}. It may, therefore, be fruitful to use \textit{HST} data to search for nuclear bars in this galaxy sample and check if they occupy a preferred region of the entropy space. i.e., $M_{\rm bh}$--$M_{\rm \star,sph}$ diagram.  While advancing to higher black hole and spheroid stellar masses reflects cumulative growth, it also reflects increased entropy along both axes.

Stellar rings \citep{1996FCPh...17...95B, 2017MNRAS.471.4027B} are often associated with orbit resonances driven by the gravitational potential of a bar \citep{1981ApJ...247...77S, 2006A&A...453...39R}. Building on \citet{1920MNRAS..80..746R}, which included a ringed system in the nebulae sequence, the third axis of the \citet{1959HDP....53..275D} classification volume codified these features. A complete ring is ascribed by the notation (r), with (\underline{r}s), (rs), and (r\underline{s}) denoting intermediate varieties advancing towards an s-shaped spiral, denoted (s). The absence of such features was handled omitting any such label. Adding to the complexity are outer rings (R) encircling a galaxy \citep{1961hag..book.....S}, nuclear rings (nr) \citep{1989woga.conf...29B}, and their partially complete forms, (R$^{\prime}$) and (nr$^{\prime}$). As noted by \citet{1989woga.conf...29B}, these designations can be problematic; for instance, some galaxies classified as (r) by \citet{1963ApJS....8...31D} are classified (s) by \citet{1981rsac.book.....S} when better image quality reveals the inner ring to be a pseudo-ring comprised of tightly wound spiral arms.

While rings and arms can be challenging to directly spot in images of galaxies whose discs are aligned close to edge-on, these features can stand out in the light profile. 
An exploration of the location of galaxies with these different ring structures in the $M_{\rm bh}$--$M_{\rm \star,sph}$ diagram is left for future work. However, it is noted that \citet{2018MNRAS.477.4116K} reported an excess of rings (and lenses) for galaxies in the `green valley'. \citet{2011ApJ...736..154M} similarly place four of five barred S0 galaxies with rings in this green zone, suggesting a link between these resonance features and the transition from the blue cloud.
Building on this, it is tempting to speculate that the lack of dust, and by inference gas, inhibits the growth of spirals or rings at the ends of these weak bars. However, it is noted that the weak spirals in the S0/a galaxies in this work reside at radii beyond the dusty gas discs, and it would be helpful to have a larger sample to confirm or refute this trend.

\subsubsection{Bar length and spiral arm class}
\label{Sec_barlength} 

\citet{1982MNRAS.201.1021E} developed a 12-division classification system for spiral-arm structure, ranging from flocculent (chaotic) to grand design (symmetric) patterns. \citet{1979ApJ...233..539K} established that grand design spirals are preferentially found in barred galaxies or those with tidal companions, implying that the bar potential or tidal interaction (recall \citealt{1852AJ......2...95A}) drives the spiral density wave \citep{1964ApJ...140..646L, 1966PNAS...55..229L} via the swing amplification mechanism \citep{1966ApJ...146..810J}.  Modulo tidally-induced grand design spirals in a galaxy about to merge, this would then appear to remove grand design spirals as a stepping stone in the evolutionary chain of galaxies.  Nonetheless, one may wonder whether some of the different arm classes reflect different evolutionary phases, possibly residing in different regions of the $M_{\rm bh}$--$M_{\rm \star,sph}$ diagram.  The division of S galaxies into up to a dozen arm classes would benefit from a larger sample; thus, expanding the current sample with directly measured black hole masses would be useful. 

In simulations, bars capture particles and grow in length over time via angular momentum exchange with the halo \citep{1972MNRAS.157....1L, 2002ApJ...569L..83A, 2003MNRAS.341.1179A}. While bar strength, $P$, was investigated here, it has not been explored whether the \textit{length} of the bars (either in kpc or relative to the disc scalelength) tracks  position in the $M_{\rm bh}$--$M_{\rm \star,sph}$ diagram. Some caution is required, as the truncation and anti-truncation of discs may complicate the latter measure, and the emergence of ansae and rings at the ends of a bar may act to somewhat shorten the measured bar length when these features are modelled as separate components.

The use of the $M_{\rm bh}$--$M_{\rm \star,sph}$  diagram does not replace the wealth of structural detail identified in the \citet{1959HDP....53..275D} classification volume and catalogued in the RC3 and the CVRHS \citep{2007dvag.book.....B}. Rather, it serves as a complementary tool. While the \citet{1959HDP....53..275D} `barrel' provides a detailed taxonomy, the mass scaling diagrams offer an evolutionary reference frame to test if these morphological features---such as arm class or ring variety---track with the hierarchical growth of galaxies. 

Finally, it is noted that while the qualitative \textit{class} of arms (flocculent vs grand design) may depend on triggers like bars, the quantitative \textit{geometry}---specifically the spiral pitch angle---is fundamentally tied to the central mass concentration. \citet{2008ApJ...678L..93S} and \citet{2017MNRAS.471.2187D} demonstrated that the $M_{\rm bh}$--(spiral pitch angle) relation has the smallest vertical root-mean-square scatter of all black hole scaling relations.\footnote{This may be a consequence of that relation's shallow slope. For a fixed scatter in the horizontal direction, the associated scatter in the vertical direction depends on the slope of the relation.} This reinforces the view that the underlying potential dictates the spiral structure, providing a link between the black hole and the bulge/disc/spiral geometry.

\subsubsection{Barred galaxies in the $M_{\rm bh}$--$\sigma$ diagram, and fundamental planes}
\label{Sec_fun}

While large-scale bars do not appear to dictate the position of galaxies in the $M_{\rm bh}$--$M_\star$ diagrams (Section~\ref{Sec_bars}), their dynamical influence in the $M_{\rm bh}$--$\sigma$ diagram \citep{2000ApJ...539L...9F, 2000ApJ...539L..13G} remains an open question. Dynamical studies indicate that bars increase the vertical (out of disc plane) velocity dispersion and drive radial motions, thereby elevating the central velocity dispersion, $\sigma$, of the host galaxy \citep{2013ApJ...765...23D, 2014MNRAS.441.1243H}. Consistent with this heating mechanism, \citet{2008ApJ...680..143G} and \citet{2009ApJ...698..812G} observed that the barred galaxies in their predominantly ETG sample tended to be offset to higher velocity dispersions than the non-barred galaxies at a given black hole mass.

However, this offset was not clearly evident in the larger sample studied by \citet{2019ApJ...887...10S}. This updated view may stem from the transient or oscillatory nature of bars. If bars dissolve and reform over cosmic time, the associated kinematic heating (elevation of $\sigma$) may also fluctuate. Consequently, galaxies might migrate to the right in the $M_{\rm bh}$--$\sigma$ diagram as a bar strengthens and recedes back toward the main relation (or remain as heated, unbarred remnants) as the bar dissolves. In a large, mixed sample of ETGs and LTGs, this complex evolutionary loop would likely manifest as increased scatter rather than a clean, distinct offset. In an effort to isolate this signal from the noise, the dissection of the $M_{\rm bh}$--$\sigma$ plane using the detailed morphological bins of the `Triangal' (e.g., separating dust-poor S0s from gas-rich spirals and S0s) shall be presented in a companion paper \citep{Graham-sigma2026}. 

Other scaling relations involving the spheroid S\'ersic index $n$ (representing the radial concentration of a galaxy's spheroid's light), size, and stellar density \citep{2001ApJ...563L..11G, 2007ApJ...655...77G, 2016ApJ...818...47S, 2016ApJ...831..134V, 2020ApJ...903...97S} warrant re-investigation. 
While these correlations have been established for broad galaxy classes, they have yet to be dissected using the detailed galaxy types and evolutionary context provided by the `Triangal' schema. Investigating potential `fundamental plane' parameters against the various galaxy morphologies, and encompassing the `Disc Down-sizing' and `Dust Attrition' sequences, may provide further quantitative constraints on galaxy speciation.

Finally, a word on the distinction between the spheroidal components used here and `pseudobulges'. The multi-component decompositions employed in this study 
explicitly separate the classical spheroid from bar-related features such as barlenses, inner discs, and X/(peanut shell)-shaped structures that are folded into the bar component. Many S~galaxies possess \textit{both} these secularly evolved features and a classical bulge.

\subsubsection{Some implications for cosmology}
\label{Sec_Cosmology}

The Introduction to this work noted
substantial untapped potential in the $M_{\rm bh}$--$M_{\rm \star}$  diagrams, which is only just beginning to be realised.
For example,
recognising that galaxy populations are not static but continually transform through accretions and mergers mapped by the `Triangal' evolutionary framework \citep{Graham-triangal} carries profound implications for precision cosmology. Specifically, understanding the shifting demographics, spatial distribution, and dust content of evolving galaxies exposes three critical intersections with cosmology: the Baryon Acoustic Oscillation (BAO) standard ruler, the abundance of Active Galactic Nuclei (AGN), and the Type Ia supernova (SN~Ia) standard candle.

The first implication concerns BAO measurements. 
Cosmological surveys frequently rely on luminous red galaxies (LRGs) to trace the large-scale structure that encodes the BAO signal, assuming their spatial distribution serves as a reliable standard ruler across cosmic time. However, this assumption is complicated by `merger bias' \citep{2025PASA...42...68G}. The LRG population is subject to ongoing evolutionary turnover: over time, existing LRGs in dense cluster environments merge to build brightest cluster galaxies (BCGs), effectively reducing their numbers. Concurrently, new LRGs --- specifically, massive dust-rich S0 galaxies --- are continually formed via lower-mass mergers in the more congenial, lower-velocity-dispersion environments of galaxy groups. These newly minted LRGs subsequently fall into the clusters over time \citep{2011MNRAS.416.3033S, 2014MNRAS.442.2131A, 2015JKAS...48..213K}. This ongoing creation, destruction, and spatial migration of LRGs alters their distribution. Consequently, this merger-driven spatial evolution could introduce a redshift-dependent bias that can skew BAO measurements, potentially masquerading as, or complicating claims of, an evolving dark energy equation-of-state.

A second, closely related implication involves the use of AGN abundance as an independent cosmological probe of dark energy \citep{2012MNRAS.420.2429L}. Because the hierarchical growth of structure is highly sensitive to the cosmic expansion history, the dark energy equation-of-state parameter, $\omega$($z$), fundamentally dictates the assembly rate of dark matter haloes and the frequency of subsequent galaxy collisions. As captured by the `Triangal' framework, gas-rich major mergers are a primary mechanism for funnelling cold gas into the nuclei of merging galaxies, thereby fuelling central supermassive black holes and triggering AGN activity. Consequently, because mergers could provide a key channel for triggering AGN activity, the evolving abundance and luminosity function of AGN across redshift serve as an observational proxy for the cosmic merger history. This allows AGN demographics to be utilised as a powerful, independent metric to test whether dark energy is a true cosmological constant or a dynamical, evolving field.

The third major implication of this morphological framework extends to the standardisation of SN~Ia, as already briefly alluded to in \citet[][section~2.4]{2024MNRAS.535..299G}. The so-called `mass step' --- whereby SNe~Ia in high-stellar-mass host galaxies appear systematically brighter after standardisation --- has become a standard empirical correction in cosmological analyses \citep[e.g.,][]{2010ApJ...715..743K, 2010MNRAS.406..782S, 2010ApJ...722..566L}. However, its physical origin remains heavily debated. \citet{2021ApJ...909...26B} proposed that this effect may be more fundamentally interpreted as a `dust step', reflecting variations in host galaxy dust rather than stellar mass itself. While this represents an important conceptual shift, the underlying driver is unlikely to be solely due to variations in dust grain size. Instead, the effect  likely arises from systematic differences in dust prevalence across distinct galaxy populations, which are not captured by a single global parameter such as stellar mass. Compounding this issue, much of the historical SN~Ia dataset has been obtained at optical wavelengths, where analyses must simultaneously account for both extrinsic dust reddening and intrinsic supernova colour variations. In standard cosmological fits, these two distinct physical phenomena are typically conflated into a single empirical colour-luminosity correction term (commonly parameterised via the colour--luminosity coefficient $\beta$), leading to fundamental degeneracies in their interpretation.

A key limitation in current approaches is the continued aggregation of fundamentally different galaxy types into overly broad categories. In particular, ETGs (E and S0) are routinely grouped together as a single `passive' or `high-mass' population, despite their markedly different evolutionary pathways and dust contents. As shown by \citet{2013ApJ...770..108C}, \citet{2020A&A...644A.176R}, and subsequent Zwicky Transient Facility (ZTF) analyses \citep[e.g.,][]{2025A&A...694A..14S}, separating E, S0, and S already reveals important differences. However, even in these studies, S0 galaxies are still treated as a homogeneous class. This masks substantial internal diversity. Figure~\ref{Fig-M-M} highlights that S0 galaxies comprise three types --- consistent with formation via (i) primeval processes, (ii) wet major mergers, and (iii) faded S galaxies \citep[e.g.,][]{Graham-triangal, 2024MNRAS.531..230G} --- and span four dust bins \citep{2023MNRAS.521.1023G}. This diversity exposes a critical blind spot in current supernova cosmology: the tendency to group heterogeneous galaxy populations in ways that obscure physically meaningful differences in dust properties.
Crucially, the position of galaxies within the $M_{\rm bh}$--$M_{\rm \star}$ diagrams, together with the morphological `dust bin' framework, provides the physically motivated host-galaxy classification required to implement this correction.

The concept of a `\emph{morphological dust step}' is therefore introduced, in which the observed luminosity offset arises from transitions between galaxy populations with intrinsically different dust content, rather than from stellar mass alone. Within this framework, the traditional mass step emerges as a low-resolution proxy for changes in galaxy morphology and dust content. This interpretation is supported by observations at longer wavelengths: in the near-infrared, where dust extinction is significantly reduced, the mass step has been reported to diminish substantially or become consistent with zero \citep[e.g.,][]{2021ApJ...923..237J, 2023arXiv230801875U, 2022MNRAS.510.3939M, 2018ApJ...869...56B}. This behaviour is naturally explained if dust, rather than stellar mass, is the primary underlying driver of the effect.

At stellar masses above $\approx 0.6\times10^{11}$ M$_\odot$ (see Figure~\ref{Fig-M-M}, right-hand panel), both dust-poor E/ES,e galaxies and dust-rich S0 galaxies are present within the same mass range. Consequently, a division based solely on stellar mass conflates these populations. The current mass step therefore primarily corrects for the offset associated with dust-poor E galaxies, while still blending ES,e systems (which resemble E galaxies) with dust-rich S0 galaxies. This mixing is expected to contribute to the observed scatter in SN~Ia luminosities. A transition from a mass-based correction to a morphology-based (and hence dust-sensitive) correction offers a pathway to reducing this scatter. Stellar mass nevertheless remains a relevant parameter, as lower-mass S0 galaxies tend to be dust-poor and may not require a morphological dust step correction.  That is, separating galaxies simply according to E versus S0 and S will group dust-poor and dust-rich S0 galaxies, muddying the standardisation process in an era of precision cosmology.

Furthermore, the SN~Ia's specific galactocentric location within its host cannot be ignored.. In dust-rich (dust $=$ Y) S0 galaxies, the dust is typically concentrated in the inner regions, while the extended outer envelopes can remain largely dust-poor. Consequently, a SN~Ia detonating in the outskirts of a dust-rich S0 galaxy may suffer minimal extrinsic extinction, experiencing an environment physically akin to a dust-poor S0 or E galaxy. This spatial heterogeneity underscores the need for future standardisation machinery to evolve beyond global host galaxy classifications. Incorporating spatially resolved, local environmental parameters --- specifically distinguishing between the dusty inner zone and the dust-poor outskirts of merger-built S0 galaxies --- will be key to accurately decouple intrinsic SN~Ia colour variations from external dust reddening. Formulation for how to do this will be presented in up-coming work.

These combined macroscopic and local considerations have important implications for both existing and future datasets. Much of the legacy SN~Ia sample used for cosmological inference has been obtained at optical wavelengths, where dust effects are non-negligible. Revisiting these samples with improved host galaxy classifications --- specifically incorporating morphological type and the associated `dust bin' framework inherent in the `Triangal' --- offers a clear pathway to reducing systematic uncertainties. At the same time, forthcoming surveys such as the Vera C.\ Rubin Observatory's 
% Large Synoptic Survey Telescope  
Legacy Survey of Space and Time 
\citep[LSST:][]{2019ApJ...873..111I}\footnote{\url{https://www.lsst.org/}}, conducted with the Simonyi Survey Telescope, 
will operate predominantly at optical wavelengths, where dust-induced biases remain significant. Consequently, the need for a physically motivated correction—such as the morphological dust step—will persist, and likely become more acute as statistical uncertainties continue to shrink.

\section*{Acknowledgements}

Work experience student Sol Oenning assisted with preliminary investigations, looked for faint (and particularly strong) spiral-like features in the S0 (and S) galaxies, and captured most of the images shown in Figures~\ref{Fig-dusty-S0} and \ref{Fig-clean-S0}.

I am grateful to David Liptai, Conrad Chan, and Robin Humble of the
Astronomy Data and Computing Services (ADACS) in Australia for keeping the
Image Reduction and Analysis Facility (IRAF) alive on the OzSTAR and `Ngarrgu
Tindebeek' (NT) computers at Swinburne University of Technology.
Some of this work was performed on the OzSTAR and NT national facility at Swinburne
University of Technology. The OzSTAR and NT program receives funding in part from the
Astronomy National Collaborative Research Infrastructure Strategy (NCRIS)
allocation provided by the Australian Government and the Victorian Higher
Education State Investment Fund (VHESIF) provided by the Victorian Government.
Publication costs were funded through the Australia and New Zealand
Institutions (Council of Australian University Librarians affiliated) Open
Access Agreement.
This work has used the NASA/IPAC Infrared Science Archive (IRSA)
and the NASA/IPAC Extragalactic Database (NED),
% \footnote{\url{https://doi.org/10.26132/NED1}},
funded by NASA and operated by the California Institute of Technology.
Funding for SDSS-III has been provided by the Alfred P. Sloan Foundation, the
Participating Institutions, the National Science Foundation, and the
U.S. Department of Energy Office of Science.
This research has also used the SAO/NASA Astrophysics Data System (ADS)
bibliographic services.
% and the {\sc Rstan} package available at \url{https://mc-stan.org/}.
%
Figure~\ref{Fig_MMsph_bar} was made with the help of TOPCAT \citep{2005ASPC..347...29T}.\footnote{\url{https://www.star.bristol.ac.uk/mbt/topcat/}}

\section{Data Availability}

The imaging data underlying this article are available in the NASA/IPAC Infrared Science Archive
and the Hubble Legacy Archive. 
Machine-readable versions of the tables are available online.

% \bibliographystyle{mn2e}
\bibliographystyle{mnras}
\bibliography{Graham-weak}{}

\appendix

\section{Greater historical context}
\label{AppdxA}

\subsection{Galaxy sequencing/staging}

Several attempts to systematically sequence galaxies arose following photographic surveys at the turn of the twentieth century. Using the Crossley 36-inch reflector, \citet{1899MNRAS..60..128K, 1900ApJ....11..325K} and successors \citep{1904LicOB...3...47P, 1918PLicO..13....9C} revealed a sheer abundance of spiral nebulae \citep{1895MNRAS..56...70R, 1899spss.book.....R}, estimating that hundreds of thousands to one million existed. This vast population necessitated new classification frameworks \citep[e.g.,][]{1908PAIKH...3..109W, 1915HelOB..15..129K, 1917ApJ....46...24P, 1918PLicO..13....9C} to supersede the designation systems of the Herschel family \citep[as described by][see p.919]{1933HDA.....5..774C} and \citet{1888MmRAS..49....1D, 1895MmRAS..51..185D}.

Building on \citet{1919pcsd.book.....J}, \citet{1920MNRAS..80..746R} developed what was essentially an E–S0/a–Sa–Sb–Sc–(R)S–Irr sequence representing stages of nebular concentration.\footnote{The term `stages' was used by \citet{1920MNRAS..80..746R} and maintained by \citet{1959HDP....53..275D} to describe the backbone of his expanded classification schema.} Central concentration, bulge prominence, spiral arm development, the presence of condensations (star-forming knots), outer rings\footnote{\citet{1920ApJ....51..276P} reports on rings in several spiral nebulae.}, and irregular structure without a definite nucleus were treated as markers of progression. Irregular galaxies were regarded as a later evolutionary stage of spirals, while the `spindle' (edge-on) classification was excluded as it was known to reflect viewing angle rather than intrinsic structure.

This sequence effectively codified a modified version of the nebular hypothesis \citep{Swedenborg1734, Kant1755, Laplace1796, Laplace1799_1825}, incorporating the idea proposed by \citet{1852AJ......2...95A} that gravitational tides from passing encounters could draw out spiral arms. These arms, discovered by William Parsons (the 3$^{\rm rd}$ Earl of Rosse; \citealt{1850b_Rosse, 1850RSPT..140..499R}), were interpreted as features emerging as nebulae evolved from amorphous elliptical systems into lenticular and spiral forms.\footnote{Using the `Leviathan' of Parsonstown, Parsons  spotted stars in some of the nebulae, leading him to reject the nebular hypothesis --- which regarded nebulae as a luminous fluid rather than a mass of stars --- and placing him at odds with the Herschels, who supported the theory.} In the absence of external perturbations, rotation alone had previously been thought to produce equatorial rings rather than spirals.\footnote{At that time, rings in nebulae \citep[e.g.,][]{1833RSPT..123..359H} were sometimes thought to condense into planets, forming solar systems.} Crucially, this interpretation --- with spirals being `pulled out’  from earlier amorphous systems --- established the `early-to-late' evolutionary direction that later became embedded in galaxy nomenclature, in contrast to the modern view in which spiral galaxies may fade or merge into more amorphous systems.

After reversing his initial ordering of nebulae \citep[as presented in][]{1922ApJ....56..162H} and partly embracing the evolutionary scenario of \citet{1919pcsd.book.....J}\footnote{To give fuller credit, \citet{1919pcsd.book.....J} had embraced the modified nebular hypothesis that was still discussed by many in regard to solar system formation  (e.g., 
\citealt{Tisserand1889_1896};
\citealt{1901ApJ....14...17C, 1916JRASC..10..473C}; 
\citealt{1905ApJ....22..165M};
\citealt[pp.~118--119]{1906PASP...18..111A}; 
\citealt{1909PA.....17..418M}).}, 
\citet{1926ApJ....64..321H} introduced the `early-type' and `late-type' terminology (borrowed from stellar classification).\footnote{Stars were referred to as `early-type' and `late-type' due to the misplaced belief that hot stars evolved into cooler stars.} 
Importantly, much of the terminology used in these classification schemes predates their formalisation. Terms such as `elliptical', `spiral', and `lenticular' were already in use to describe the visual appearance of nebulae \citep{1844ccob.book.....S, 1860shcc.book.....S, 1881cco..book.....S, 1852AJ......2...95A, 1902Obs....25..321G, 1863AReg....1...49.}. This historical inheritance meant that some morphological labels carried implicit physical interpretations, including evolutionary connotations, which later became somewhat embedded in classification frameworks.
Although the nebular evolutionary interpretation had largely fallen from favour by the mid-1920s \citep{1971JHA.....2..109H}, \citet{1926ApJ....64..321H} retained the terminology while withdrawing its explicit evolutionary meaning prior to publication.

\citet{1926ApJ....64..321H} adopted much of the Jeans–Reynolds sequence backbone but introduced a key modification: the bifurcation of spiral galaxies into barred and unbarred families, giving rise to the Tuning Fork diagram. A competing perspective later emerged in the form of the \citet{1976ApJ...206..883V} `Trident', which retained the E–S0–S backbone but regarded spiral strength as more evolutionarily informative than bar development. Neither framework, however, directly incorporates cumulative measures of galaxy growth, leaving open the question of whether these morphological features trace  
short-lived states or long-lived features that might, nonetheless, be better regarded as embellishments rather than tracers of speciation and growth 
\citep[e.g.,][]{2005MNRAS.364L..18B, 2014MNRAS.439..623G, 2019MNRAS.489..116S, 2024A&A...684A.179R}.

\subsection{Galaxy Naming/Typing}

The terminology used in modern galaxy classification predates its formal codification and originates from nineteenth- and early twentieth-century visual descriptions of nebulae. The term `elliptical' was already well established by the time of the popular series “A Cycle of Celestial Objects” \citep{1844ccob.book.....S, 1860shcc.book.....S, 1881cco..book.....S}, and was used by \citet{1852AJ......2...95A} and others \citep[e.g.,][]{1902Obs....25..321G} to describe featureless nebulae. 
In parallel, \citet{1847raom.book.....H} assigned such nebulae a rating from 1 (round) to 5 (highly elongated), a precursor to the axis-ratio ellipticity classifications later employed by \citet{1917ApJ....46...24P} and others.
Similarly, \citet{1863AReg....1...49.} described nebulae as oval, lenticular, spiral, and `bizarre', cementing some of the terminology that would later be adopted in galaxy classification.

Following \citet{1863AReg....1...49.}, \citet{1919pcsd.book.....J} used the term `lenticular' to describe objects intermediate between `elliptical' and `spiral' nebulae. \citet{1920MNRAS..80..746R} explicitly applied this designation to the Andromeda nebula (his Stage~II object), which he considered to be a remote stellar system \citep[see also][]{1925MNRAS..85.1014R}. Although \citet{1936rene.book.....H} is often credited with introducing the lenticular class, his work largely formalised an already established intermediary category.\footnote{Hubble’s reliance on Reynolds’ work appears significant. \citet{1913MNRAS..74..132R} introduced a light profile model for the dusty S0 Andromeda nebula. However, this model would later be popularly referred to as the `Hubble model' or ‘Hubble Law’ \citep[e.g.,][]{1981ApJ...251...61C, 1983ApJ...264..337S} or at least attributed to Hubble \citep[e.g.,][]{2018PASA...35...31P, 2021MNRAS.508.1870S}, despite \citet{1930ApJ....71..231H} {\it reproducing} Reynolds' model without accreditation. It seems fair to say that John Henry Reynolds \citep[1874--1949:][]{1950MNRAS.110..131J, 1950Obs....70...30.} deserves more recognition for his contributions to astronomy in general, and galaxy morphology specifically, than he receives. 
 Indeed, Reynolds’ contribution extended to the very hardware that shaped these classification systems. He financed the 30-inch telescope at Helwan Observatory (Egypt), used by \citet{1915HelOB..15..129K} to identify what may have been the first galaxy bars \citep[later symbolised as $\phi$ 
 by][]{1918PLicO..13....9C}. Reynolds later donated this telescope \citep{2011AntAs...5...36M} to Mount Stromlo Observatory (Australia), where it became the primary instrument used by the de Vaucouleurs in the 1950s to refine their ideas about galaxy classifications discussed herein \citep[e.g.,][]{1959HDP....53..275D, 1964rcbg.book.....D, 1991rc3..book.....D}.
The Reynolds telescope was the largest optical reflector in the Southern Hemisphere until the
1950 and 1955 opening of the 74-inch telescopes in Pretoria, South Africa, and
at Mount Stromlo Observatory, respectively.
}

\citet{1926ApJ....64..321H} adopted much of the existing terminology and sequences, referring to Stage~I systems of \citet{1920MNRAS..80..746R} as `Elliptical', Stages III–V as Sa–Sc\footnote{\citet[][p.324]{1926ApJ....64..321H} even used the same galaxy (NGC~5457)
as \citet[][p.747]{1920MNRAS..80..746R} to illustrate the late-type Stage V (aka Sc) spiral
galaxies.}, and Stage~VII as Irregular, as Reynolds named them.  Notably, he excluded the Stage~II systems (lenticulars), effectively prioritising the presence or absence of bars over the transitional S0/a and S0 classes highlighted by \citet{1919pcsd.book.....J} and \citet{1920MNRAS..80..746R}, and later emphasised by \citet{1976ApJ...206..883V} and \citet{2011MNRAS.416.1680C}.
Another type of galaxy, now confirmed as transitional, is the ETG with an intermediate-scale disc, designated ES by \citet{1966ApJ...146...28L}. Her work appeared after the landmark study by  \citet{1959HDP....53..275D}, and therefore was not included in it, likely affecting exposure and awareness of ES galaxies.

\subsection{The modern framework}\label{Sec_Modern}

Galaxy morphology has long served as a descriptive classification tool, yet its connection to the underlying physical pathways of galaxy speciation has remained somewhat ambiguous. 
While the above century-old labels have endured, the understanding of physics has shifted dramatically. 
Today, the disconnect between nomenclature and physical origin (for example, the name `lenticular' now refers to galaxies with three different physical origins) motivates a reassessment of how morphological features are used in evolutionary narratives. The `Triangal' framework, derived empirically from the distribution of galaxies in the $M_{\rm bh}$--$M_{\rm \star,sph}$ diagram rather than from visual appearance alone, offers a means to reconcile classical morphology with modern evolutionary understanding. By examining where barred and weakly spiralled galaxies fall within this framework, one can test whether long-standing morphological distinctions correspond to distinct evolutionary pathways or merely reflect transient or slowly evolving (quasi-static) structural states. 
Clarifying whether bar or spiral strength traces galaxy evolution has direct implications for how morphology is interpreted in both observations and simulations.

A localised, largely unidirectional inflow of gas driven by mutual gravitational attraction can lead to a low-mass spheroidal (triaxial) galaxy in a monolithic-like collapse \citep{1962ApJ...136..748E} if the gas is dense enough for the stars to form from condensations in the cooling clouds of gas \citep{1902RSPTA.199....1J, 1978MNRAS.183..341W} prior to galaxy gas-disc formation. Modern cosmological simulations have revisited such dissipative collapse scenarios, demonstrating that rapid, centrally concentrated star formation during early gas-rich phases can likewise produce compact quasi-spheroidal systems \citep[e.g.,][]{2018MNRAS.473.4077P, 2022ApJ...937...15P}.
Such triaxial galaxies may have some rotation if the initial gas reservoir did \citep[e.g.,][]{1980MNRAS.193..189F, 1997ApJ...482..659D}, and some (perhaps some primeval ultra-diffuse galaxies: UDGs) may even resemble thick discs.  Gas clouds drawn in from further afield will take longer to come in from the cold 
and may be channelled along streams \citep{2009Natur.457..451D, 2009MNRAS.396.2332K}. Gas that arrives with higher angular momentum may lead to the formation of larger, flatter, disc-like stellar structures \citep[e.g.,][]{1998MNRAS.295..319M, 2004ApJ...613L..41N, 2017MNRAS.465.1241W}. Subsequent perturbations of these discs, from the accretion of lumpy gas and minor mergers, can foster spiral patterns 
\citep{1966ApJ...146..810J, 1984ApJ...282...61S, 2011Natur.477..301P, 2014PASA...31...35D, 2015ApJ...808L...8D, 2022MNRAS.512..366A}, rejuvenate star formation \citep{2007ApJS..173..538T, 2008A&ARv..15..189S}, and contribute to bulge growth, 
in part through dynamical heating of the disc and redistribution of the (infalling and host galaxy) stars \citep[e.g.,][]{1993ApJ...403...74Q, 1996ApJ...460..121W}.  While the ‘Triangal’ captures the broad morphological transformations driven by these processes, it does not yet account for the emergence of bars \citep{2012ApJ...757...60K, 2019MNRAS.483.2721P, 2022MNRAS.515.1524Z, 2025A&A...698A..20R} or weak spirals. This raises the question of whether these features mark a definitive evolutionary stage or simply come and go during the existence of discs.

Although tidally-induced arms are well-documented (e.g., the Antennae galaxies), and \citet{Roche:1850} overflow
can build tidal bridges and transfer material \citep[e.g.,][]{1970SvA....13..968T, 2007ApJ...658..345H, 2017ApJ...834...16M}, not all spiral patterns are `grand design' density waves induced by an external tidal perturbation \citep{1964ApJ...140..646L, 1966PNAS...55..229L}.  Many spiral arms form from smaller gravitational perturbations in pre-existing discs \citep[e.g.,][]{1966ApJ...146..810J, 1984PhR...114..319A, 1984ApJ...282...61S, 1985ApJ...288..438E, 2002A&A...394L..35B}.   Ongoing accretion (and gas recycling) and mergers are nowadays also known to conduct the galactic symphony. 
 
It is also known that disc instabilities (induced by external gravitational perturbations, accretion events, or intrinsic forces if the dark matter halo is insufficient to stabilise the disc) can lead to the formation of bars \citep{1964ApJ...139.1217T, 1973ApJ...186..467O, 1986MNRAS.221..213A, 1987MNRAS.228..635N, 1990A&A...230...37G, 2002A&A...392...83B, 2004MNRAS.347..220B, 2013seg..book..305A, 2018ApJ...857....6L, 2018MNRAS.473.2608Z, 2018MNRAS.479.5214Z, 2025MNRAS.539.2262C}. 
However, a significant fraction of disc galaxies remain unbarred, particularly among lower-mass, late-type spirals (Sc, Sd). These systems often possess dynamically `hotter' and thicker stellar discs \citep{2004ApJ...608..189D, 2006AJ....131..226Y, 2008ApJ...682.1004Y, 2017AandA...597A..48F} with higher gas fractions \citep{2010ApJ...719.1470V, 2013MNRAS.429.1949A, 2013ApJ...766...34D, 2018MNRAS.474.5372E, 2024MNRAS.527.3366K, 2022MNRAS.514.1006I, 2023ApJ...947...80B, 2024ApJ...968...86B}. In the context of the `Triangal', these late-type spirals may represent the early stages of disc growth where gas is cooling into a plane within a triaxial dwarf host. Ongoing accretion and minor mergers may advance these systems from late-type spiral galaxies to early-type spiral galaxies, with bars being a transient feature at times.  
Strong and long bars are known to be more prevalent in the early-type spiral galaxies (with more massive bulges) than in late-type spiral galaxies \citep{2000AJ....119..536E, 2011MNRAS.415.3308G, 2011MNRAS.411.2026M}. 
Bars are not just a feature of spiral galaxies, but are also observed in S0 galaxies \citep[e.g.,][]{1959HDP....53..275D, 2004ApJ...607..103L, 2005MNRAS.362.1319L}.  

Major mergers of dust-rich spiral galaxies can destroy the more fragile spiral patterns and lead to the development of dusty S0 galaxies with dynamically hotter discs and more massive bulges \citep{1996ApJ...471..115B, 1998ApJ...502L.133B, 2014A&A...570A.103B, 2015A&A...579L...2Q, 2018A&A...617A.113E}. 
Inflowing gas that lost orbital angular momentum in the collision can fuel central starbursts and make more compact remnants, or create the smaller embedded discs seen in ES galaxies \citep{1966ApJ...146...28L, 2019MNRAS.487.4995G}, while dry (gas-poor) mergers can puff up stellar systems \citep{2017MNRAS.465.1241W} and/or deposit material to create disc-like structures having a large radial extent about the system's new centre of mass \citep{2024MNRAS.535..299G}.
Further dry (gas-poor) major mergers can dynamically heat the stellar system, leading to the formation of an elliptical galaxy \citep{2006ApJ...636L..81N}. 

The ‘Triangal’ framework (Graham 2023c) effectively maps many of the above distinct physical processes to specific locations in the $M_{\rm bh}$--$M_{\rm \star,sph}$ diagram. Unlike preceding schemata,
including the galaxy classification grid in \citet{2019MNRAS.487.4995G} that captures a variety of early-type galaxy morphology, 
 the `Triangal' captures the three distinct S0 formation pathways: primeval systems formed via gravitational collapse (the true `early-type' galaxies); faded spirals resulting from environmental stripping of gas; and remnants of major mergers. 
The `Triangal' arose from the distribution of galaxy types in the $M_{\rm bh}$--$M_{\rm \star,sph}$ diagram \citep{Graham-triangal}, rather than from a cladogram. 
It reveals that commonly used morphological labels do not uniquely encode galaxy formation pathways, but instead conflate systems with distinct physical origins.

This study investigates what insights the $M_{\rm bh}$--$M_{\rm \star,sph}$ (and $M_{\rm bh}$--$M_{\rm \star,gal}$) diagram may offer regarding the presence and relevance of bar and spiral strength in the evolutionary progression of galaxies. 
For instance, might bars form before some spiral patterns \citep{1979MNRAS.187..101L}, thereby preferentially occurring in galaxies near the boundary of the distribution of low-mass dust-poor S0 galaxies and the S galaxy sequence in the $M_{\rm bh}$--$M_{\rm \star,sph}$ diagram, or perhaps they are associated with later stages of evolution, occurring between the S and high-mass dust-rich S0 galaxy sequences. Might the galaxy distribution reveal evidence of weak spirals at effectively earlier phases (in lower-mass disc galaxies) than regular spirals, or will weak spirals associated with fading spiral galaxies at a later stage of evolution be found?

By anchoring classical morphological features to the bivariate distribution of galaxies in the $M_{\rm bh}$--$M_{\rm \star,sph}$ plane, this work explores how bar strength and spiral strength may be tied to galaxy evolution. In doing so, it offers a unifying context --- provided by a physically motivated black hole–spheroid mass framework --- in which to reassess the apparent clash between the Tuning Fork and the Trident, and to clarify the evolutionary status of barred galaxies, weakly barred systems, and S0/a galaxies with faint spiral structure.

\section{Modelled light profiles}
\label{App_fits}

\begin{figure*}
\begin{center}
\includegraphics[trim=0.0cm 0cm 0.0cm 0cm, height=0.26\textwidth, angle=0]{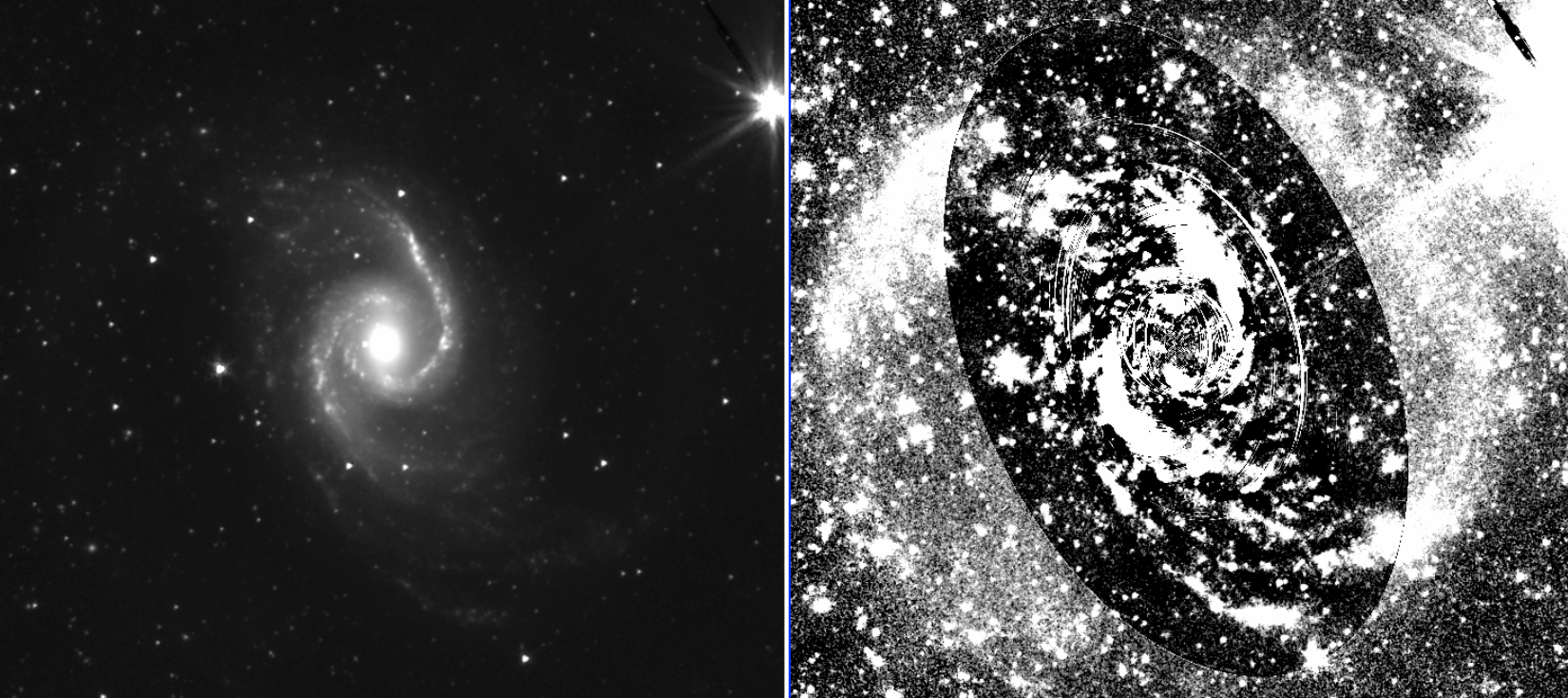}
\includegraphics[trim=0.0cm 0cm 0.0cm 0cm, height=0.26\textwidth, angle=0]{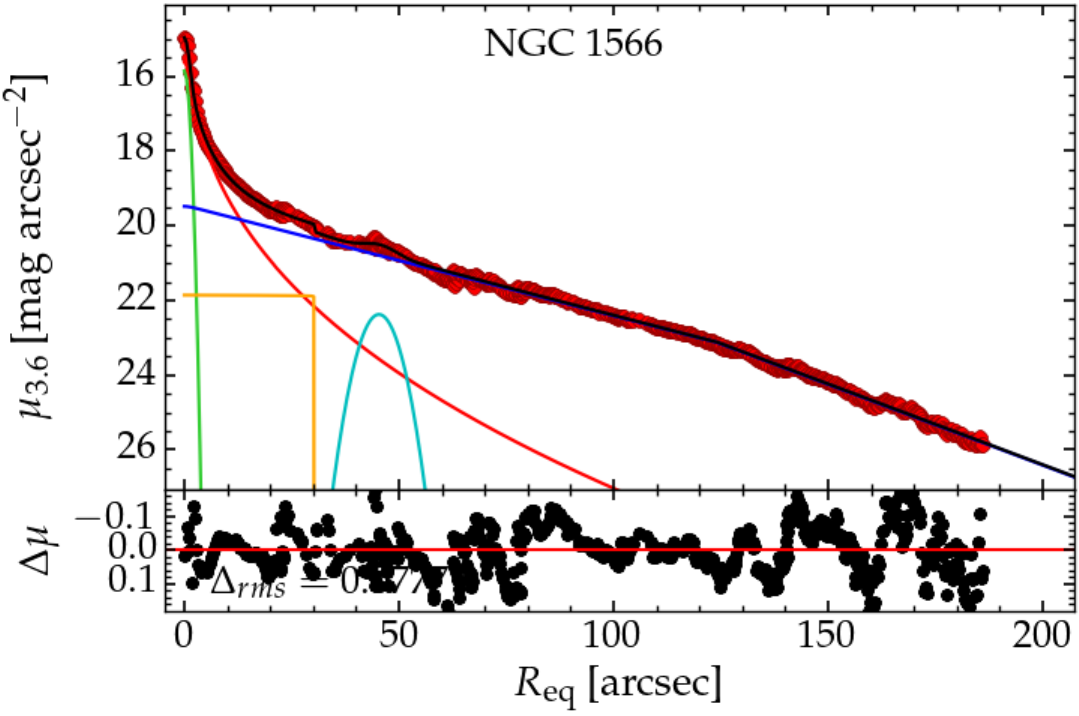}
\caption{NGC~1566.  \textit{SST} 3.6~$\mu$m image (left) and residual image (middle) after subtraction of the truncated galaxy model, which has an equivalent-axis, i.e., geometric-mean axis, radial extent of $\sim$188$\arcsec$ (corresponding to a major-axis radius of 243$\arcsec$).
Right: A Gaussian component (cyan curve) has been added at a geometric-mean axis radius of $\sim$45$\arcsec$ to account for the spiral arms in the `truncated disc' (bent blue line). In addition to the S\'ersic bulge (red) and faint Ferrers' bar (orange), a point source \citep[green:][]{2014MNRAS.441.3570G} is included for the nuclear star cluster and weak AGN. 
The galaxy's apparent 3.6~$\mu$m magnitude is 9.37~mag (AB).  
The $B/T$ ratio is given in Table~\ref{Table-extra}. 
}
\label{Fig_N1566}
\end{center}
\end{figure*}

\begin{figure*}
\begin{center}
\includegraphics[trim=0.0cm 0cm 0.0cm 0cm, height=0.26\textwidth, angle=0]{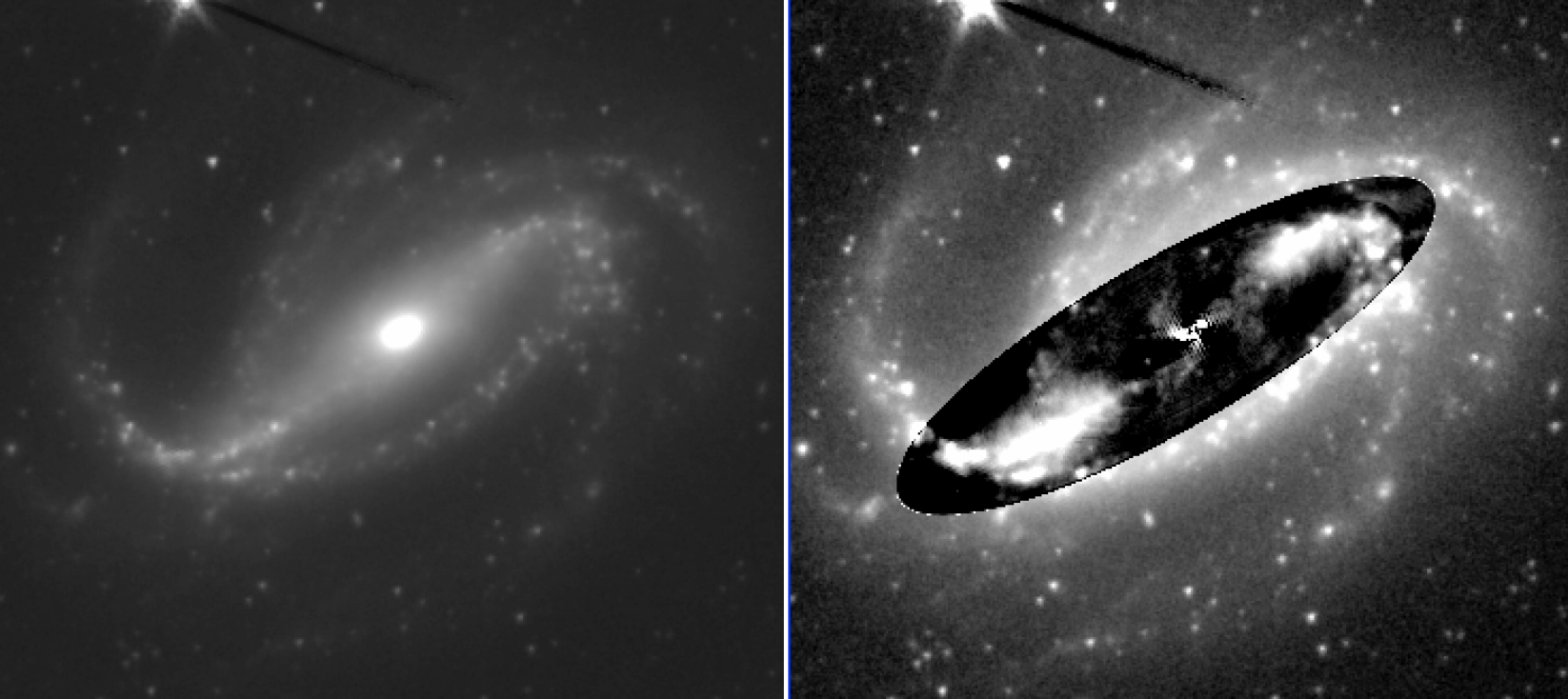}
\includegraphics[trim=0.0cm 0cm 0.0cm 0cm, height=0.26\textwidth, angle=0]{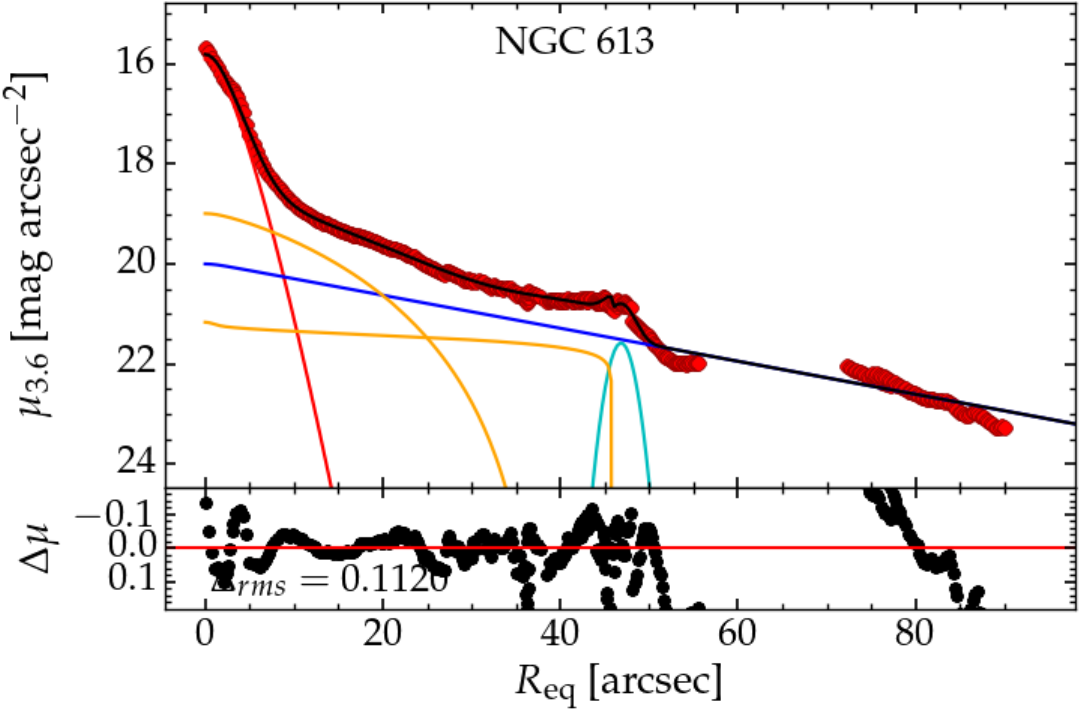}
\caption{Similar to Figure~\ref{Fig_N1566} but for NGC~613.  The residual image was created using a model extending to a geometric-mean axis radius of $\sim$56$\arcsec$ (corresponding to a major-axis radius of $\sim$108$\arcsec$).
Right: A Gaussian component (cyan) has been added at a geometric-mean axis radius of $\sim$45$\arcsec$ to account for the head of the spiral arms at the end of the bar, which is shown as two orange components due to the presence of an X-shaped barlens. A S\'ersic bulge (red) and a non-truncated exponential disc (blue) roughly approximate the disc light.  Dust and a star-forming nuclear ring \citep{2008AJ....135..479B, 2014MNRAS.438..329F} adds some complexities that have been ironed over here with the S\'ersic bulge.
The galaxy magnitude is 9.74~mag.  See the text in Appendix~\ref{App_fits} for further discussion. 
} 
\label{Fig_N613}
\end{center}
\end{figure*}

\begin{figure*}
\begin{center}
\includegraphics[trim=0.0cm 0cm 0.0cm 0cm, height=0.25\textwidth, angle=0]{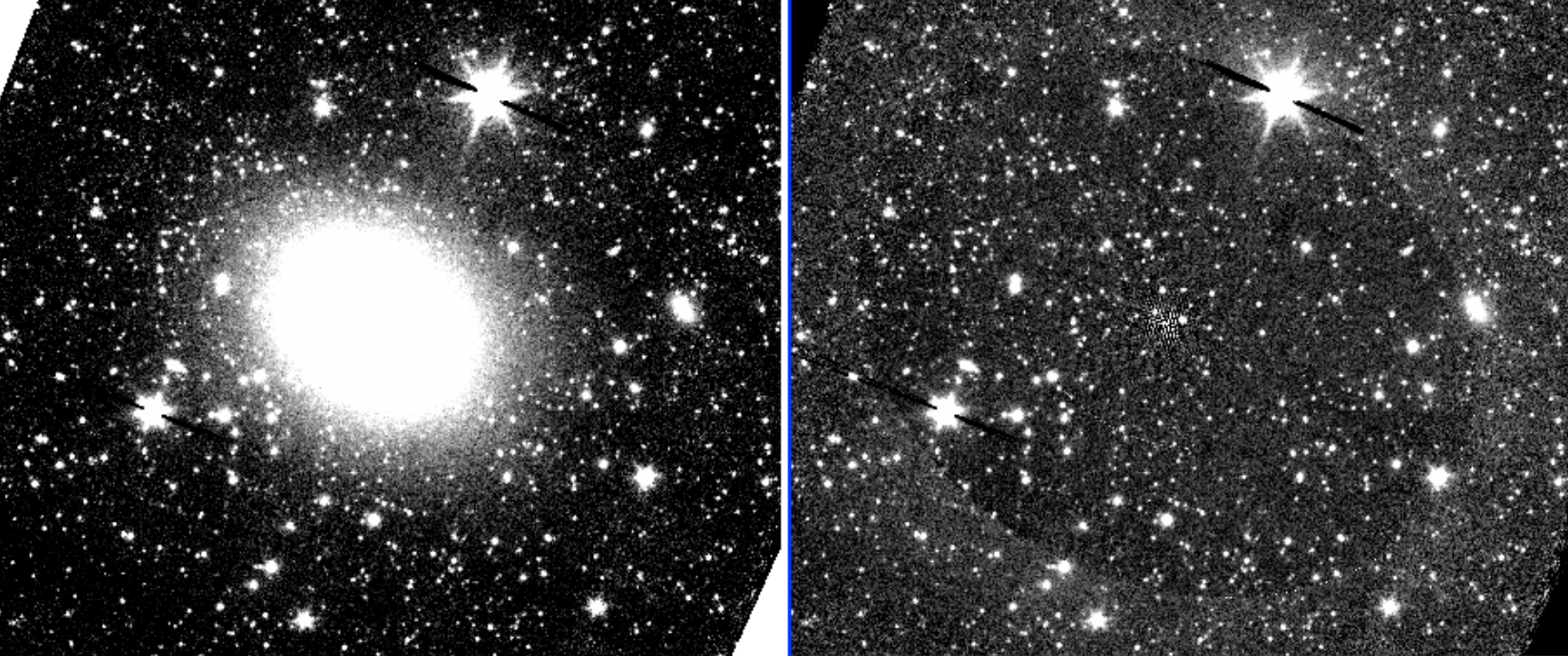}
\includegraphics[trim=0.0cm 0cm 0.0cm 0cm, height=0.25\textwidth, angle=0]{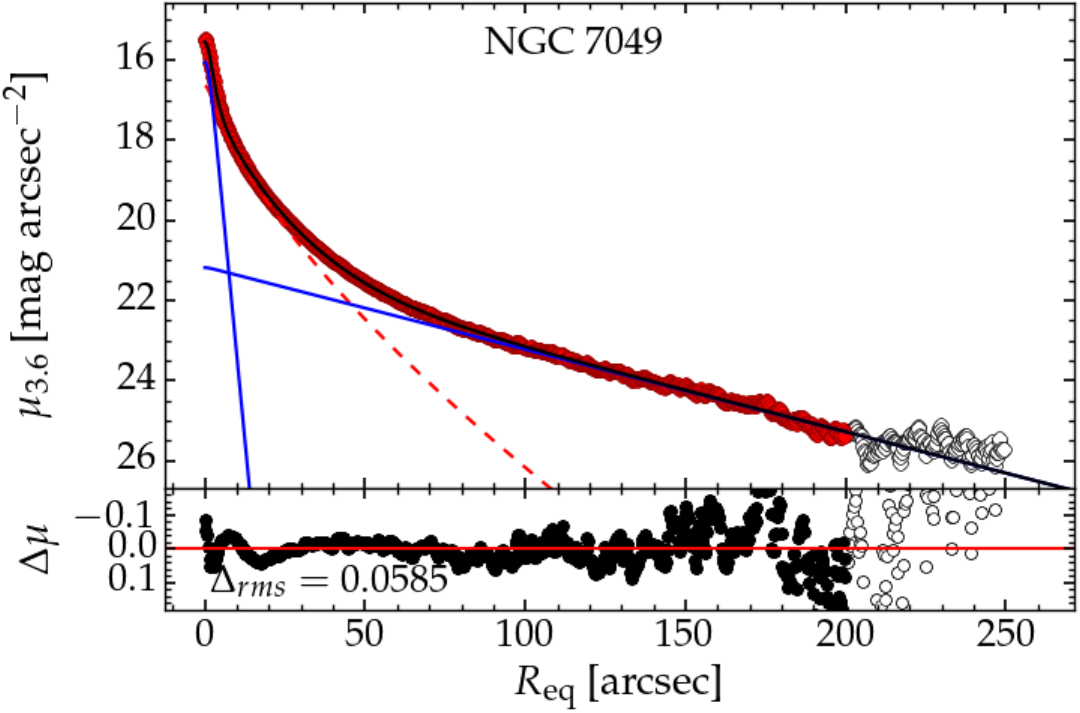}
\caption{Similar to Figure~\ref{Fig_N1566} but for NGC~7049.  The residual image was created using a model extending to a geometric-mean axis radius of $\sim$240$\arcsec$ (corresponding to a major-axis radius of $\sim$374$\arcsec$).
Right: A S\'ersic bulge (dashed red curve) plus large-scale exponential disc (blue line) and inner exponential function \citep[steep blue line: ][]{2019A&A...625A..62T} have been used to describe the light profile.
The galaxy magnitude is 9.57~mag. 
}
\label{Fig_N7049}
\end{center}
\end{figure*}

\begin{figure*}
\begin{center}
\includegraphics[trim=0.0cm 0cm 0.0cm 0cm, height=0.27\textwidth, angle=0]{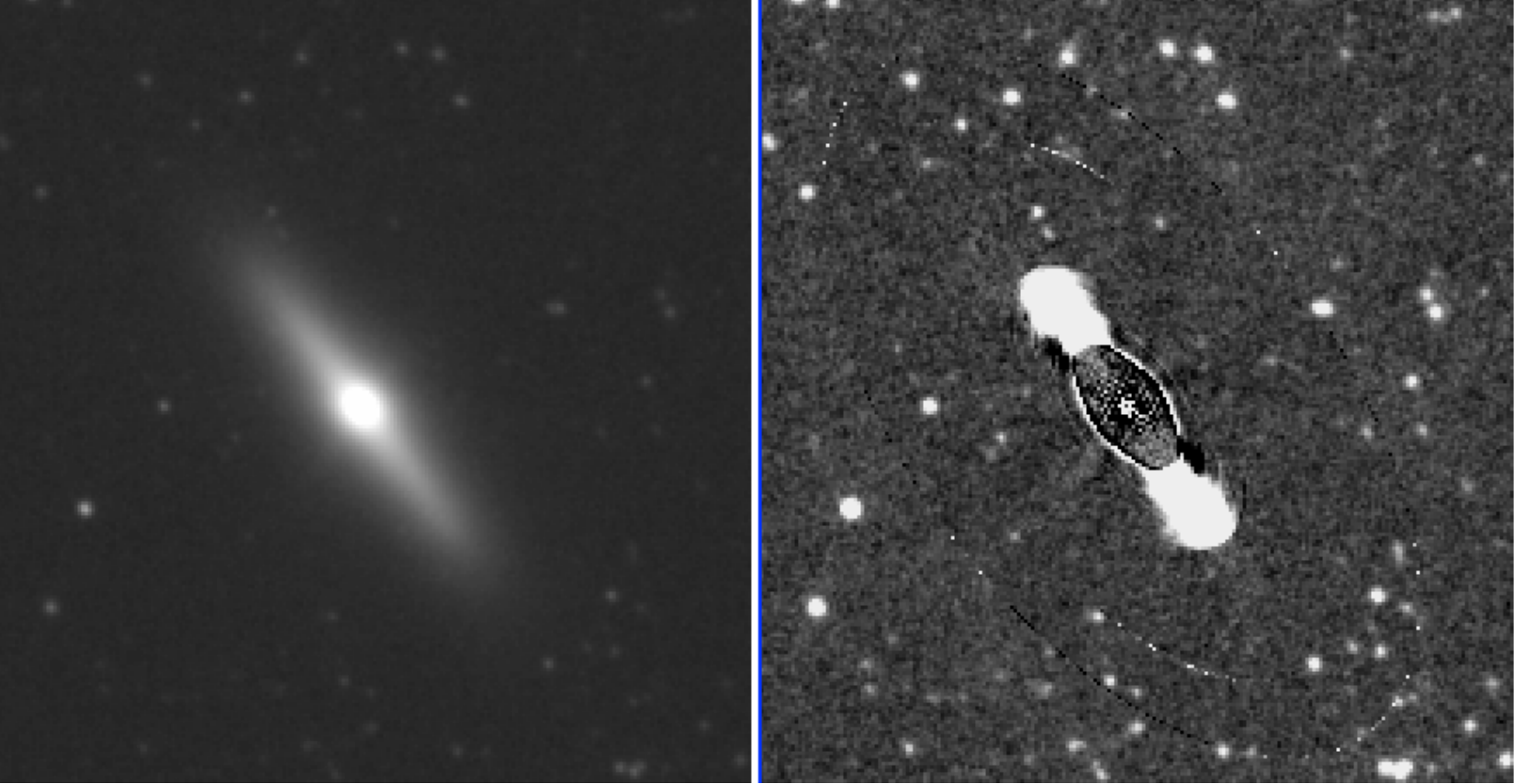}
\includegraphics[trim=0.0cm 0cm 0.0cm 0cm, height=0.27\textwidth, angle=0]{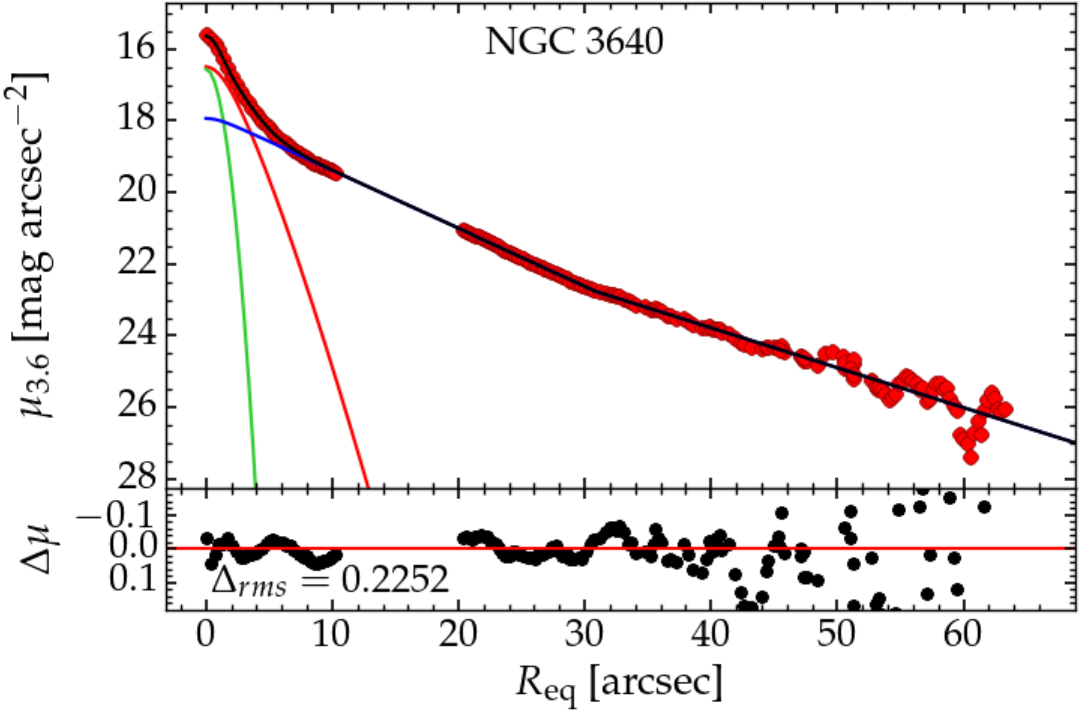}
\caption{Similar to Figure~\ref{Fig_N1566} but for the merger NGC~3640 \citep{2009AJ....138.1417T}.  The residual image was created using a model extending to a geometric-mean axis radius of $\sim$64$\arcsec$ (corresponding to a major-axis radius of $\sim$117$\arcsec$).  Some of the disc was not captured in the modelling, and therefore the data from $22-48\arcsec$ (geometric-mean axis) is excluded from the light profile.
Right: A point-source (green), S\'ersic bulge (red), and anti-truncated disc (blue).  Excluding the point-source component from the modelling increases the $B/T$ ratio from 0.19 to 0.24.
The galaxy magnitude is 11.32~mag.
} 
\label{Fig_N3640}
\end{center}
\end{figure*}

\begin{figure*}
\begin{center}
\includegraphics[trim=0.0cm 0cm 0.0cm 0cm, height=0.27\textwidth, angle=0]{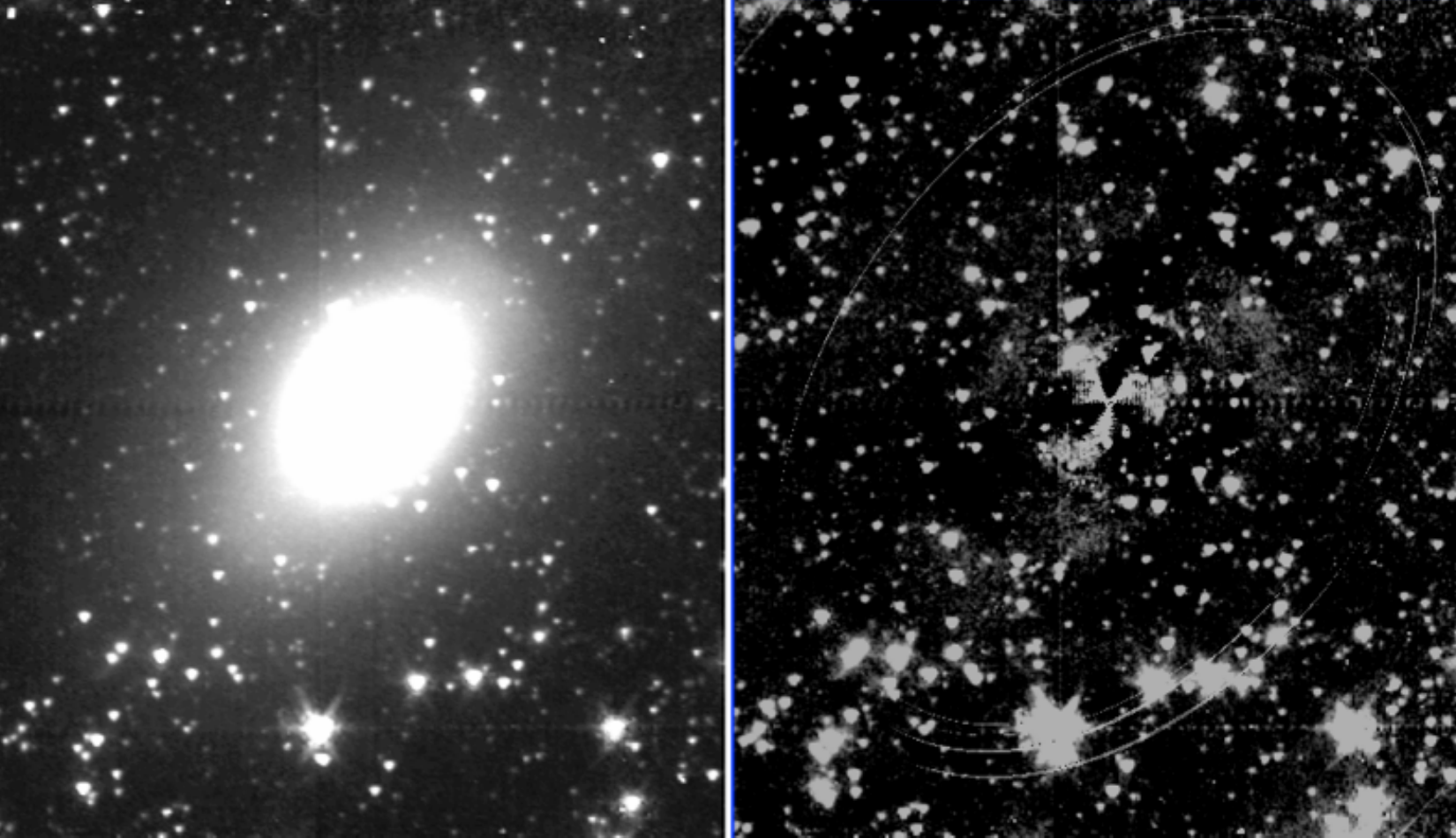}
\includegraphics[trim=0.0cm 0cm 0.0cm 0cm, height=0.27\textwidth, angle=0]{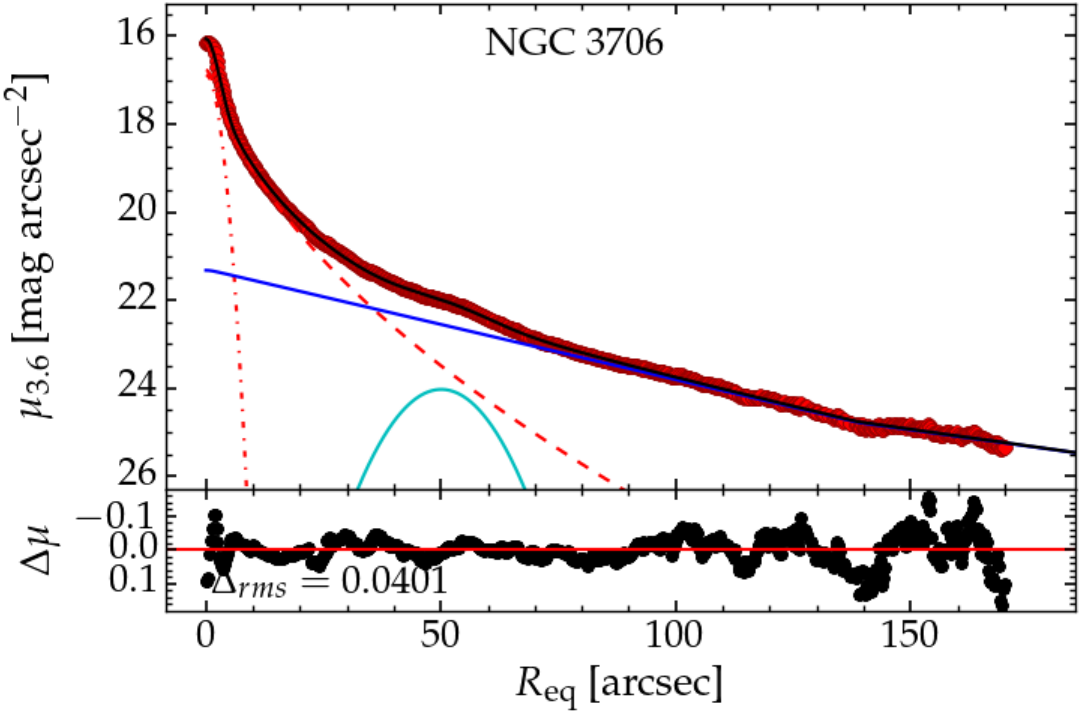}
\caption{Similar to Figure~\ref{Fig_N1566} but for NGC~3706.  The image size is roughly $283\arcsec$ wide by $343\arcsec$ high.
Right: Components are an inner S\'ersic for the $\sim$0$\farcs$7 nuclear disc \citep[dot-dash:][]{2014ApJ...781..112G}, a S\'ersic bulge (dashed red curve), an anti-truncated disc (blue), and a Gaussian (cyan) for a shelf-like feature in the image and light profile whose contribution peaks at $R_{\rm major}\approx70\arcsec$. 
The faint three-pronged spiral at the nucleus (middle panel) is an artifact, not a real feature. 
The galaxy magnitude is 10.14~mag. 
}
\label{Fig_N3706}
\end{center}
\end{figure*}

\begin{figure*}
\begin{center}
\includegraphics[trim=0.0cm 0cm 0.0cm 0cm, height=0.27\textwidth, angle=0]{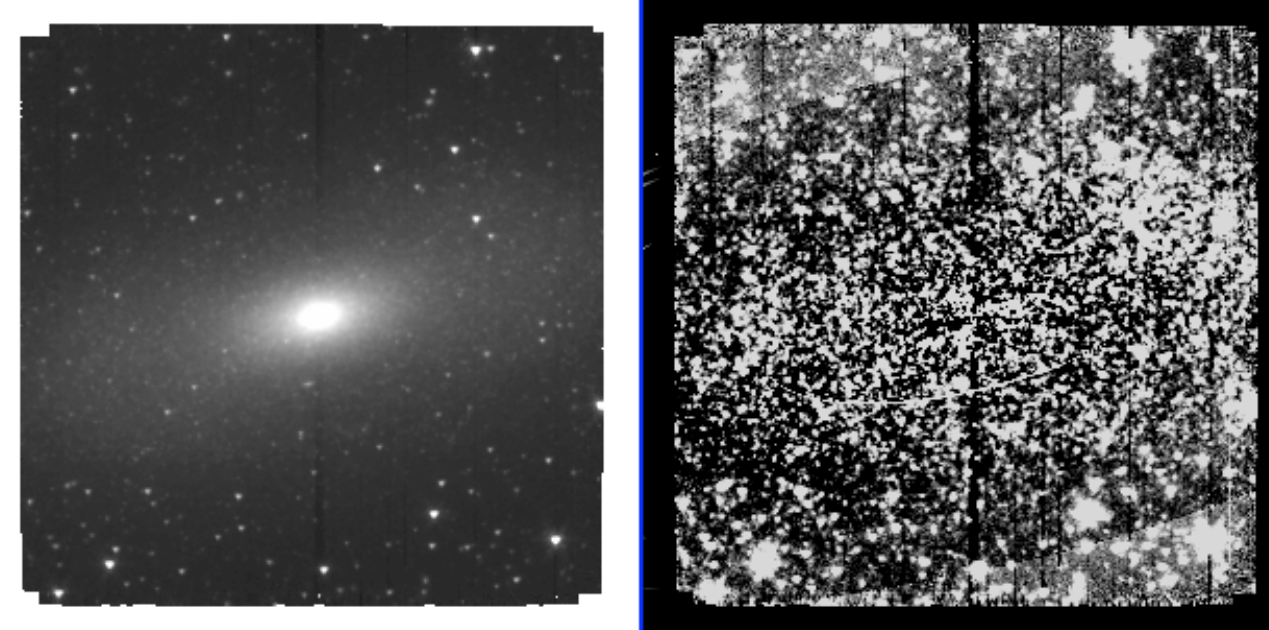}
\includegraphics[trim=0.0cm 0cm 0.0cm 0cm, height=0.27\textwidth, angle=0]{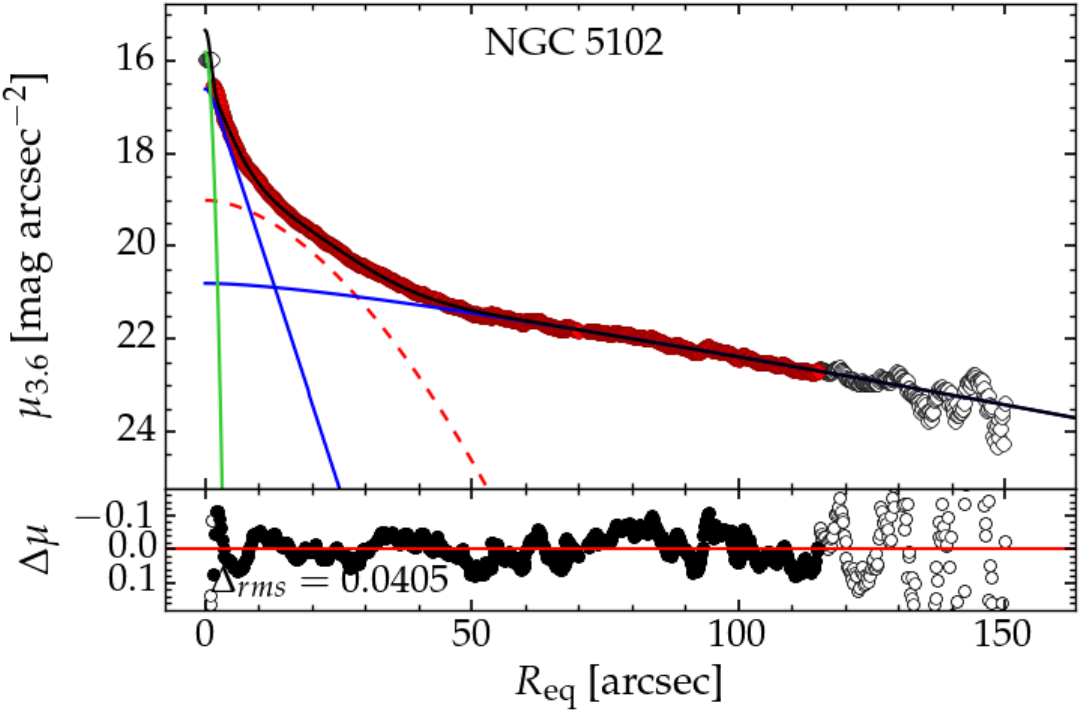}
\caption{Similar to Figure~\ref{Fig_N1566} but for NGC~5102.  The image size is roughly $333\times333\arcsec$.
Right:  A S\'ersic bulge (red), an inclined disc (blue), and a central star cluster point-source \citep[green:][]{2018ApJ...858..118N} and inner exponential disc \citep[blue:][]{2017MNRAS.464.4789M} describe the profile.  
The galaxy magnitude is 9.41~mag. 
} 
\label{Fig_N5102}
\end{center}
\end{figure*}

\begin{figure*}
\begin{center}
\includegraphics[trim=0.0cm 0cm 0.0cm 0cm, height=0.26\textwidth, angle=0]{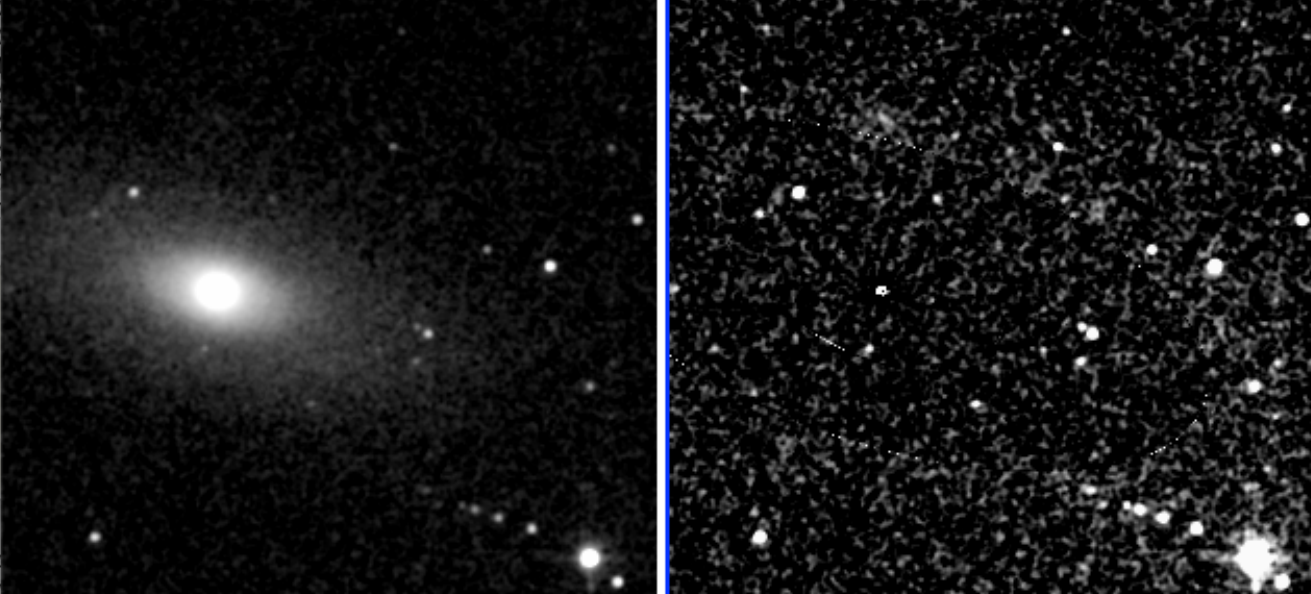}
\includegraphics[trim=0.0cm 0cm 0.0cm 0cm, height=0.26\textwidth, angle=0]{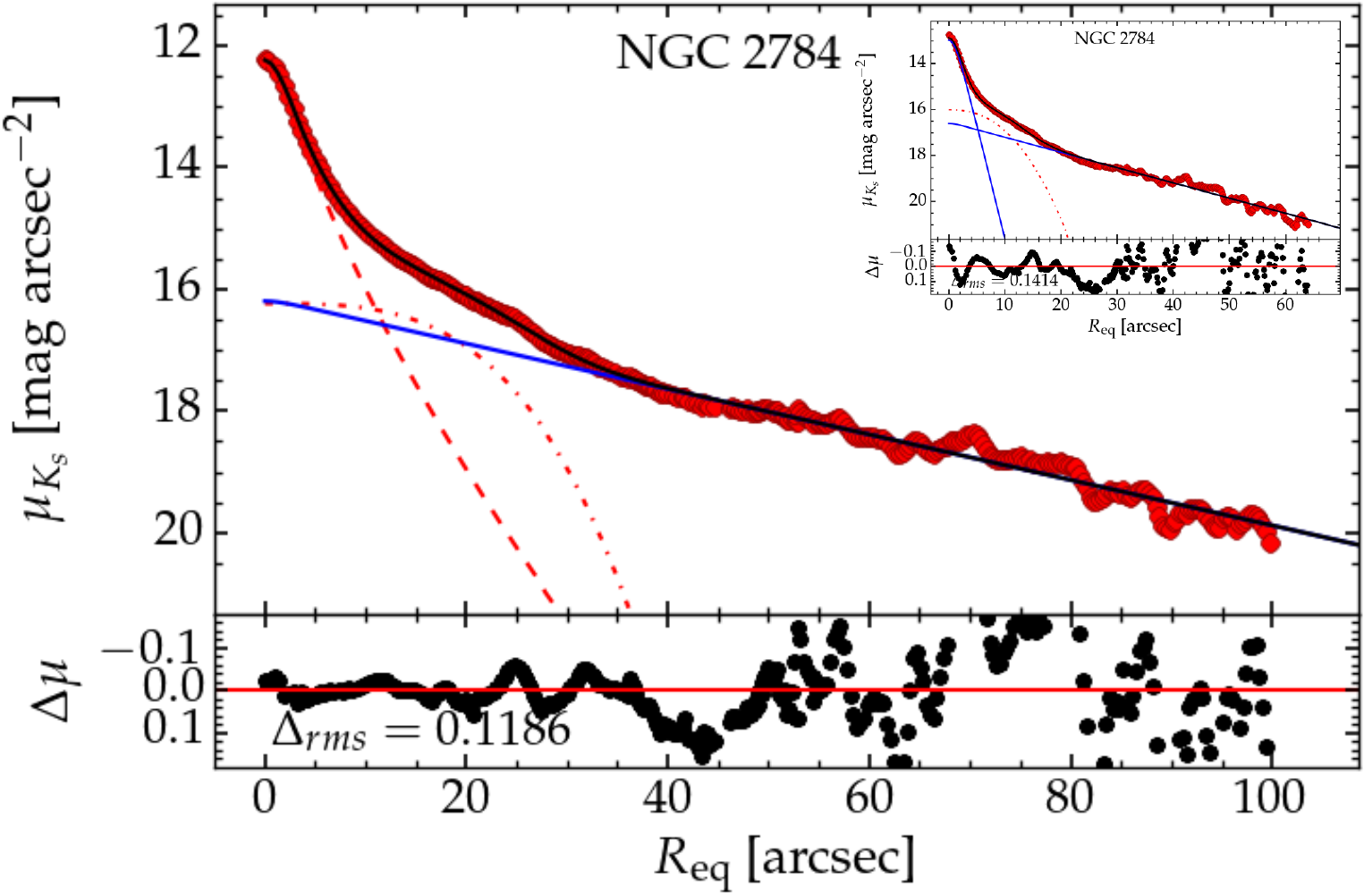}
\caption{\textit{2MASS} $K_s$-band light profile for NGC~2784.  The image size is roughly $318\times306\arcsec$.
Right: S\'ersic bulge (dashed red curve) surrounded by a S\'ersic oval/lens \citep[dotted red line:][]{2018ApJ...862..100G}, and a large-scale exponential disc (blue line).
The fraction of bulge light (0.23) is greater than that of the oval/lens (0.15).
However, if the `lens' model has been fit to what is actually the bulge, and the bulge model has been fit to what is actually an inner disc, then one desires the results from fitting an inner exponential plus a S\'ersic bulge plus a large-scale disc, which yields a similar $B/T$ ratio of 0.20. 
The galaxy $K_s$-band magnitude is 6.30~mag (Vega). 
} 
\label{Fig_N2784}
\end{center}
\end{figure*}

\begin{figure*}
\begin{center}
\includegraphics[trim=0.0cm 0cm 0.0cm 0cm, height=0.25\textwidth, angle=0]{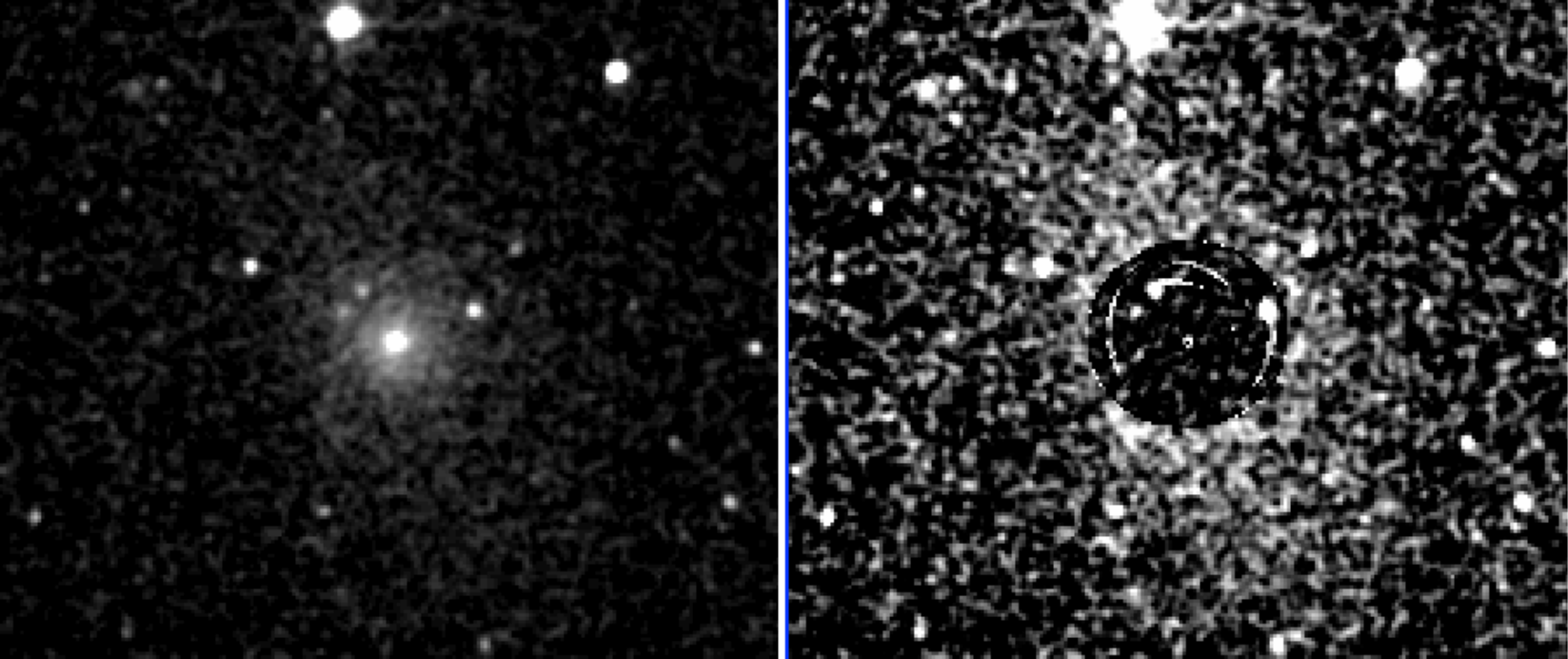}
\includegraphics[trim=0.0cm 0cm 0.0cm 0cm, height=0.25\textwidth, angle=0]{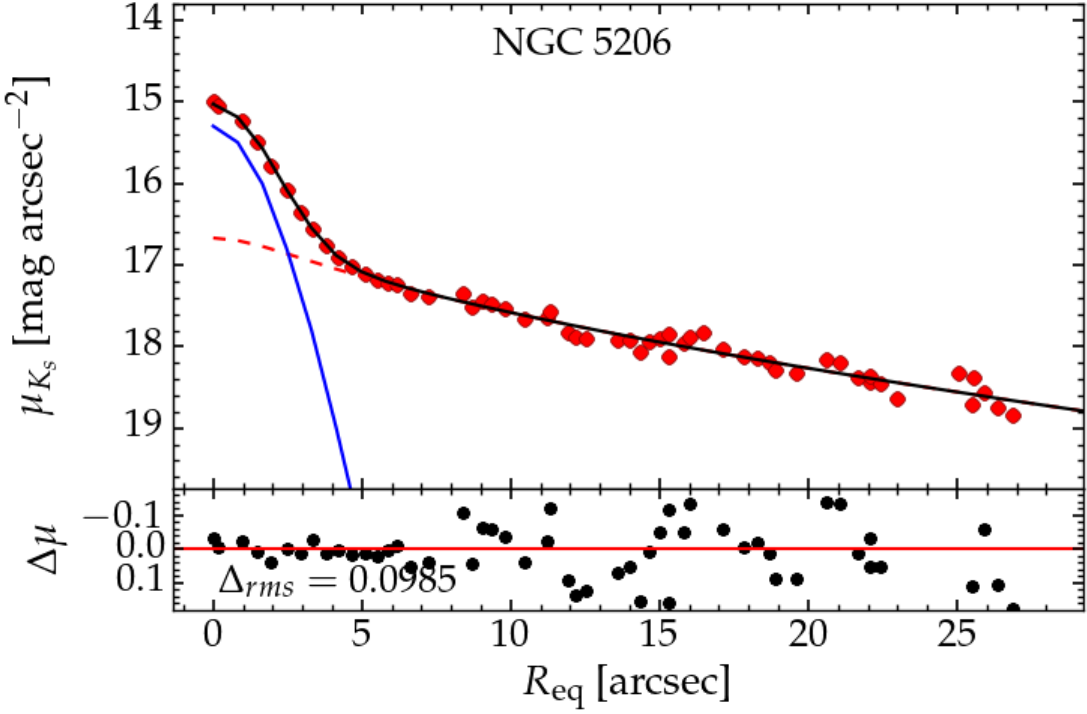}
\caption{Similar to Figure~\ref{Fig_N2784} but for NGC~5206.  The image size is roughly $223\times192\arcsec$.
Right:  A S\'ersic bulge/galaxy (dashed red curve) plus an inner exponential \citep[blue:][]{2018ApJ...858..118N} describe the profile.
The galaxy $K_s$-band magnitude is 8.40~mag.
}
\label{Fig_N5206}
\end{center}
\end{figure*}

New decompositions of galaxy light profiles are provided for eight galaxies (Figures~\ref{Fig_N1566}--\ref{Fig_N5206}).
The data and the process are described in \citet{2019ApJ...873...85D} and \citet{2019ApJ...876..155S}.  The relatively new IRAF task
{\sc Isofit} \citep{2015ApJ...810..120C} was used to extract geometric mean (aka 
equivalent-axis) light profiles, which were modelled using a Python~3 version of the \textsc{Profiler} code \citep{2016PASA...33...62C}.\footnote{The \textsc{Profiler} code was updated to Python3 by the author for this work.} 
The results are shown here, in AB mag for the \textit{SST} data and Vega mag for the \textit{2MASS} data,
and the spheroid and galaxy stellar masses are given in Table~\ref{Table-extra}. 

Notes: The bar component (fit with a Ferrers function in Figure~\ref{Fig_N613}) for NGC~613 is arguably somewhat longer than the actual bar.  At the same time, the Gaussian function fit there is arguably not broad enough to capture the width of the spiral arms emanating from near the end of the actual bar. 
Considerable stellar knots, and perhaps some bar light, can be seen in the residual image. Including this light would act to reduce the slight `sag' in the light profile from $\sim$ 25-40$\arcsec$.  The model bar component, which contains 15 per cent of the galaxy light, would be roughly half as bright again if these stellar knots that appear to trace the spiral arm were considered part of the actual bar. The shorter barlens contains roughly 12 per cent of the galaxy light.
A second, shorter Ferrers function was used to capture/describe this barlens.  Although NGC~613 may have a truncated disc, as evidenced by the steeper decline in the data beyond $\sim$70$\arcsec$, too many parameter degeneracies prevented its quantification. 

`Lenses' and `ovals' are generally interpreted as disc-related stellar structures formed through secular evolution. While frequently associated with bars, lenses are also observed in unbarred galaxies, consistent with either bar dissolution or formation through weaker non-axisymmetric disc instabilities \citep{2009ApJ...692L..34L, 2015MNRAS.454.3843A}.    
A questionable oval or lens is seen in NGC~2784 (Figure~\ref{Fig_N2784}, but the author cannot rule out the possibility that this might be the bulge component of the galaxy.

NGC~3706 (Figure~\ref{Fig_N3706}) also displays a possible oval/lens, although the abrupt edge --- typical of such components --- primarily resides in just one quadrant of the galaxy image, and the decomposition does not favour a Ferrers or low-$n$ S\'ersic function (with their abrupt decline), which would be typical of a lens.

%%%%%%%%%%%%%%%%%%%%%%%%%%%%%%%%%%%%%%%%%%%%%%%%%%  
% Don't change these lines 
\bsp    % typesetting comment 
\label{lastpage}
\end{document}